\documentclass[a4paper,fleqn]{cas-dc}
\usepackage[numbers]{natbib}

\def\tsc#1{\csdef{#1}{\textsc{\lowercase{#1}}\xspace}}
\tsc{WGM}
\tsc{QE}
\tsc{EP}
\tsc{PMS}
\tsc{BEC}
\tsc{DE}

\usepackage[utf8]{inputenc} 
\usepackage[T1]{fontenc}    
\usepackage{hyperref}       
\usepackage{amsmath, amssymb, amsfonts}
\usepackage{xcolor}
\usepackage{cleveref}
\usepackage{xfrac} 
\usepackage[numbers]{natbib}
\hypersetup{
  colorlinks   = true, 
  urlcolor     = ForestGreen, 
  linkcolor    = BrickRed, 
  citecolor   = blue 
}
\usepackage{nicefrac}       
\usepackage{comment}

\usepackage{subfigure} 

\usepackage{framed,multirow}

\usepackage{algorithm}
\usepackage{algpseudocode}
\usepackage{graphicx}
\usepackage{mathrsfs}

\usepackage{latexsym}
\usepackage{bm}
\usepackage{amssymb}
\usepackage{array}
\usepackage{multirow}
\usepackage[english]{babel}

\usepackage{footnote}
\usepackage{tabularx}
\makesavenoteenv{tabular}
\makesavenoteenv{table}
\usepackage{relsize}

\usepackage{cleveref}
\crefname{equation}{Eq.}{Eqs.}
\crefname{table}{Table}{Tables}
\crefname{figure}{Fig.}{Figs.}
\crefname{section}{Section}{Sections}
\crefname{subsection}{Section}{Secs.}
\Crefname{figure}{Figs.}{Figs.}
\Crefname{algorithm}{Algorithm}{Algorithms}

\usepackage{caption}

\usepackage{mathtools}
\usepackage{xfrac}
\usepackage{amsthm}
\newtheorem{remark}{Remark}
\newtheorem{lemma}{Lemma}

\usepackage[dvipsnames]{xcolor}
\definecolor{newcolor}{rgb}{.8,.349,.1}

\newcommand{\largesfrac}[2]{{\large\nicefrac{#1}{#2}}}

\newcolumntype{C}[1]{>{\centering\arraybackslash}p{#1}}

\begin{document}
\let\WriteBookmarks\relax
\def\floatpagepagefraction{1}
\def\textpagefraction{.001}


\shortauthors{E. Eshra et~al.}

\title [mode = title]{A direct importance sampling-based framework for rare event uncertainty quantification in non-Gaussian spaces}

\author[1]{Elsayed Eshra}

\cormark[1]

\ead{eme5375@psu.edu}

\author[1]{Konstantinos G. Papakonstantinou}
\cormark[1]
\ead{kpapakon@psu.edu}
\author[2]{Hamed Nikbakht}

\affiliation[1]{organization={The Pennsylvania State University},
    city={University Park, PA 16802},
    country={USA}}

\affiliation[2]{organization={Moody's},
    city={Newark, CA 94560},
    country={USA}}

\cortext[cor1]{Corresponding authors.}


\begin{abstract}
This work introduces a novel framework for precisely and efficiently estimating rare event probabilities in complex, high-dimensional non-Gaussian spaces, building on our foundational Approximate Sampling Target with Post-processing Adjustment (ASTPA) approach. An unnormalized sampling target is first constructed and sampled, relaxing the optimal importance sampling distribution and appropriately designed for non-Gaussian spaces.  Post-sampling, its normalizing constant is estimated using a stable inverse importance sampling procedure, employing an importance sampling density based on the already available samples. The sought probability is then computed based on the estimates evaluated in these two stages. The proposed estimator is theoretically analyzed, proving its unbiasedness and deriving its analytical coefficient of variation. To sample the constructed target, we resort to our developed Quasi-Newton mass preconditioned Hamiltonian MCMC (QNp-HMCMC) and we prove that it converges to the correct stationary target distribution. To avoid the challenging task of tuning the trajectory length in complex spaces, QNp-HMCMC is effectively utilized in this work with a single-step integration. We thus show the equivalence of QNp-HMCMC with single-step implementation to a unique and efficient preconditioned Metropolis-adjusted Langevin algorithm (MALA). An optimization approach is also leveraged to initiate QNp-HMCMC effectively, and the implementation of the developed framework in bounded spaces is eventually discussed. A series of diverse problems involving high dimensionality (several hundred random variables), strong nonlinearity, and non-Gaussianity is presented, showcasing the capabilities and efficiency of the suggested framework and demonstrating its advantages compared to relevant state-of-the-art sampling methods.
\end{abstract}




\maketitle

\section{Introduction}
\noindent This work introduces a new framework for directly estimating rare event probabilities in complex, high-dimensional non-Gaussian spaces. Rare event uncertainty quantification is a pivotal component in contemporary decision-making across diverse domains. It is particularly crucial for assessing and mitigating risks tied to infrequent yet highly impactful rare event occurrences, such as structural failures induced by extreme loads like earthquakes and hurricanes. This field, commonly referred to as reliability estimation in engineering, has a long history, and comprehensive insights into its complexities can be read in \citep{ditlevsen1996structural,nikolaidisengineering,lemaire2009structural,au2014engineering,melchers2018structural,der2022structural}. In many practical applications, rare event probabilities are fortunately exceptionally low, often within the range of $10^{-4}$ to even $10^{-9} $ and lower, posing significant numerical and mathematical challenges. Further complicating these challenges is the reliance on computationally expensive models of the involved systems with high-dimensional random variable spaces. Achieving accurate estimates while minimizing model evaluations is thus a critical aspect.\par

Numerous existing solution techniques are specifically designed for Gaussian spaces and typically seek to transform non-Gaussian random variables to Gaussian ones. However, when such Gaussian transformations are not possible, these techniques often encounter challenges in non-Gaussian spaces due to potential asymmetries, complex dependencies, intricate geometries, and constrained spaces. Approximation methods, such as First Order Reliability Method (FORM) \citep{rackwitz2001reliability,breitung201540}, Second Order Reliability Method (SORM) \citep{breitung1984asymptotic}, and Large deviation theory (LDT) \citep{asmussen2007stochastic,dembo2009large, grafke2019numerical}, provide asymptotic expressions for estimating rare event probabilities \citep{breitung2006asymptotic}. In general, these methods lose accuracy and/or practicality for arbitrary non-Gaussian distributions, and may also exhibit considerable errors in problems with moderately to highly nonlinear performance functions \citep{rackwitz2001reliability,valdebenito2010role}. Sampling methods can tackle the problem in its utmost generality, however, with a higher computational cost. In addressing the known limitations of the crude Monte Carlo approach \cite{bai2023curse}, numerous sophisticated sampling methods have been proposed. Among others, the Subset Simulation (SuS) approach, originally presented in \citep{au2001estimation}, has been widely used and continuously enhanced for Gaussian spaces \citep{zuev2011modified,zuev2012bayesian,au2016rare,papaioannou2015mcmc},  given its potential to handle high-dimensional problems, despite some limitations \citep{breitung2019geometry}. Several more recent works have also shown enhanced performance in non-Gaussian spaces by mainly employing advanced samplers within SuS, such as Hamiltonian MCMC \citep{neal2011mcmc,wang2019hamiltonian}, Riemannian Manifold Hamiltonian Monte Carlo (RMHMC) \citep{girolami2011riemann,chen2022riemannian}, Hamiltonian Neural Networks (HNN) \cite{thaler2024reliability}, and affine invariant ensemble MCMC sampler \citep{goodman2010ensemble,shields2021subset}. It should be noted that Subset Simulation also has similarities with multilevel splitting methods \citep{guyader2011simulation,cerou2012sequential,walter2015moving}. Another sampling method capable of efficiently estimating rare event probabilities is importance sampling (IS), a well-known variance-reduction approach. Since this work builds upon the theoretically rigorous foundations of importance sampling, we provide a comprehensive review and discussion of IS techniques for rare event estimation in \cref{rare_event_est}. Combining surrogate models, approximating computationally expensive original models, with sampling-based approaches has also demonstrated significant savings in computational cost across various applications \citep{huang2016assessing, marelli2018active, moustapha2022active}; however, their applicability to high-dimensional, complex problems still faces practical challenges.\par

In this paper, we introduce a novel framework for efficiently and precisely estimating rare event probabilities directly in non-Gaussian spaces, building on our foundational Approximate
Sampling Target with Post-processing Adjustment (ASTPA) approach, recently developed for Gaussian spaces in \citep{Papakon2023HMCMC}. As shown in this work, the developed framework demonstrates excellent performance in challenging high-dimensional non-Gaussian scenarios, thereby overcoming a significant limitation of numerous existing methods. In essence, rare event probability estimation is a normalizing constant estimation problem for the joint distribution of the involved random variables truncated over the rare event domain, also known as the optimal sampling target; a demanding problem traditionally approached through sequential techniques. Conversely, the ASTPA approach directly and innovatively addresses this challenge by decomposing the problem into two less demanding estimation problems. Hence, the proposed methodology first constructs an unnormalized sampling target, appropriately designed for non-Gaussian spaces, relaxing the optimal importance sampling distribution. To sample this constructed target adequately, our Quasi-Newton mass preconditioned Hamiltonian MCMC (QNp-HMCMC) algorithm is employed, particularly suitable for complex spaces, as discussed later. The obtained samples are subsequently utilized not only to compute a shifted estimate of the sought probability but also to guide the second ASTPA stage. Post-sampling, the normalizing constant of the approximate sampling target is computed through the inverse importance sampling (IIS) technique \citep{Papakon2023HMCMC}, which utilizes an importance sampling density (ISD) fitted based on the samples drawn in the first stage. A practical approach augmenting the IIS stability in challenging non-Gaussian settings, by avoiding anomalous samples, is also devised in this work. The rare event probability is eventually computed by utilizing this computed normalizing constant to correct the shifted estimate of the first stage. The statistical properties of the proposed estimator are thoroughly derived, including its proven unbiasedness and analytical coefficient of variation (C.o.V). Notably, the derived analytical C.o.V expression, accompanied by an applied implementation technique based on the effective sample size, has demonstrated accurate agreement with numerical results.

Hamiltonian Markov Chain Monte Carlo (HMCMC) is one of the most successful and powerful MCMC algorithms \citep{neal2011mcmc,hoffman2021adaptive}, based, as the name suggests, on Hamiltonian dynamics \citep{duane1987hybrid}. Its performance largely depends on tuning three parameters: the mass matrix, the simulation step size, and the length of the simulated trajectories. Its efficiency in high-dimensional complex spaces can be significantly improved by selecting an appropriate mass matrix \citep{girolami2011riemann, zhang2011quasi, fu2016quasi}. Therefore, in this work, we rigorously analyze a Quasi-Newton mass preconditioned HMCMC (QNp-HMCMC) sampling algorithm, an approach we initially conceptualized in our earlier work \citep{Papakon2023HMCMC,nikbakht2019HMCMC}. QNp-HMCMC effectively and uniquely leverages the geometric information of the target distribution. During the burn-in stage, a skew-symmetric preconditioning scheme for the Hamiltonian dynamics is adopted \citep{ma2015complete}, incorporating Hessian information merely approximated based on the already available gradients. The estimated Hessian is subsequently utilized to provide an informed, preconditioned mass matrix, better describing the scale and correlation of the involved random variables. To reinforce its theoretical basis, we prove here that simulating the skew-symmetric preconditioned dynamics in QNp-HMCMC leads to the correct stationary target distribution. To effectively implement QNp-HMCMC, we provide detailed recommendations for approximating the Hessian information in complex non-Gaussian spaces. The simulation step size in QNp-HMCMC is automatically tuned using the dual averaging algorithm \citep{hoffman2014no,nesterov2009primal}. Equipped with the preconditioned dynamics and mass matrix, QNp-HMCMC is characterized by large informed sampling steps, enabling it to work efficiently even with single-step simulated Hamiltonian dynamics, avoiding the challenging task of tuning its trajectory length in general, intricate domains, and enabling increased computational efficiency by minimizing the number of model calls (gradient evaluations). As proven in this work, this QNp-HMCMC version with a single-step implementation is equivalent to an original preconditioned Metropolis-adjusted Langevin algorithm (MALA). Finally, an optimization approach is also devised in this paper to initiate the QNp-HMCMC sampling effectively.

This paper is structured as follows. \cref{rare_event_est} offers a comprehensive discussion on existing importance sampling-based methods for rare event probability estimation, highlighting their common characteristics and, notably, their connections and differences with our ASTPA approach. The developed framework is subsequently presented in \cref{ASTPA_sec}, and \cref{Sampling_target} discusses in detail the single-step QNp-HMCMC algorithm. \cref{Discovery} introduces an effective optimization approach for rare event domain discovery, specifically using the Adam optimizer \citep{kingma2014adam}, aiming to provide good initial states for the MCMC sampler. Given that it is more practical for MCMC and HMCMC algorithms to operate in unconstrained spaces, \cref{bounded} discusses transforming bounded random variables to unconstrained ones. A summary of the QNp-HMCMC-based ASTPA framework for non-Gaussian domains is then presented in \cref{ASTPA_summary}. To fully examine the capabilities of the proposed methodology, in \cref{Numerical_results} its performance is demonstrated and successfully compared against the state-of-the-art Subset Simulation, and related variants, in a series of challenging low- and high-dimensional non-Gaussian problems. The paper then concludes in \cref{conclusion_sec}.

\section{Rare Event Probability Estimation}\label{rare_event_est}
\noindent Let $\bm{X}$ be a continuous random vector taking values in $\mathcal{X} \subseteq \mathbb{R}^d$ and having a joint probability density function (PDF) $\pi_{\bm{X}}$. We are interested in rare events $\mathcal{F} \coloneqq \{ \bm{x} \,:\, g(\bm{x}) \leq 0\}$, where $g: \mathcal{X} \rightarrow \mathbb{R}$ is a performance function, also known as the limit-state function, defining the occurrence of a rare event. In the context of reliability estimation, which considers failures as rare events, the limit state function can be indicatively formulated as $g(\bm{X})=\lambda - \mathcal{M}(\bm{X})$, where $\mathcal{M}(\bm{X})$ denotes a model output (a quantity of interest). Here, the predefined threshold $\lambda$ determines the rarity of the failure event by controlling when $g(\bm{X})$ starts to take non-positive values, i.e., rare event occurrence. In this work, we aim to estimate the rare event probability $p_\mathcal{F}$:
\begin{equation}  \label{p_f}
p_\mathcal{F}=  \int_{\mathcal{F}} \pi_{\bm{X}}(\bm{x}) d\bm{x} = 
  \int_{\mathcal{X}} I_{\mathcal{F}} (\bm{x})\,\pi_{\bm{X}}(\bm{x}) d\bm{x} = {\mathop{\mathbb{E}}}_{\pi_{\bm{X}}} [ I_{\mathcal{F}} (\bm{X})] 
\end{equation} 
where $I_{\mathcal{F}}: \mathcal{X} \rightarrow \{0, 1\}$ is the indicator function, i.e., $I_{\mathcal{F}} (\bm{x})=1$ if $\bm{x} \in \mathcal{F}$, and $I_{\mathcal{F}} (\bm{x})=0$ otherwise, and $\mathop{\mathbb{E}}$ is the expectation operator.\par

Our objective in this work is to estimate the described integration in \cref{p_f} under these challenging yet realistic settings: \textbf{\textit{(i)}} the analytical calculation of \cref{p_f} is generally intractable, as the limit-state function can be complicated, e.g., involve the solution of a differential equation; \textbf{\textit{(ii)}} the computational effort for evaluating $I_{\mathcal{F}}$ for each realization $\bm{x}$ is assumed to be quite significant, often relying on computationally intensive models, necessitating the minimization of such function evaluations (model calls); \textbf{\textit{(iii)}} the rare event probability $p_\mathcal{F}$ is extremely low, e.g., in order of $p_\mathcal{F} \sim 10^{-4} - 10^{-9}$; \textbf{\textit{(iv)}} the random variable space, $\mathcal{X} \subseteq \mathbb{R}^{d}$, is assumed to be high-dimensional, e.g., $d=10^{2}$ and more; \textbf{\textit{(v)}} the joint probability density, $\pi_{\bm{X}}(\bm{x})$, is non-Gaussian, with transformations to the Gaussian domain assumed inapplicable, necessitating \cref{p_f} evaluation directly in non-Gaussian spaces.\par
Under these settings, several sampling methods become highly inefficient and fail to address the problem effectively. For instance, the standard Monte Carlo estimator of \cref{p_f} using $\{\bm{x}_i\}_1^N$ draws from $\pi_{\bm{X}}$, $\hat{p}_\mathcal{F} = \frac{1}{N} \sum_{i=1}^N  I_{\mathcal{F}} (\bm{x}_i)$, has a coefficient of variation $\sqrt{(1-p_\mathcal{F})/(N p_\mathcal{F})}$, rendering this estimate inefficient by requiring a prohibitively large number of samples to accurately quantify small probabilities; see setting \textbf{\textit{(iii)}} above. Importance sampling (IS), as a variance-reduction method, can efficiently compute the integral in \cref{p_f}, through sampling from an appropriate importance sampling density (ISD), $h(\boldsymbol{\bm{X}})$. The ISD places greater importance on rare event domains in the random variable space compared to $\pi_{\bm{X}}(\bm{x})$, satisfying the sufficient condition $\text{supp}(I_{\mathcal{F}}\,\pi_{\bm{X}}) \subseteq \text{supp}(h)$, with $\text{supp}(\cdot)$ denoting the support of the function. This results in an increased number of rare event samples, thereby reducing the number of required model calls. Eq.~\eqref{p_f} can then be written as:
\begin{equation} \label{p_f_I_S}
p_\mathcal{F}= \int_{\mathcal{X}} I_{\mathcal{F}} (\bm{x})\, \dfrac{\pi_{\bm{X}}(\bm{x})}{h(\bm{x})} h(\bm{x}) d\bm{x} = {\mathop{\mathbb{E}}}_{h} \big[ I_{\mathcal{F}} (\bm{X}) \dfrac{\pi_{\bm{X}}(\bm{X})}{h(\bm{X})}\big] 
\end{equation} 
leading to the unbiased IS estimator: 
\begin{equation} \label{p_f_I_S_estimate}
\hat{p}_\mathcal{F}= \frac{1}{N} \sum_{i=1}^N  I_{\mathcal{F}} (\bm{x}_i)\dfrac{\pi_{\bm{X}}(\bm{x}_i)}{h(\bm{x}_i)}
\end{equation} 

The efficiency of IS relies heavily on the careful selection of both the ISD and the sampling algorithm. A theoretically optimal ISD $h^*$, providing zero-variance IS estimator,  can be given as \citep{ang1992optimal}:
\begin{equation}\label{opt_I_S_density}
\begin{aligned}
&h^*(\bm{X}) = \dfrac{1}{p_\mathcal{F}} I_{\mathcal{F}}(\bm{X}) \pi_{\bm{X}}(\bm{X})
\end{aligned}
\end{equation}
Yet, it is evident that this ISD cannot be utilized due to the fact that the normalizing constant in Eq.~\eqref{opt_I_S_density} is the sought target probability $p_\mathcal{F}$. Notably, finding a near-optimal ISD, approximating $h^*$, is an ongoing research pursuit. The existing approaches can be largely classified into two categories. The first directly follows the IS estimator in \cref{p_f_I_S_estimate} and seeks to approximate $h^*$ by a parametric or non-parametric probability density function (PDF), $h$. An early approach in this category utilizes densities centered around one or more design points \citep{schueller1987critical}, provided by an initial FORM analysis, a powerful, practical process, yet with certain limitations for high-dimensional, complex spaces, as discussed in \citep{au2001estimation}. A more recent popular approach is the cross entropy-based importance sampling (CE-IS). It alternatively constructs a parametric ISD $h$ by minimizing the Kullback-Leibler (KL) divergence
between $h^*$ and a chosen parametric family of probability distributions \citep{kurtz2013cross, wang2016cross, papaioannou2019improved, uribe2021cross, demange2023variational}. Recent CE-IS variants incorporate information from influential rare event points, e.g., design points, to effectively initialize the cross entropy process \citep{tong2023large,chiron2023failure}. 
Within this first category of methods, several different noteworthy approaches have also been proposed to approximate $h^*$, e.g., \citep{au1999new,au2003important,tabandeh2022review,ehre2023stein, cui2024deep}.

Before delving into the second category of methods, we define ISD $h= \nicefrac{\tilde{h}(\bm{X})}{C_h}$, where $\tilde{h}$ is an unnormalized importance distribution, and $C_h=\int_\mathcal{X} \tilde{h} (\bm{x}) d\bm{x}$ is its normalizing constant. Following this definition, $\tilde{h}^*=I_{\mathcal{F}}(\bm{X}) \pi_{\bm{X}}(\bm{X})$ denotes an unnormalized optimal importance sampling distribution, and $C_{h^*}=p_\mathcal{F}$ is its normalizing constant. Thus, in the second category of methods, sequential variants of importance sampling are employed, mainly utilizing a number of intermediate unnormalized distributions. These unnormalized distributions are generally more flexible than the PDFs usually utilized within the first category, as discussed above, thus being capable of providing a better approximation of $h^*$. Employing an unnormalized importance sampling distribution $\tilde{h}$, \cref{p_f_I_S} can be written as:
\begin{equation} \label{p_f_I_S_seq}
p_\mathcal{F}=  {\mathop{\mathbb{E}}}_{h} \big[ I_{\mathcal{F}} (\bm{X}) \dfrac{\pi_{\bm{X}}(\bm{X})}{h(\bm{X})}\big] = C_{h} \,{\mathop{\mathbb{E}}}_{h} \big[ I_{\mathcal{F}} (\bm{X}) \dfrac{\pi_{\bm{X}}(\bm{X})}{\tilde{h}(\bm{X})}\big]
\end{equation} 
Existing methods currently compute the normalizing constant, $C_h$, using consecutive intermediate unnormalized distributions $\tilde{h}_i$'s, $i=1,2,...,n$, following \cref{C_h_seq}. These distributions should be designed to link between $\pi_{\bm{X}}$ and $h^*$, with $\tilde{h}_1$ and $\tilde{h}_n=\tilde{h}$ being the closest to $\pi_{\bm{X}}$ and $h^*$, respectively.
\begin{equation} \label{C_h_seq}
\dfrac{C_{h_n}}{C_\pi} = \dfrac{C_{h_1}}{C_\pi}   \prod_{i=2}^n \dfrac{C_{h_i}}{C_{h_
{i-1}}}
\end{equation} 
where $C_{h_i}$ is the normalizing constant of $h_i$, and $C_\pi=1.0$, typically, is the normalizing constant of the original distribution $\pi_{\bm{X}}$. $C_\pi$ is left in \cref{C_h_seq} for generality purposes, particularly to cover scenarios when an unnormalized original distribution $\tilde{\pi}_{\bm{X}}$ is instead provided (substitute $\pi_{\bm{X}}=\nicefrac{\tilde{\pi}_{\bm{X}}}{C_\pi}$ in \cref{p_f_I_S_seq} to verify that).  \cref{C_h_seq} can be computed using the sequential importance sampling (SIS) approach as \citep{del2006sequential,papaioannou2016sequential}: 
\begin{equation}
\dfrac{C_{h_i}}{C_{h_{i-1}}} = {\mathop{\mathbb{E}}}_{h_{i-1}} \Bigl[ \dfrac{\tilde{h}_i(\bm{X})}{\tilde{h}_{i-1}(\bm{X})} \Bigr]
\end{equation}
Alternatively, bridge sampling can evaluate \cref{C_h_seq} as \citep{bennett1976efficient,meng1996simulating,sinha2020neural}: 
\begin{equation}
\begin{aligned}
&\dfrac{C_{h_i}}{C_{h_{i-1}}} =  \dfrac{\mathop{\mathbb{E}}_{h_{i-1}} [ \tilde{h}_{i}(\bm{X})h_i^B(\bm{X})]}{\mathop{\mathbb{E}}_{h_{i}} [\tilde{h}_{i-1}(\bm{X})h_i^B(\bm{X})]},\text{or} \\ 
&\dfrac{C_{h_i}}{C_{h_{i-1}}} = \dfrac{\nicefrac{C_{h_i^B}}{C_{h_{i-1}}}}{\nicefrac{C_{h_i^B}}{C_{h_{i}}}} =  \dfrac{\mathop{\mathbb{E}}_{h_{i-1}} [ \nicefrac{h_i^B(\bm{X})}{\tilde{h}_{i-1}(\bm{X})}]}{\mathop{\mathbb{E}}_{h_{i}} [ \nicefrac{h_i^B(\bm{X})}{\tilde{h}_{i}(\bm{X})}]}
\end{aligned}
\end{equation}
where $h_i^B(\bm{X})$ is a bridge distribution \citep{meng1996simulating} between $\tilde{h}_{i}$ and $\tilde{h}_{i-1}$, and $C_{h_i^B}$ is its associated normalizing constant. This latter technique aims to provide a more accurate estimate even if the two consecutive distributions, $\tilde{h}_{i}$ and $\tilde{h}_{i-1}$, are not close enough. \cref{C_h_seq} can also be computed using path sampling \citep{gelman1998simulating,johansen2005sequential}. A key to the success of these sequential methods is to design nearby intermediate distributions and to sufficiently sample them to accurately compute \cref{C_h_seq}, a non-trivial and computationally demanding task. A discussion on methods within this second category can be found in \cite {xian2024relaxation}. Interestingly, the subset simulation method \cite{au2001estimation} can also be seen as a special case of this second category, wherein the ratio $\nicefrac{C_{h_i}}{C_{h_{i-1}}}$ is the conditional probability between every two consecutive intermediate events, $ \mathcal{F}_i$ and $\mathcal{F}_{i-1}$.  

In this work, we are originally estimating \cref{p_f_I_S_seq} in a non-sequential manner, using a single unnormalized distribution $\tilde{h}$, approximating $h^*$, i.e., $n=1$. Therefore, the proposed approach significantly reduces the computational cost associated with employing multiple intermediate distributions. The discovery of the rare event domain is alternatively achieved through the utilization of advanced samplers and optimization techniques, as detailed in \cref{Discovery}. It should also be noted that the discussed importance sampling-based methods in the current literature are mostly tailored for Gaussian spaces. Consequently, the ability of our proposed framework to operate directly in non-Gaussian spaces is an additional substantial benefit.

\section{Approximate Sampling Target with Post-processing Adjustment (ASTPA) in Non-Gaussian Spaces}\label{ASTPA_sec}

\noindent ASTPA estimates the rare event probability in \cref{p_f} through an innovative non-sequential importance sampling variant evaluating \cref{p_f_I_S_seq}. To this end, ASTPA constructs a single unnormalized approximate sampling target $\tilde{h}$, relaxing the optimal ISD in \cref{opt_I_S_density}. Post-sampling, its normalizing constant is computed utilizing our devised inverse importance sampling. For clarity, we rewrite \cref{p_f_I_S_seq} as: 
\begin{equation} \label{p_f_I_S_ASTPA}
p_\mathcal{F}=  {\mathop{\mathbb{E}}}_{h} \big[ I_{\mathcal{F}} (\bm{X}) \dfrac{\pi_{\bm{X}}(\bm{X})}{h(\bm{X})}\big] = C_{h} \,{\mathop{\mathbb{E}}}_{h} \big[ I_{\mathcal{F}} (\bm{X}) \dfrac{\pi_{\bm{X}}(\bm{X})}{\tilde{h}(\bm{X})}\big] = C_h \, \tilde{p}_\mathcal{F}
\end{equation} 
Computing $p_\mathcal{F}$ is thus decomposed into two generally simpler problems. The first involves constructing $\tilde{h}$ and sampling $\{\bm{x}_i\}_{i=1}^N \sim h$\footnote{$\{\bm{x}_i\}_{i=1}^N \sim h$ is equivalent to $\{\bm{x}_i\}_{i=1}^N \sim \tilde{h}$.},  to compute the unbiased expectation of the weighted indicator function (shifted probability estimate) as:
\begin{equation} \label{p_f_tilde_ASTPA}
\tilde{p}_\mathcal{F} = \,{\mathop{\mathbb{E}}}_{h} \big[ I_{\mathcal{F}} (\bm{X}) \dfrac{\pi_{\bm{X}}(\bm{X})}{\tilde{h}(\bm{X})}\big] \approx \hat{\tilde{p}}_\mathcal{F} = \dfrac{1}{N} \sum_{i=1}^{N} I_{\mathcal{F}} (\bm{x}_i) \dfrac{\pi_{\bm{X}}(\bm{x}_i)}{\tilde{h}(\bm{x}_i)} 
\end{equation} 
An empirical estimation for the normalizing constant, $\hat{C}_h$, is then computed using the inverse importance sampling, as discussed in \cref{IIS}. The ASTPA estimator for the sought rare event probability can then be computed as:  
\begin{equation} \label{p_f_ASTPA}
\hat{p}_\mathcal{F} = \hat{\tilde{p}}_\mathcal{F} \,\, \hat{C}_h 
\end{equation} 

\begin{figure*}[t!]
 \centering
 \vspace*{-0.5in}
  \begin{tabular}{ccccc}
   \vspace*{-0.2in}
    \hspace*{-0.1in}
   &\includegraphics[width=.245\textwidth,keepaspectratio]{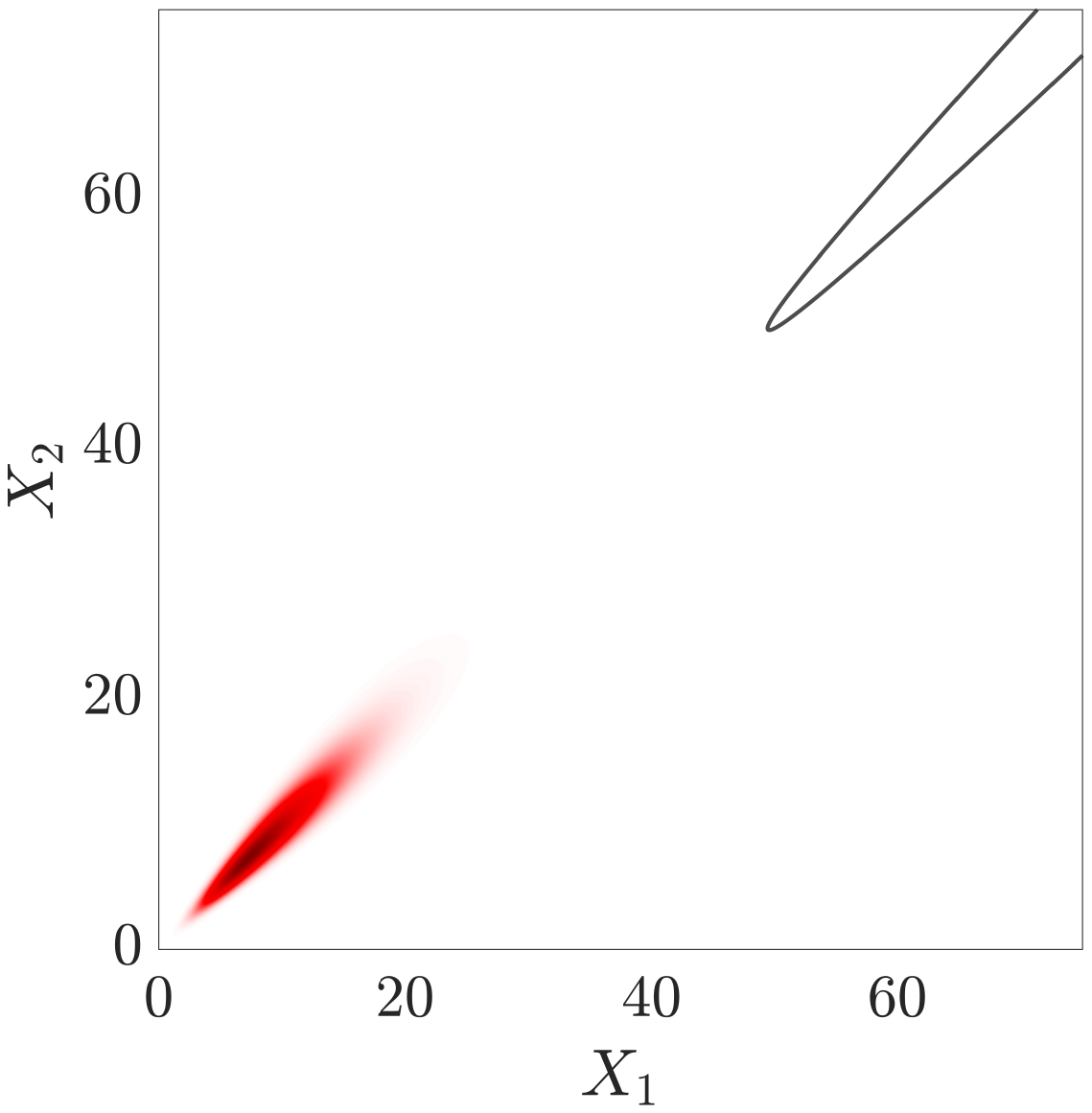}&    \hspace*{-0.2in}
   \includegraphics[width=.245\textwidth,keepaspectratio]{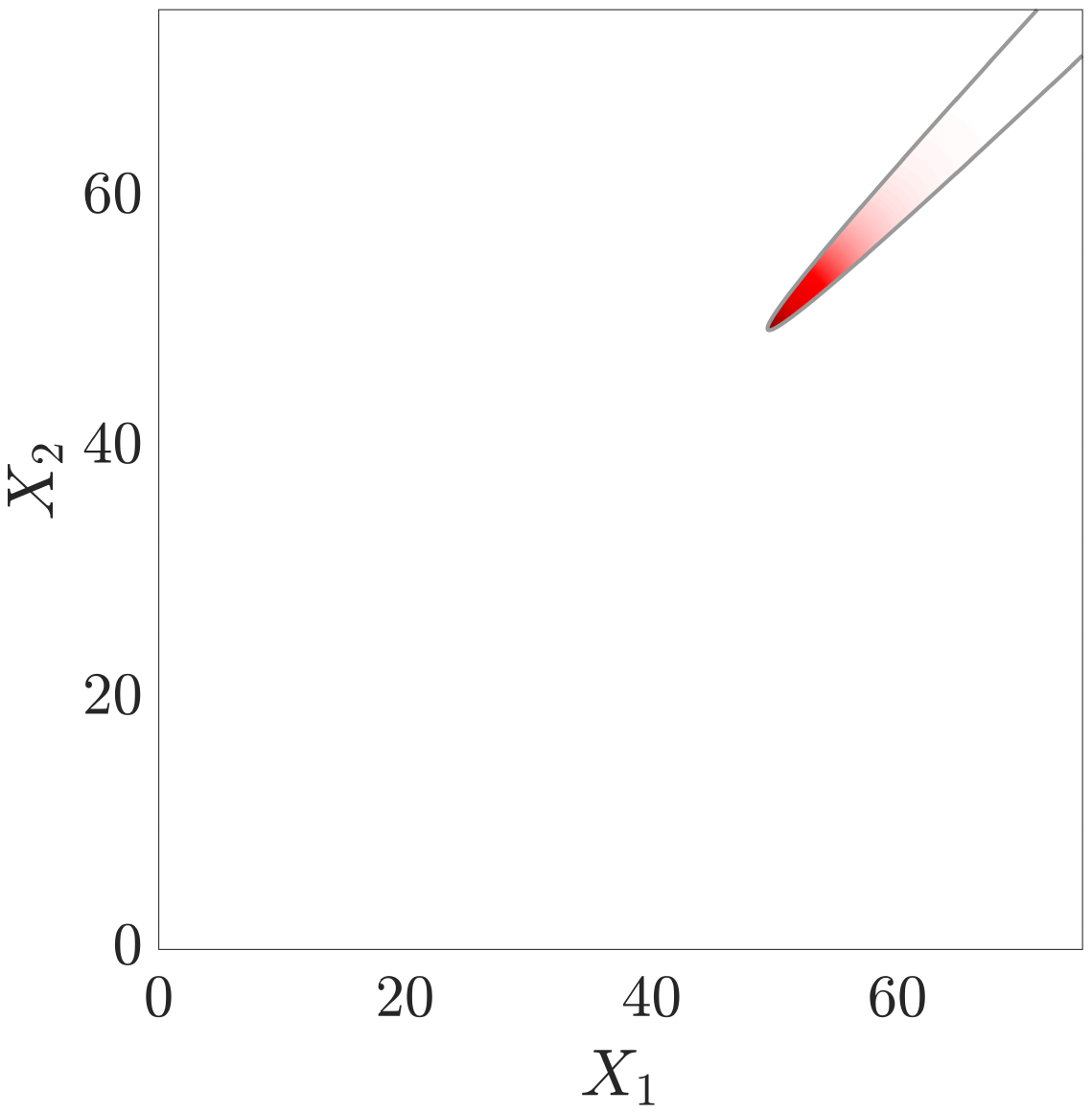}&  &\\&
   \hspace*{-0.2in}{\fontsize{8.5pt}{7.2} (a) $\pi_{\bm{X}}$}&\hspace*{-0.05in}{\fontsize{8.5pt}{7.2} (b) $h^*(\bm{X})$}&\vspace*{0.1in}\\
    \cline{1-4}\vspace*{-0.35in}\\
   \hspace*{-0.2in}
   \includegraphics[width=.245\textwidth,keepaspectratio]{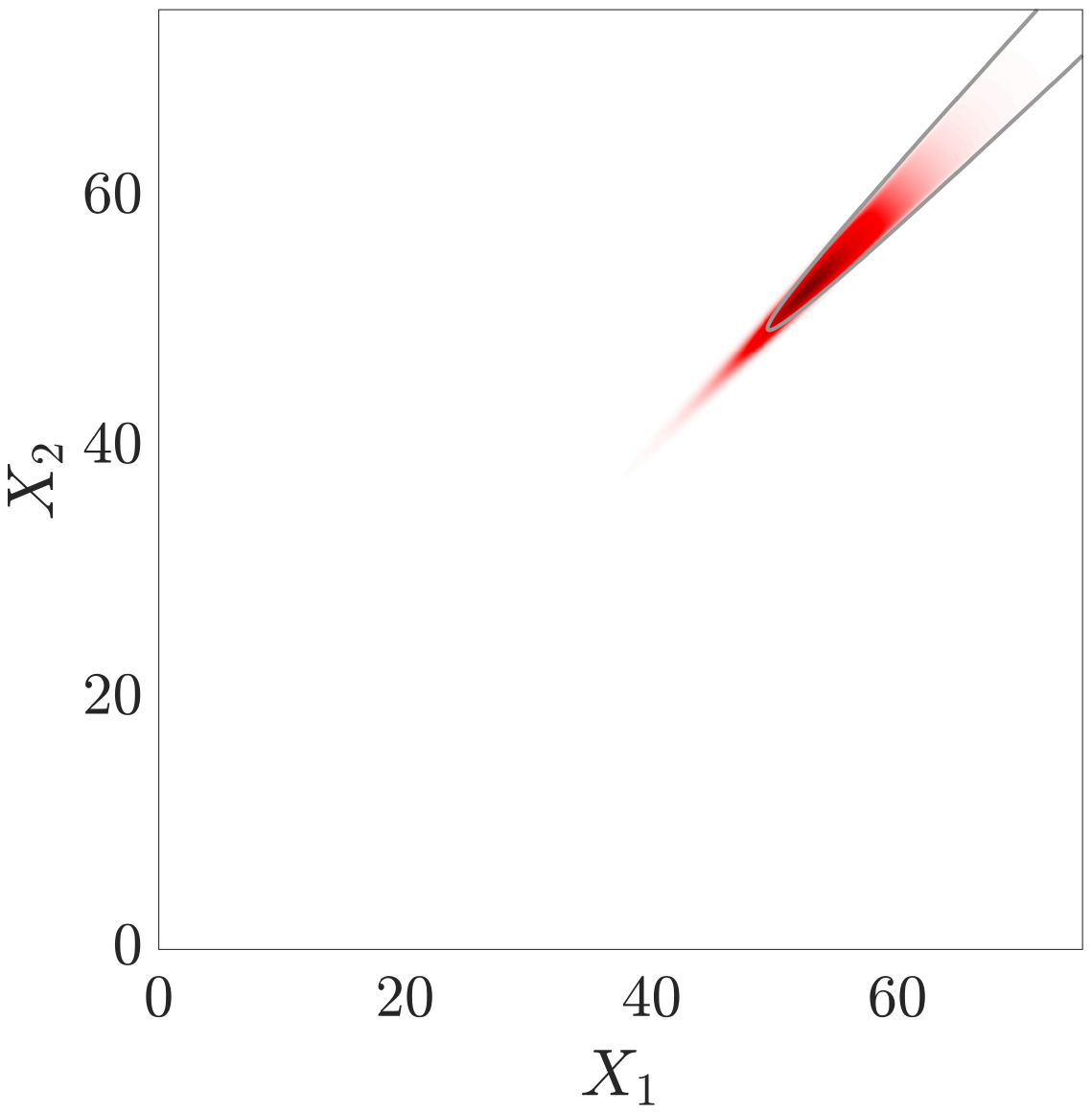}&
   \hspace*{-0.2in}
      \includegraphics[width=.245\textwidth,keepaspectratio]{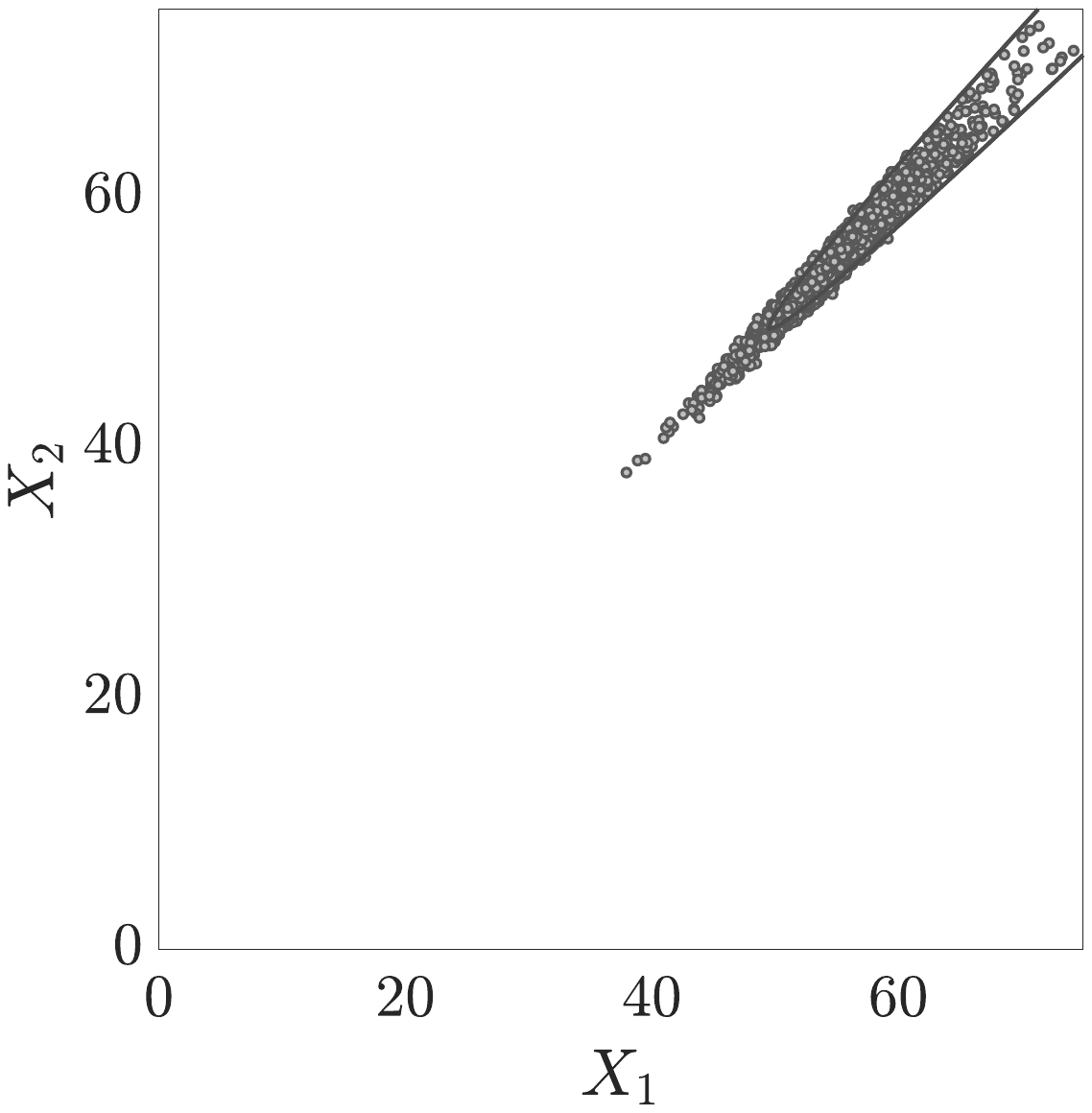}& \hspace*{-0.2in} 
         \includegraphics[width=.245\textwidth,keepaspectratio]{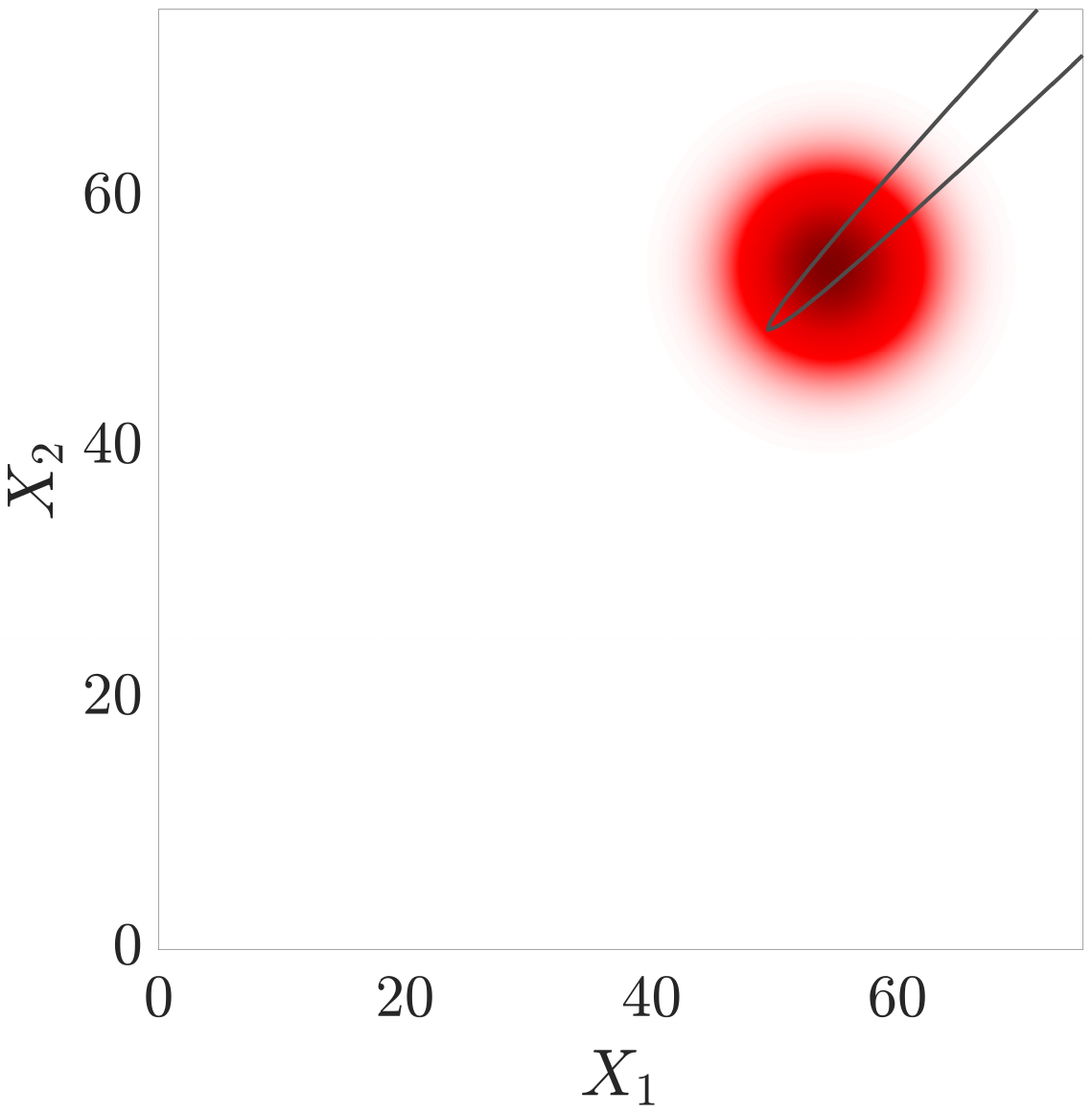}& \hspace*{-0.2in}
   \includegraphics[width=.245\textwidth,keepaspectratio]{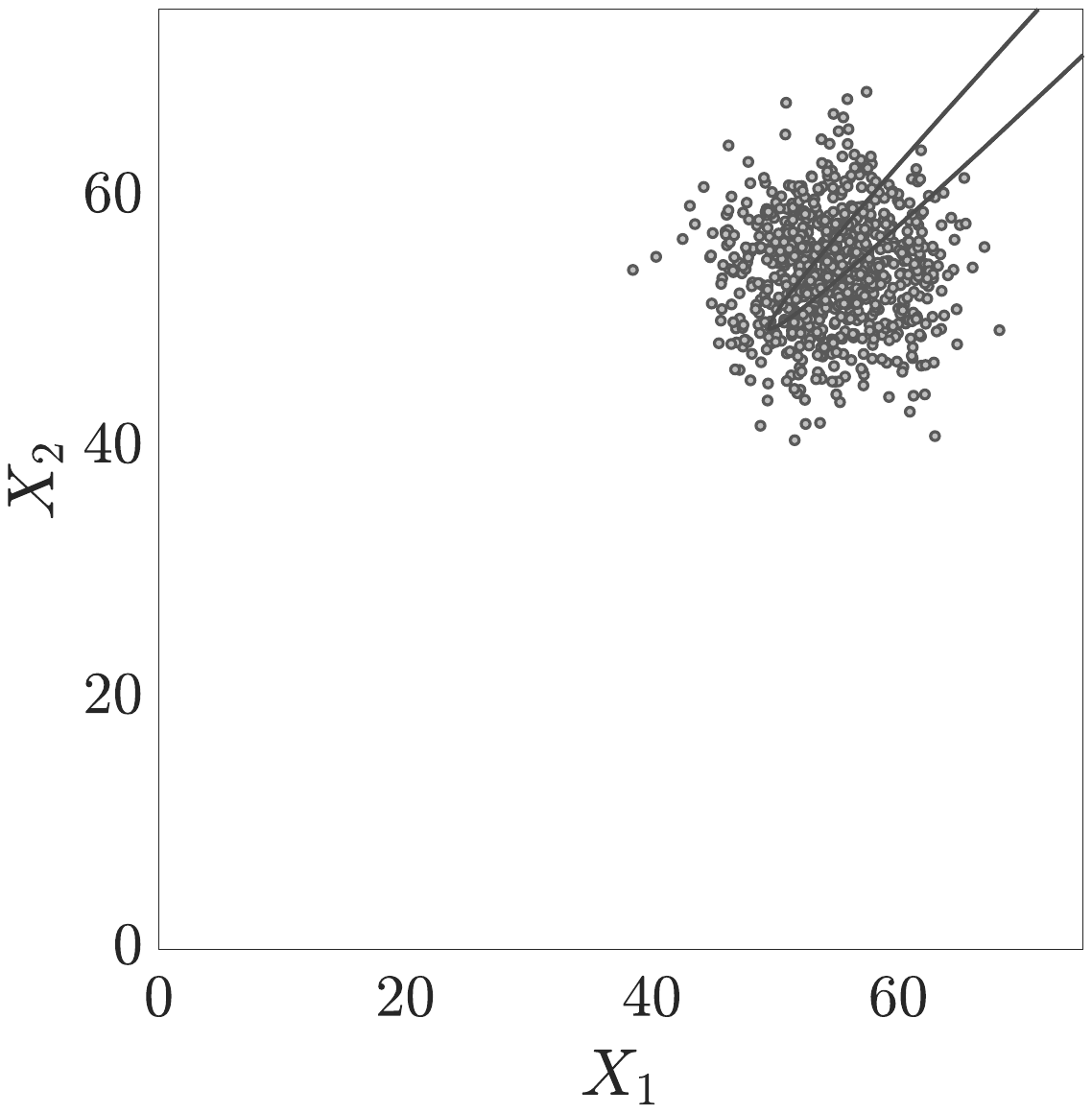}
\vspace*{-0.25in}\\
    \hspace*{-0.25in}
 \vspace*{+0.05in}
    {\fontsize{8.5pt}{7.2} (c) $\tilde{h}(\bm{X})$}&
    \hspace*{-0.25in}
    {\fontsize{8.5pt}{7.2}  (d) Samples $\sim \tilde{h}(\bm{X})$}&
    \hspace*{-0.25in}
   {\fontsize{8.5pt}{7.2}  (e) $Q(\bm{X})$}& \hspace*{-0.25in}
   {\fontsize{8.5pt}{7.2}  (f) Samples $\sim Q(\bm{X})$}\\
    \vspace*{-0.1in}\\
   \vspace*{-0.2in}
 \end{tabular}
 \caption{Outlining the ASTPA framework; (a) shows an original bivariate joint distribution $\pi_{\bm{X}}$ in red, and a limit state function at $g(\bm{X})=0$ shown using a gray line, with the rare event domain being inside $g(\bm{X})=0$, as detailed in \cref{Corr_Gumbel}, 
 (b) visualizes the optimal importance sampling density, $h^*$, in \cref{opt_I_S_density}, (c) depicts the constructed approximate sampling target, $\tilde{h}$, in \cref{Tar_ASTPA}, (d) showcases samples from this target, subsequently used to compute $\hat{\tilde{p}}_\mathcal{F}$ in \cref{p_f_tilde_ASTPA}, and (e) and (f) demonstrate the inverse importance sampling procedure to compute the normalizing constant $\hat{C}_h$ in \cref{IIS_est_hat}. Specifically, (e) and (f) present a crudely fitted $Q(\bm{X})$ and its samples, respectively. Another choice for $Q$ is also shown in \cref{IIS_acc_Q_fig}. The sought probability is eventually computed using \cref{p_f_ASTPA}.}
 \label{ASTPA_framework_fig}
\end{figure*}

\cref{ASTPA_framework_fig} concisely portrays the proposed framework using a correlated Gumbel distribution and a quadratic limit state function, as detailed in \cref{Corr_Gumbel}, with $p_{\mathcal{F}}\sim 2.51 \times10^{-7}$. The construction of the approximate sampling target in ASTPA is discussed in \cref{Targ_form}, including its recommended parameters for problems described in non-Gaussian spaces. Then, we present the inverse importance sampling approach in \cref{IIS}. Subsequently, \cref{analCOV} thoroughly discusses the statistical properties of the proposed estimator presented in \cref{p_f_ASTPA}. 

\subsection{Target distribution formulation}\label{Targ_form}

\noindent As discussed earlier, the basic idea is to construct a near-optimal sampling target distribution $\tilde{h}$, placing higher importance on the rare event domain $\mathcal{F}$. Similar to methods utilizing sequential variants of IS (see \cref{rare_event_est} for details), the approximate sampling target in ASTPA is constructed by approximating the indicator function $I_\mathcal{F}$ with a smooth function, namely a likelihood function $\ell_{g_{\bm{X}}}$. This likelihood function is chosen as a \textit{logistic} cumulative distribution function (CDF) of the negative of a scaled limit state function, $F_{\text{CDF}}\left(\largesfrac{-g(\bm{X})}{g_c}\right)$, with $g_c$ being a scaling constant:
\begin{equation}
\begin{aligned}
\ell_{g_{\bm{X}}} &=  \, F_{\text{CDF}}\bigg  (\dfrac{-g(\bm{X})}{g_c}\bigg\arrowvert\ \mu_{g} , \ \sigma\bigg ) \\ &=  \, \Bigg (1 + \exp\bigg (\dfrac{(\frac{g(\bm{X})}{g_{c}})+\mu_{g}}{(\frac{\sqrt{3}}{\pi})\sigma} \bigg ) \Bigg )^{-1}
\label{likelihood}
\end{aligned}
\end{equation}
where $\mu_{g}$ and $\sigma$ are the mean and dispersion factor of the logistic CDF, respectively; their recommended values are subsequently discussed. Our numerical experiments in this work and \citep{Papakon2023HMCMC} have underscored the effectiveness of this chosen approximation of $ I_\mathcal{F}$. It should also be noted, however, that there exist numerous choices for approximating $I_\mathcal{F}$ in the literature, e.g., standard normal CDF \citep{papaioannou2016sequential,papaioannou2019improved}, and exponential tilting barrier \citep{sinha2020neural}. The approximate sampling target is ASTPA is then expressed as: 
\begin{equation}
\begin{aligned}
\tilde{h}(\bm{X}) &= \ell_{g_{\bm{X}}} \, \pi_{\bm{X}}(\bm{X}) \\&=  \, \Bigg (1 + \exp\bigg (\dfrac{(\frac{g(\bm{X})}{g_{c}})+\mu_{g}}{(\frac{\sqrt{3}}{\pi})\sigma} \bigg ) \Bigg )^{-1} \, \pi_{\bm{X}}(\bm{X})
\label{Tar_ASTPA}
\end{aligned}
\end{equation}
The scaling $\largesfrac{g(\bm{X})}{g_{c}}$ is suggested to handle the varying magnitudes that different limit-state functions can exhibit. It is implemented in a way ensuring that the input to the logistic CDF $(\largesfrac{g(\bm{X})}{g_{c}})$ at an influential point with respect to the original distribution $\pi_{\bm{X}}$ consistently falls within a predefined range, regardless the magnitude of $g(\bm{X})$. In this work, we use the mean value of the original distribution, $\boldsymbol{\mu}_{\pi_{\bm{X}}}$, to represent this influential point, albeit other points could also serve this purpose. To make the input to the logistic CDF, when evaluated at $\bm{x}=\boldsymbol{\mu}_{\pi_{\bm{X}}}$, contained in a desired range of $[10\, \, 20]$, we appropriately define $g_c$ as:
\begin{equation}
\begin{aligned}
&g_{c}=
\begin{cases}
\frac{g(\boldsymbol{\mu}_{\pi_{\bm{X}}})}{q}, q \in [10 \, \,20] \quad \text{if}  \, \bigg ( \big (g(\boldsymbol{\mu}_{\pi_{\bm{X}}})>20 \big ) \bigcup \\ \qquad \qquad \qquad \qquad  \qquad \quad  \big (0<g(\boldsymbol{\mu}_{\pi_{\bm{X}}})< 10 \big ) \bigg ) \\
1, \,\, \text{otherwise}
\end{cases}\label{g_c}
\end{aligned}
\end{equation}
For example, if $g(\boldsymbol{\mu}_{\pi_{\bm{X}}})=10^4$, this scaling ensures that the input at $\bm{x}=\boldsymbol{\mu}_{\pi_{\bm{X}}}$ equals $q$, i.e., $\largesfrac{g(\boldsymbol{\mu}_{\pi_{\bm{X}}})}{g_{c}}=\largesfrac{g(\boldsymbol{\mu}_{\pi_{\bm{X}}})}{(\sfrac{g(\boldsymbol{\mu}_{\pi_{\bm{X}}})}{q}})=q$. Fine-tuning values of the scaling constant, $q$, within the recommended range is not usually necessary. Our empirical investigation
has confirmed the effectiveness of this proposed scaling technique and demonstrated that the recommended range for $q$ is practical for the studied non-Gaussian problems.

The dispersion factor $\sigma$ of the utilized logistic CDF, $F_{\text{CDF}}$, determines the spread of the likelihood function, thus controlling the shape of the constructed target $\tilde{h}$. For non-Gaussian spaces, $\sigma$ values are recommended in the range  $[0.1 \, \,0.6]$. Fine-tuning higher decimal values of $\sigma$ in that range is not usually necessary. Lower values, $[0.1 \, \,0.4]$, usually work efficiently, whereas a higher $\sigma$ value within the recommended range may be needed for cases involving a large number of (strongly) nonlinear random variables.\par
For the mean parameter, $\mu_{g}$, of the logistic CDF, $F_{\text{CDF}}$, we are interested in generally locating it inside the rare event domain, $g(\bm{X})<0$, to enhance the sampling efficiency. As such, we are describing $\mu_{g}$ through a percentile, $p$, of the logistic CDF and its quantile function, $\Upsilon_{g}$:
\begin{equation}
\Upsilon_{g} (p;\mu_{g},\sigma) = \mu_{g} + (\frac{\sqrt{3}}{\pi} \sigma) \ln \bigg (\frac{p}{1-p} \bigg )
\end{equation}
By placing a chosen percentile $p$ of the logistic CDF on the limit-state surface at $g(\bm{X}) = 0$ $(\Upsilon_{g} (p;\mu_{g},\sigma) = 0)$, the mean parameter $\mu_{g}$ can be computed as:
\begin{equation}
\mu_{g} =  -(\frac{\sqrt{3}}{\pi} \sigma) \ln \bigg (\frac{p}{1-p} \bigg )  \label{mu_g}
\end{equation}
Based on our empirical investigation, we found that using the $p10$ percentile ($p=0.1$) yields good efficiency. Therefore, it is used in all our experiments in this work. Substituting $p=0.1$ in \cref{mu_g}, the mean parameter is then defined as a function of the dispersion factor, as $\mu_{g} \approx 1.21\,\sigma$.

\begin{figure*}[t!]
 \vspace*{-0.7in}
 \centering
  \begin{tabular}{cccc}
   \vspace*{-0.7in}
   \includegraphics[width=.325\textwidth,keepaspectratio]{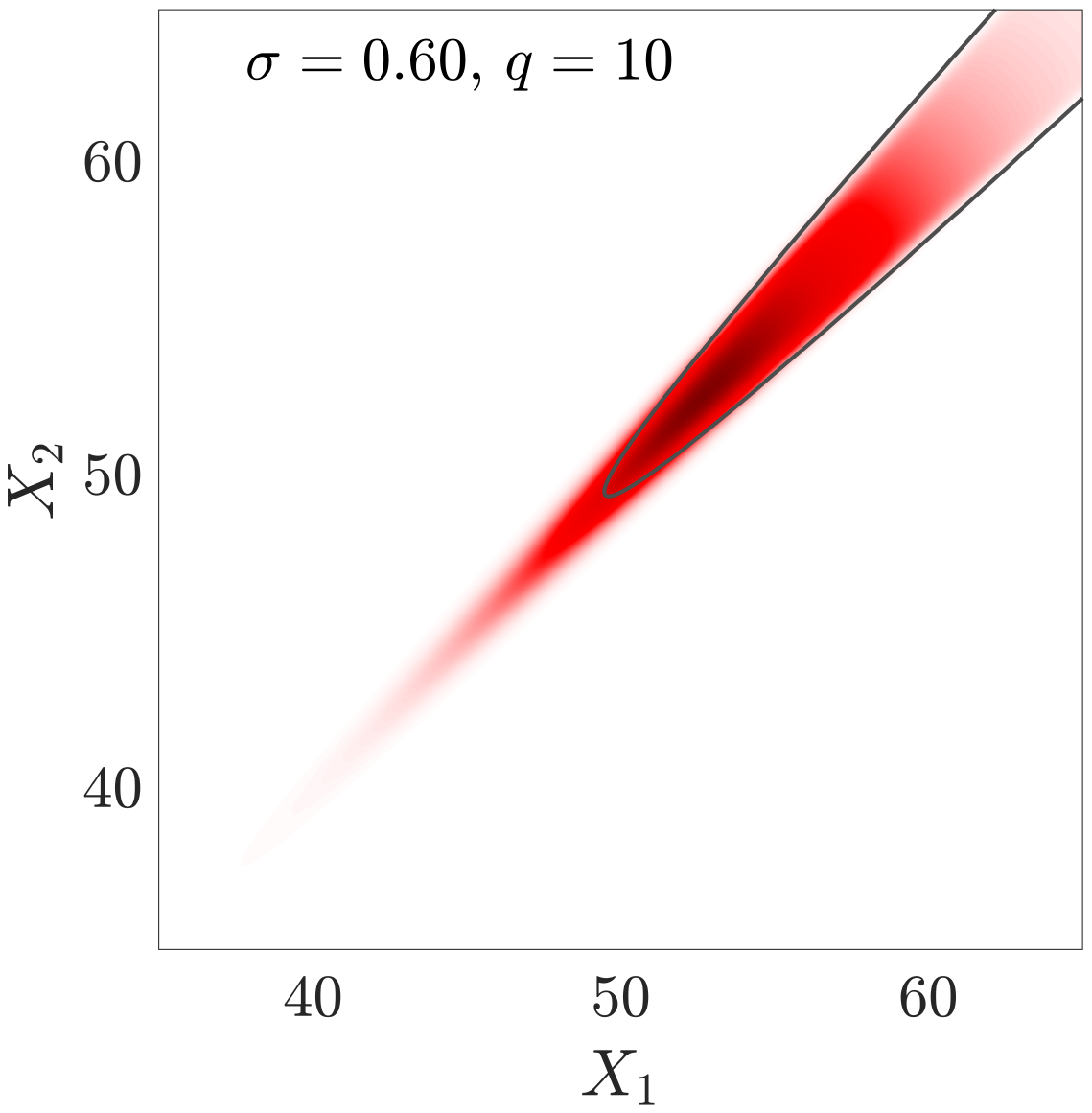}& \hspace*{-0.2in}
   \includegraphics[width=.325\textwidth,keepaspectratio]{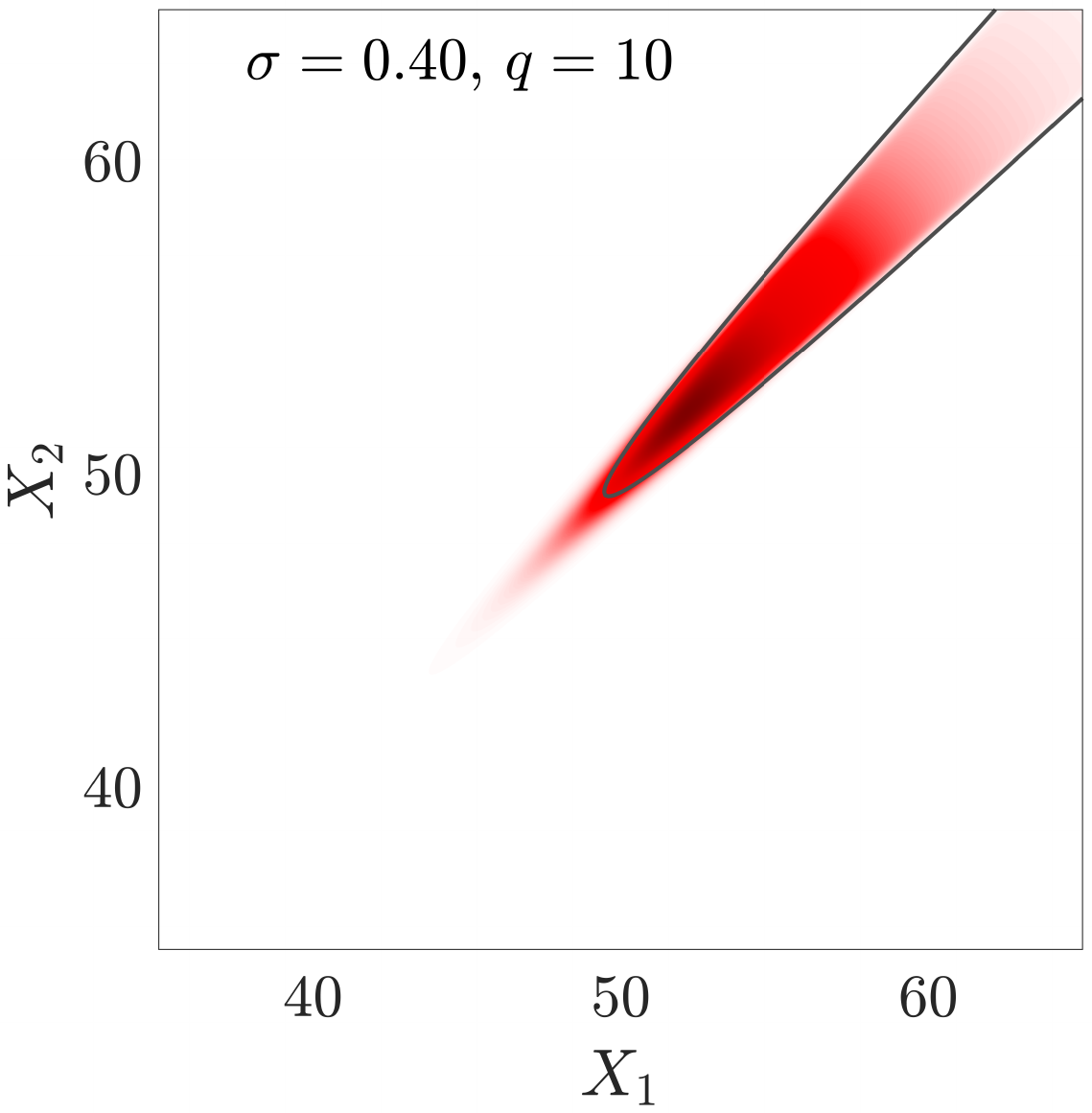}& \hspace*{-0.2in}
      \includegraphics[width=.325\textwidth,keepaspectratio]{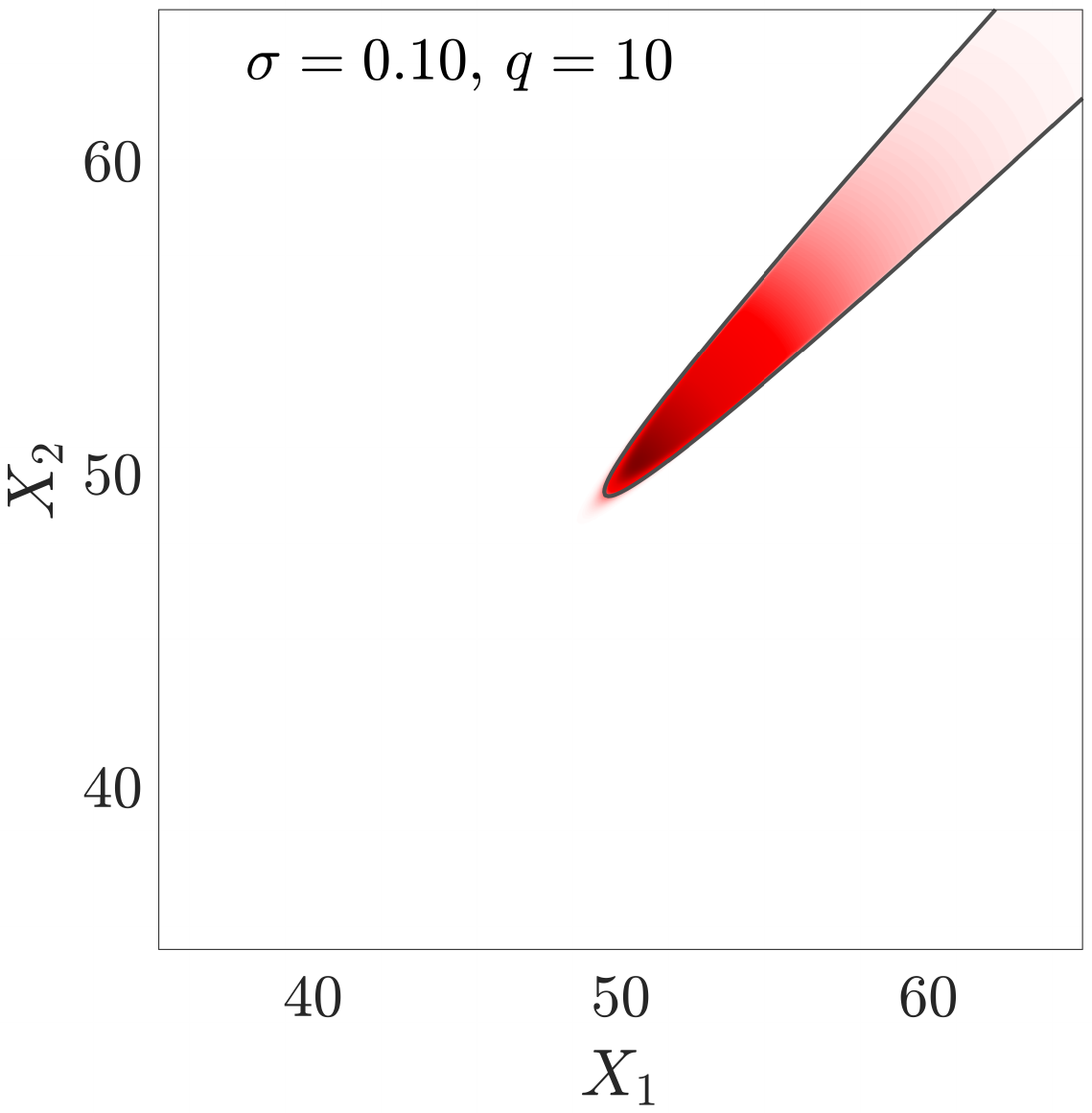}& \hspace*{-0.2in} 
\\ \vspace{-0.4in}
    \includegraphics[width=.325\textwidth,keepaspectratio]{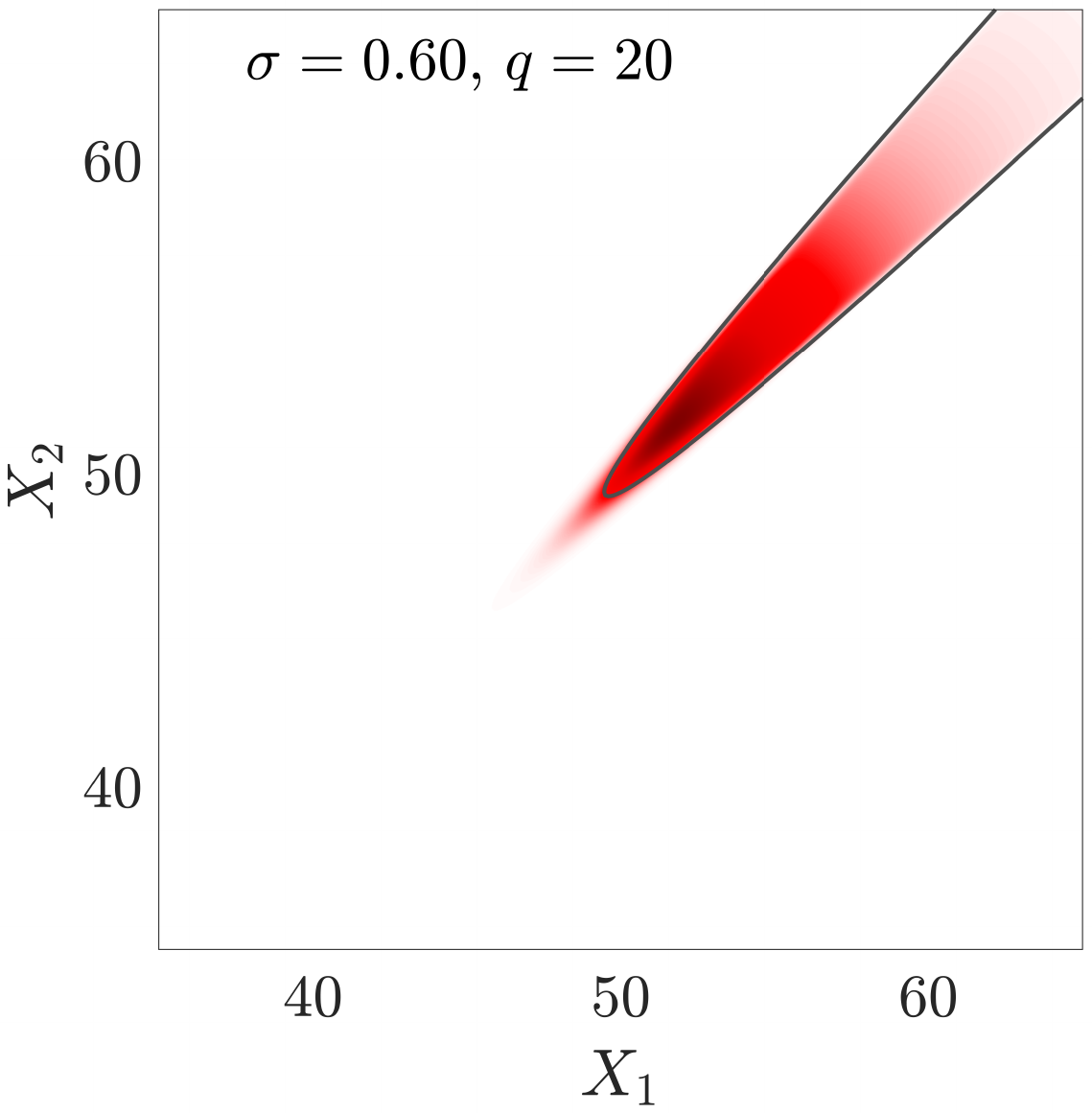}& \hspace*{-0.2in}
   \includegraphics[width=.325\textwidth,keepaspectratio]{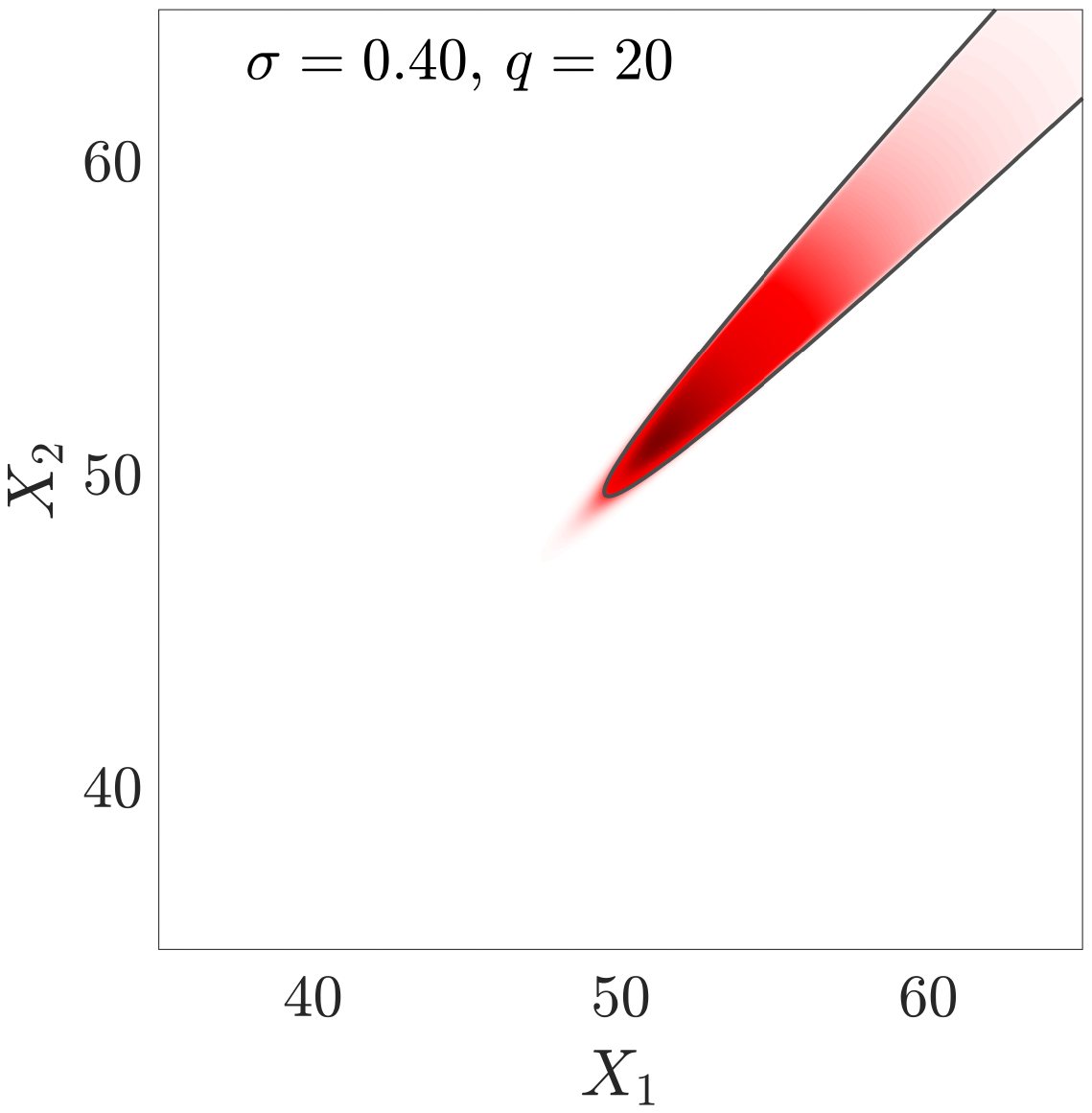}& \hspace*{-0.2in}
   \includegraphics[width=.325\textwidth,keepaspectratio]{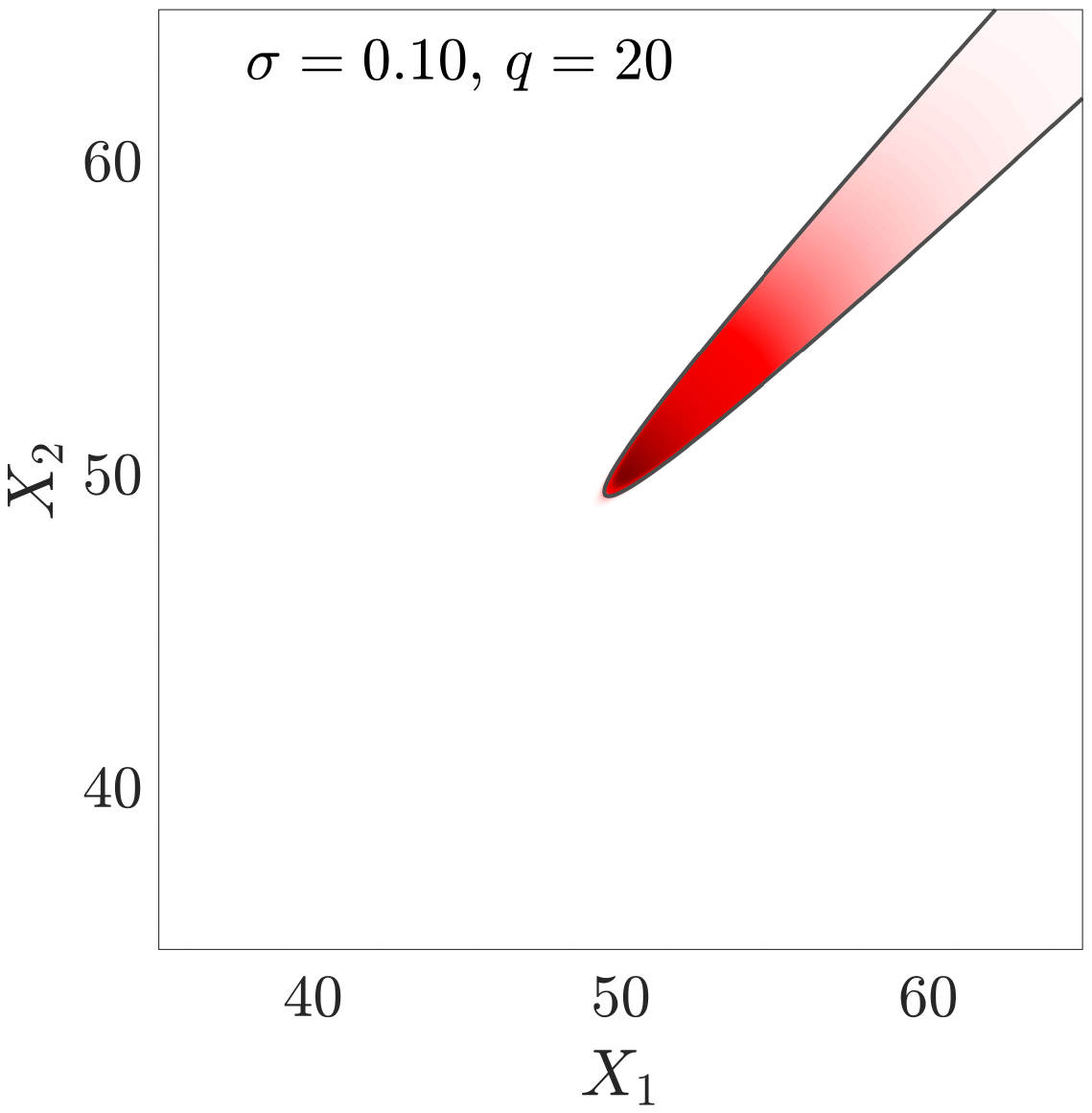}& \hspace*{-0.2in}
 \end{tabular}
 \caption{Effect of the likelihood dispersion factor, $\sigma$,  and the scaling constant, $q$, on the target distribution for a quadratic limit-state function, shown in gray at $g(\bm{X})=0$, with a correlated Gumbel distribution, as detailed in \cref{Corr_Gumbel}.}
 \label{fig_effect_on_target}
\end{figure*}

In \cref{fig_effect_on_target}, we illustrate how the likelihood dispersion factor, $\sigma$, and the scaling constant, $q$, influence the shape of the target distribution, based on a correlated Gumbel distribution, as detailed in \cref{Corr_Gumbel}. As can be seen, decreasing $\sigma$ and/or increasing $q$ results in a more concentrated target distribution inside the rare event domain. Although an increasing number of rare event samples can be obtained by pushing the target further inside the rare event domain, i.e., being very close to the optimal one  $(I_{\mathcal{F}} \pi_{\bm{X}})$, that may also complicate the sampling process and the computation of the normalizing constant $\hat{C}_h$, as discussed in \cref{sec_impact_on_p_f}.

\subsection{Inverse importance sampling}\label{IIS} 
\noindent Inverse importance sampling (IIS) is a general technique to compute normalizing constants, demonstrating exceptional performance in the context of rare event probability estimation. Given an unnormalized distribution $\tilde{h}$, and samples $\{\bm{x}_i\}_{i=1}^N \sim h$, inverse importance sampling first fits an ISD $Q(\bm{x})$ based on the samples $\{\bm{x}_i\}_{i=1}^N \sim h$; details of $Q(\bm{x})$ are discussed below. IIS then estimates the normalizing constant $C_h$ as:
\begin{equation} \label{IIS_est}
C_h = \int_{\mathcal{X}} \tilde{h}(\bm{x}) d\bm{x} = \int_{\mathcal{X}} \dfrac{\tilde{h}(\bm{x})}{Q(\bm{x})}Q(\bm{x}) d\bm{x}\,=\mathbb{E}_{Q} \big[ \dfrac{\tilde{h}(\bm{X})}{Q(\bm{X})}] 
\end{equation} 
By drawing $\{\bm{x}^\prime\}_{i=1}^M$ i.i.d. samples from $Q$, the unbiased normalizing constant estimator can be computed as:
\begin{equation} \label{IIS_est_hat}
\hat{C}_h = \dfrac{1}{M}\sum_{i=1}^{M} \dfrac{\tilde{h}(\bm{x}_i^\prime)}{Q(\bm{x}_i^\prime)}
\end{equation}

\begin{figure*}[t!]
\centering
\vspace*{-0.5in}
\begin{tabular}{cc}
\includegraphics[trim=0cm 3.5cm 0cm 3cm,width=0.3\textwidth]{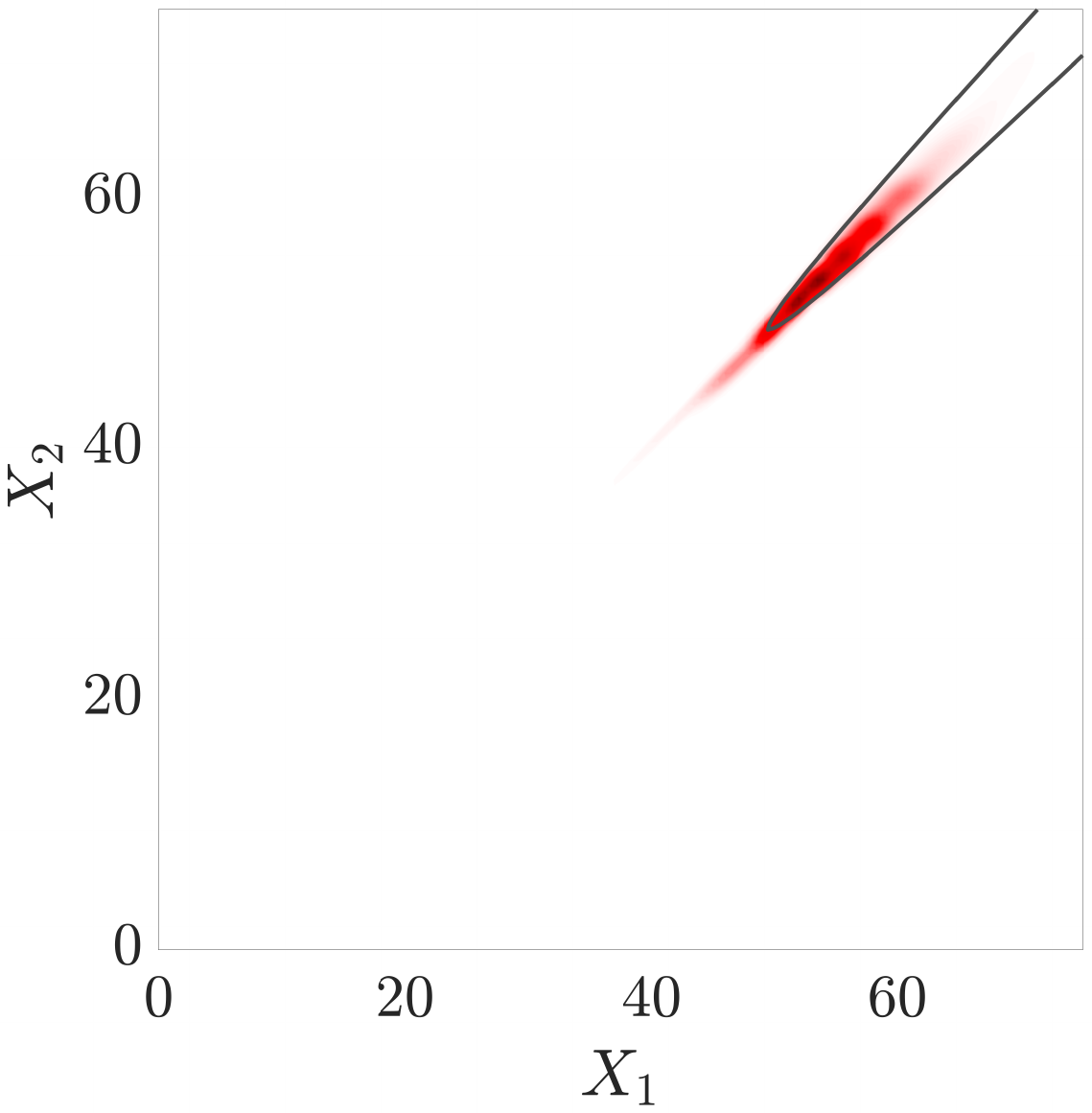}&
\includegraphics[trim=0cm 3.5cm 0cm 3cm,width=0.3\textwidth]{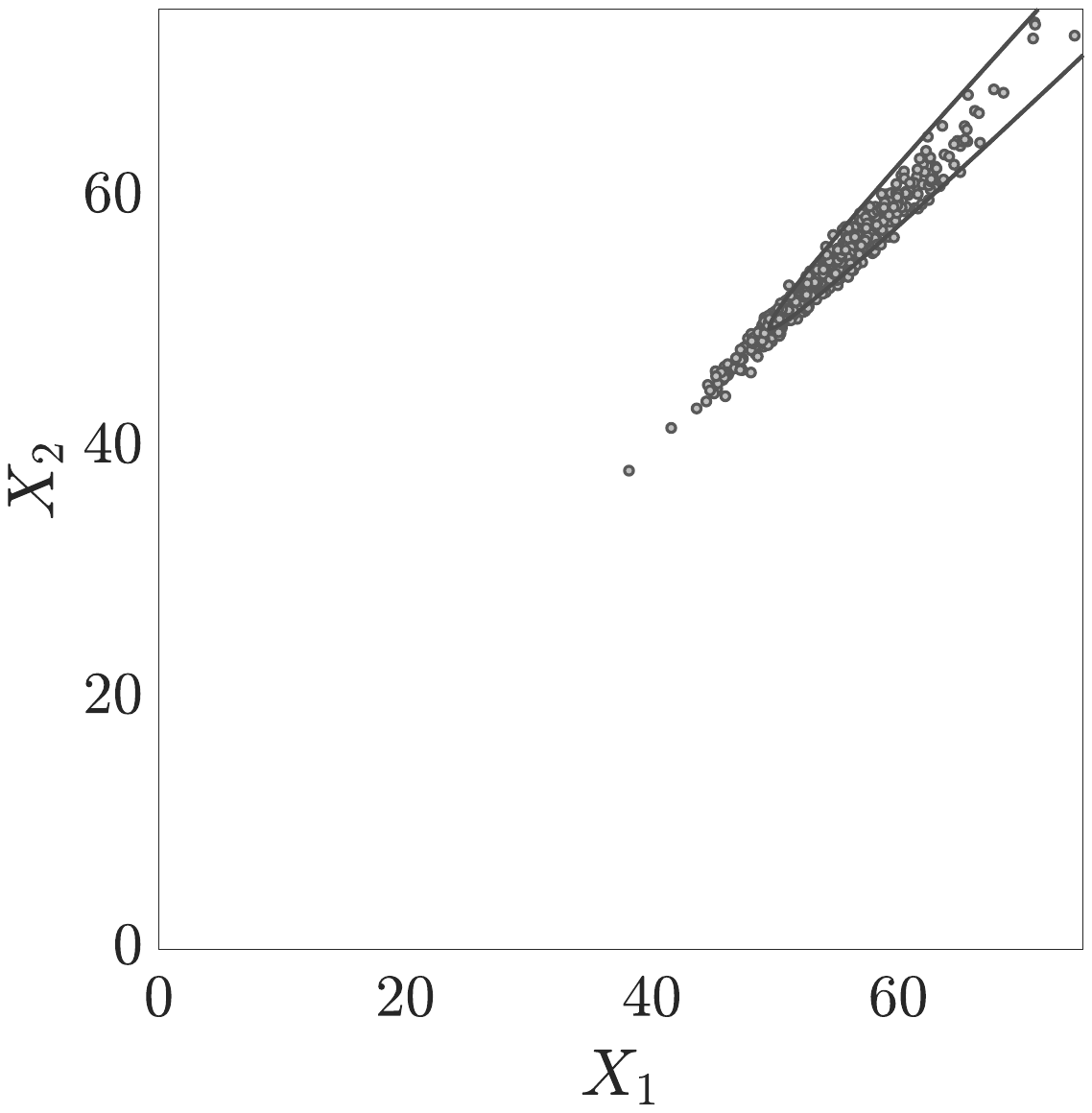}\\
 \hspace*{-0.25in}
   {\fontsize{8.5pt}{7.2}  (a) $Q(\bm{X})$}&
   \hspace*{-0.25in}
   {\fontsize{8.5pt}{7.2}  (b) Samples $\sim Q(\bm{X})$}
\end{tabular}
\caption{Showcases the inverse importance sampling technique for the same illustrative example in \cref{ASTPA_framework_fig}, but here with an accurately fitted GMM, $Q$, with ten Gaussian components having full covariance matrices, as shown in (a), where its samples are showcased in (b).\vspace{-0.1in}}\label{IIS_acc_Q_fig}
\end{figure*}

In this work, the ISD $Q(.)$ is a computed Gaussian Mixture Model (GMM), based again on the already available samples, $\{\bm{x}_i\}_{i=1}^N$, and the generic Expectation-Maximization (EM) algorithm \citep{mclachlan2000finite}. A typical GMM expression can be then given by:
\begin{align}
Q(\boldsymbol{\bm{X}})= \sum_{k=1}^{\bm{K}} w_{k} \phi(\bm{X}\, ;\,\boldsymbol{\mu}_{k},\boldsymbol{\Sigma}_{k})
\end{align}       
where $\phi(.)$ is the PDF, $w_{k}$ is the weight, $\boldsymbol{\mu}_{k}$ is the mean vector and $\boldsymbol{\Sigma}_{k}$ is the covariance matrix of the $k^{th}$ Gaussian component, that can all be estimated, for all components, by the EM algorithm. In low-dimensional spaces, where $d<20$, we employ a GMM with a large number $(\sim 10)$ of Gaussian components that have full covariance matrices. This approach aims to accurately approximate the target distribution $\tilde{h}$ with an ISD $Q(.)$, thus improving the IIS estimator in \cref{IIS_est_hat}. In contrast, for higher dimensions, we fit Gaussian Mixture Models (GMMs) with diagonal covariance matrices. That is to reduce the number of GMM parameters to be estimated, thereby mitigating the scalability issues commonly associated with GMMs. Specifically, for high-dimensional examples discussed in this paper, we utilize a GMM fitted with a \textit{single component and a diagonal covariance matrix}, i.e., a multivariate independent Gaussian density. This method has demonstrated exceptional efficacy in calculating the normalizing constant $C_h$ for challenging examples, even in spaces with dimensions as high as 500. Notably, this IIS technique can generally still work effectively even using a crudely fitted $Q$. We demonstrate this feature by employing two different choices for the ISD, $Q$, to quantify the target probability in the illustrative example depicted in \cref{ASTPA_framework_fig}, with $p_{\mathcal{F}}\sim 2.51 \times10^{-7}$. The first $Q$ is an accurate GMM with ten Gaussian components having full covariance matrices, as depicted in \cref{IIS_acc_Q_fig}
, which yields $\mathop{\mathbb{E}}[\hat{p}_{\mathcal{F}}]$ of $2.52\times10^{-7}$ and a coefficient of variation (C.o.V) of $0.03$. On the other hand, using the same number of total model calls ($4,048$), the crudely fitted GMM in \cref{ASTPA_framework_fig}(e) still provides very good performance with an estimated probability of $2.49\times10^{-7}$ and a C.o.V of $0.10$. The ASTPA estimator of the rare event probability can finally be computed as:
\begin{equation} \label{p_f_ASTPA_final}
\begin{aligned}
p_\mathcal{F}&= \,{\mathop{\mathbb{E}}}_{h} \big[ I_{\mathcal{F}} (\bm{X}) \dfrac{\pi_{\bm{X}}(\bm{X})}{\tilde{h}(\bm{X})}\big] \,\mathbb{E}_{Q} \big[ \dfrac{\tilde{h}(\bm{X})}{Q(\bm{X})}]  \\ &\approx \hat{\tilde{p}}_\mathcal{F} \, \hat{C}_h  =  \bigg(\dfrac{1}{N} \sum_{i=1}^{N} I_{\mathcal{F}} (\bm{x}_i) \dfrac{\pi_{\bm{X}}(\bm{x}_i)}{\tilde{h}(\bm{x}_i)}\bigg) \bigg( \dfrac{1}{M}\sum_{i=1}^{M} \dfrac{\tilde{h}(\bm{x}_i^\prime)}{Q(\bm{x}_i^\prime)}\bigg)
\end{aligned}
\end{equation} 
where $\{\bm{x}_i\}_{i=1}^N$ and $\{\bm{x}_i^\prime\}_{i=1}^M$ are samples from $h$ and $Q$, respectively. 

\begin{remark}
    We introduce an effective approach to enhance the robustness of the IIS estimator $\hat{C}_h$ against the impact of rarely observed outliers or anomalous samples. Initially, the sample set $\{\bm{x}_i^\prime\}_{i=1}^M \sim Q$ is split into two subsets of equal size: $\{\bm{x}_i^\prime\}_{i=1}^{M/2}$ and $\{\bm{x}_i^\prime\}_{i=M/2+1}^{M}$, given that $M$ is an even number.  Subsequently, \cref{IIS_est_hat} is computed separately for each subset, yielding two estimates, $\hat{C}_h^1$ and $\hat{C}_h^2$. If these two estimates fall within a largely reasonable range of each other, specifically if $\largesfrac{1}{3} \leq \largesfrac{\hat{C}_h^1}{\hat{C}_h^2} \leq 3$, we then compute the final IIS estimate as their average, $\hat{C}_h= 0.5(\hat{C}_h^1+\hat{C}_h^2)$. This is equivalent to computing \cref{IIS_est_hat} using the entire set $\{\bm{x}_i^\prime\}_{i=1}^M$. However, in rare cases, the estimates diverge beyond this threshold, indicating potential instability, and a conservative approach is thus adopted. In these instances, the final estimator $\hat{C}_h$ is determined as the minimum of the two estimates, $\hat{C}_h = \min(\hat{C}_h^1, \hat{C}_h^2)$. This method has been successfully applied in all examples presented in this work.
\end{remark}

\subsection{Statistical properties of the ASTPA estimator} \label{analCOV}
\noindent In this section, we show the unbiasedness and the analytical Coefficient of Variation (C.o.V) of the ASTPA estimator, $\hat{p}_\mathcal{F}= \hat{\tilde{p}}_\mathcal{F} \, \hat{C}_h$, in \cref{p_f_ASTPA_final}. Assuming $\hat{\tilde{p}}_{\mathcal{F}}$ and $\hat{C}_h$ are independent, the expected value of their product can be expressed as:
\begin{equation}
\mathop{\mathbb{E}}[\hat{p}_{\mathcal{F}}] = \mathop{\mathbb{E}}[\hat{\tilde{p}}_\mathcal{F} \, \hat{C}_h] = \mathop{\mathbb{E}}[\hat{\tilde{p}}_\mathcal{F}] \mathop{\mathbb{E}}[\hat{C}_h]
\end{equation}
Considering their formulation in \cref{p_f_tilde_ASTPA,IIS_est_hat}, $\mathop{\mathbb{E}}[\hat{\tilde{p}}_\mathcal{F}]$ and $\mathop{\mathbb{E}}[\hat{C}_h]$ are unbiased estimators of  ${\tilde{p}}_\mathcal{F}$ and $ {C}_h$, respectively, i.e.,  $\mathop{\mathbb{E}}[\hat{\tilde{p}}_\mathcal{F}]={\tilde{p}}_\mathcal{F}$ and $\mathop{\mathbb{E}}[\hat{C}_h]=C_h$. The ASTPA estimator is thus unbiased:
\begin{equation}
\mathop{\mathbb{E}}[\hat{p}_{\mathcal{F}}] = \mathop{\mathbb{E}}[\hat{\tilde{p}}_\mathcal{F} \, \hat{C}_h] = \tilde{p}_\mathcal{F}\, C_h
\end{equation}
The variance ($\widehat{\text{Var}}$) of the ASTPA estimator $\hat{p}_{\mathcal{F}}$ can then be computed as:
\begin{equation}
\widehat{\text{Var}}(\hat{p}_{\mathcal{F}}) = (\hat{\tilde{p}}_{\mathcal{F}})^2 \,\widehat{\text{Var}}(\hat{C}_h) + (\hat{C}_h)^2 \widehat{\text{Var}}(\hat{\tilde{p}}_{\mathcal{F}}) + \widehat{\text{Var}}(\hat{\tilde{p}}_{\mathcal{F}})\widehat{\text{Var}}(\hat{C}_h)
\end{equation}
where $\hat{\tilde{p}}_{\mathcal{F}}$ and $\hat{C}_h$ can be computed according to \cref{p_f_tilde_ASTPA,IIS_est_hat}, respectively. $\widehat{\text{Var}}(\hat{\tilde{p}}_{\mathcal{F}})$ and $\widehat{\text{Var}}(\hat{C}_h)$ can thus be estimated as follows: 
\begin{align}
\widehat{\text{Var}}(\hat{\tilde{p}}_{\mathcal{F}}) = \frac{1}{N_{s}(N_{s}-1)} \sum_{i=1}^{N_{s}} \bigg ( \dfrac{I_{\mathcal{F}} (\bm{x}_i)\,\pi_{\bm{X}}(\bm{x}_i)}{\tilde{h}(\bm{x}_i)}  - \hat{\tilde{p}}_\mathcal{F} \bigg )^{2}
\label{var_IS_p_f_tilde}
\end{align}
\begin{align}
\widehat{\text{Var}}(\hat{C}_h) = \frac{1}{M(M-1)} \sum_{i=1}^{M} \Bigg (\dfrac{\tilde{h}(\bm{x}_i^\prime)}{Q(\bm{x}_i^\prime)} -\hat{C}_{h} \Bigg )^{2}
\label{var_IIS_c_h_tilde}
\end{align}
where $\{\bm{x}_i\}_{i=1}^{N_s}$ and $\{\bm{x}_i^\prime\}_{i=1}^M$ are samples from $h$ and $Q$, respectively. $N_{s}$ denotes the number of used Markov chain samples, taking into account the fact that the samples are not independent and identically distributed (i.i.d) in this case. 
The analytical Coefficient of Variation (C.o.V) can then be provided as: 
\begin{align}
\textrm{C.o.V} \approx \dfrac{\sqrt{\widehat{\text{Var}}(\hat{p}_{\mathcal{F}})}} {\hat{p}_\mathcal{F}} \label{cov_ASTPA}
\end{align}

In this work, the thinning process reduces the sample size from $N$ to $N_s$ based on the effective sample size (ESS) of the sample set $\{\bm{x}_i\}_{i=1}^{N}$. We select every $j^{th}$ sample for thinning, where $j$, an integer, is determined by:
\begin{equation}
    j=\left\lfloor\dfrac{N}{4 \cdot \text{ESS}_{\text{min}}}\right\rfloor
    \label{thining}
\end{equation}
where $\text{ESS}_{\text{min}}$ represents the minimum effective sample size across all dimensions and is computed as implemented in  TensorFlow Probability software \cite{dillon2017tensorflow, tensorflow2024ess}. To ensure $j$ remains within practical bounds for the thinning process, it is constrained to the set $\{3, 4, 5, \ldots, 30\}$, with any calculated value of $j$ outside this range being adjusted to the nearest bound within it. The analytical C.o.V computed according to this approach showed good agreement with the sample C.o.V in all our numerical examples.

\begin{remark}
The unbiasedness of the ASTPA estimator is based on assuming independence between $\hat{\tilde{p}}_{\mathcal{F}}$ and $\hat{C}_h$. If dependence exists, the expected value of their product becomes:
\begin{equation}\label{bias_cov}
\begin{aligned}
\mathop{\mathbb{E}}[\hat{p}_{\mathcal{F}}] &= \mathop{\mathbb{E}}[\hat{\tilde{p}}_\mathcal{F} \, \hat{C}_h] = \mathop{\mathbb{E}}[\hat{\tilde{p}}_\mathcal{F}] \mathop{\mathbb{E}}[\hat{C}_h]+\mathrm{Cov}(\hat{\tilde{p}}_\mathcal{F},\hat{C}_h)\\ &=\tilde{p}_\mathcal{F}\, C_h+\mathrm{Cov}(\hat{\tilde{p}}_\mathcal{F},\hat{C}_h)
\end{aligned}
\end{equation}
where $\mathrm{Cov}(\hat{\tilde{p}}_\mathcal{F},\hat{C}_h)$ is the covariance between $\hat{\tilde{p}}_\mathcal{F}$ and $\hat{C}_h$. This dependence could arise from using the sample set $\{\bm{x}_i\}_{i=1}^{N}$ to fit the ISD $Q$, leading to a potential bias in the estimator.  Despite this potential for bias, our empirical investigations using a finite sample size reveal that any bias of the ASTPA estimator is minimal. One approach to ensure independence, if wanted, is to fit the ISD $Q$ on an additional sample set $\{\bm{x}_i^{\prime\prime}\}_{i=1}^{N} \sim h$, separate from the one used to estimate $\hat{\tilde{p}}_{\mathcal{F}}$. Our empirical findings, however, suggest that such steps are generally unnecessary.
\end{remark}

\subsection{Impact of the constructed sampling target on the estimated probability}\label{sec_impact_on_p_f}
\begin{figure*}[t!]
 \centering
  \begin{tabular}{ccccc}
   \vspace*{-0.3in}
    \hspace{0.3in}
   \includegraphics[width=.24\textwidth,keepaspectratio]{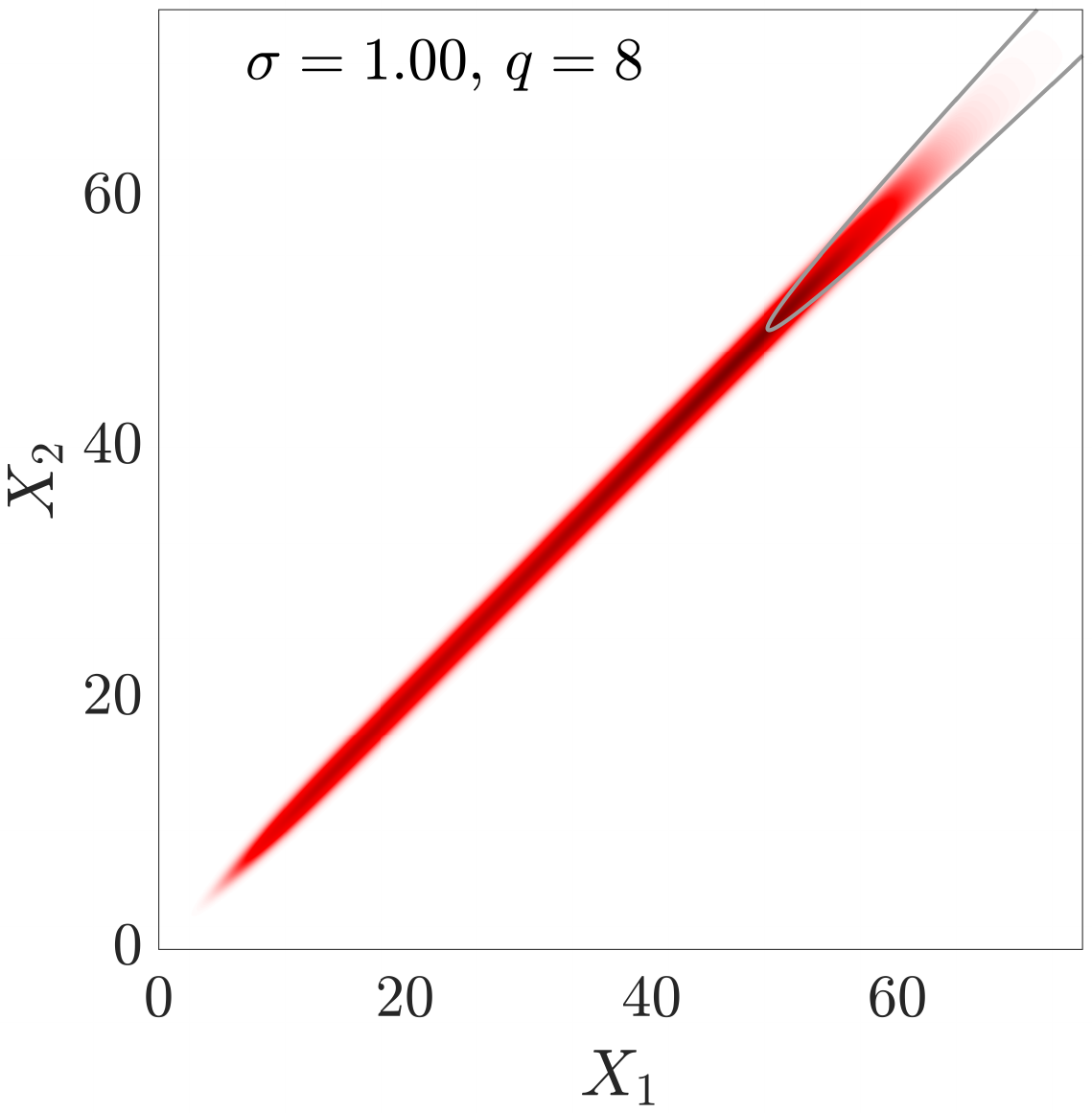}& \hspace*{-0.275in}
   \includegraphics[width=.24\textwidth,keepaspectratio]{Figures/sigma_0.6_q_10.pdf}& \hspace*{-0.275in} 
         \includegraphics[width=.24\textwidth,keepaspectratio]{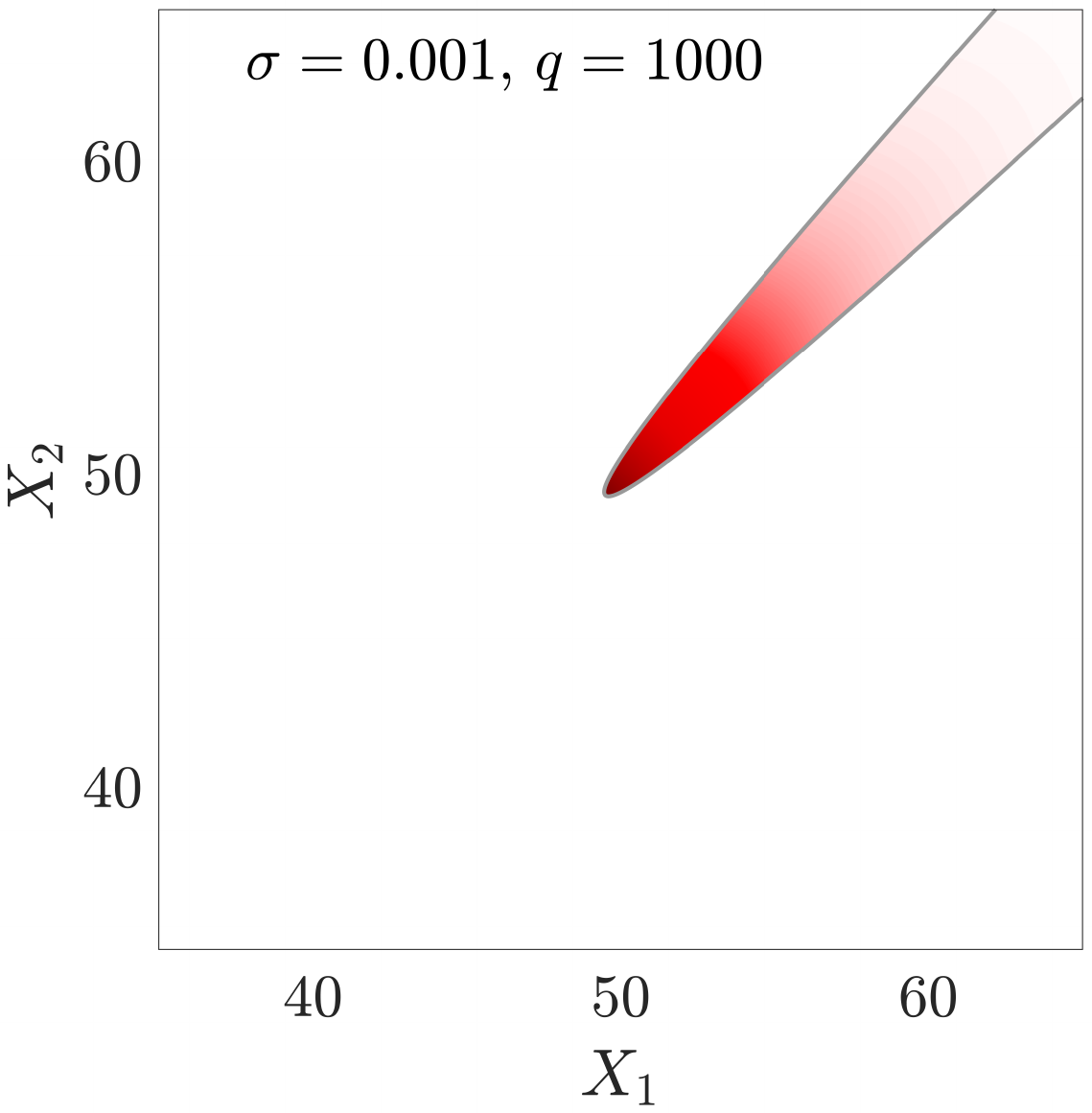}& \hspace*{-0.275in}
\\
    \vspace*{-0.1in}\\
    \hspace{0.35in}
    {\fontsize{8.5pt}{7.2} (a)}&
    \hspace*{-0.25in}
    {\fontsize{8.5pt}{7.2} (b)}&
    \hspace*{-0.25in}
   {\fontsize{8.5pt}{7.2} (c)}
      \\\hline\noalign{\smallskip}

      \hspace{-1.5in}{\fontsize{8.5pt}{7.2}C.o.V$(\hat{\tilde{p}}_{\mathcal{F}})$}&\hspace{-3.1in}{\fontsize{8.5pt}{7.2}0.40} & \hspace{-3.1in}{\fontsize{8.5pt}{7.2}0.03}  & \hspace{-1.5in}{\fontsize{8.5pt}{7.2}0.00}  \\   \hspace{-1.5in}{\fontsize{8.5pt}{7.2}C.o.V$(\hat{C}_h)$}&\hspace{-3.1in}{\fontsize{8.5pt}{7.2}0.13} & \hspace{-3.1in}{\fontsize{8.5pt}{7.2}0.02}  & \hspace{-1.5in}{\fontsize{8.5pt}{7.2}0.13}\\
           \hspace{-1.5in}{\fontsize{8.5pt}{7.2}C.o.V$(\hat{p}_{\mathcal{F}})$}&\hspace{-3.1in}{\fontsize{8.5pt}{7.2}0.40} & \hspace{-3.1in}{\fontsize{8.5pt}{7.2}0.03} & \hspace{-1.5in}{\fontsize{8.5pt}{7.2}0.13}\\
                  \hspace{-1.5in}{\fontsize{8.5pt}{7.2} $\mathop{\mathbb{E}}[\hat{p}_{\mathcal{F}}]$}  &\hspace{-3.1in}{\fontsize{8.5pt}{7.2}2.38E-7} & \hspace{-3.1in}{\fontsize{8.5pt}{7.2}2.52E-7} &  \hspace{-1.5in}{\fontsize{8.5pt}{7.2}2.33E-7} \\\hline
 \end{tabular}
 \caption{Impact of the constructed sampling target on the accuracy of the proposed estimator; reference $p_{\mathcal{F}}\sim 2.51 \times10^{-7}$.}
 \label{Effect_prob_fig}
\end{figure*}

\noindent In this section, we illustrate how constructing the sampling target, $h$, influences the estimated probability $p_\mathcal{F}$. To this end, we consider three cases: (a) a highly relaxed target using $\sigma$ and $q$ values outside the recommended ranges, (b) a target constructed with $\sigma$ and $q$ values within the recommended ranges, and (c) a target highly concentrated inside the rare event domain, i.e., very close to the theoretically optimal ISD in \cref{opt_I_S_density}. We again consider the correlated Gumbel distribution example, as detailed in \cref{Corr_Gumbel}, with $p_{\mathcal{F}}\sim 2.51 \times10^{-7}$. \cref{Effect_prob_fig} depicts these three constructed targets, accompanied by the corresponding $\sigma$ and $q$ values. The impact on the estimator accuracy is judged using the sampling coefficients of variation (C.o.V) of the normalizing constant $\hat{C}_h$ (\cref{IIS_est_hat}), the shifted probability estimate $\hat{\tilde{p}}_{\mathcal{F}}$ (\cref{p_f_tilde_ASTPA}), and the sought probability $\hat{p}_{\mathcal{F}}$ (\cref{p_f_ASTPA_final}), as well as the expected values of $\hat{p}_{\mathcal{F}}$. \cref{Effect_prob_fig} shows the results obtained using a total number of model calls $\sim 4,000$ in all cases and based on 100 independent simulations.

In \cref{Effect_prob_fig}(a), the highly relaxed target is mainly located outside the rare event domain, thus resulting in a low number of rare event samples. Therefore, the C.o.V of $\hat{\tilde{p}}_{\mathcal{F}}$ becomes high since only a few samples with $I_{\mathcal{F}}=1$ appear in \cref{p_f_tilde_ASTPA}. Conversely, the case shown in \cref{Effect_prob_fig}(b) demonstrates excellent performance with low coefficients of variation and $\mathbb{E}[\hat{p}_{\mathcal{F}}]$ very close to the reference value. On the other hand, the highly concentrated target in \cref{Effect_prob_fig}(c), constructed using $\sigma$ and $q$ values outside the recommended range,  presents interesting results with an approximately zero C.o.V for $\hat{\tilde{p}}_{\mathcal{F}}$, and a comparatively higher C.o.V of $\hat{C}_h$. The reason for the former is that almost all the samples yield $I_{\mathcal{F}}=1$, therefore $\hat{\tilde{p}}_{\mathcal{F}} \approx 1.0$ regardless of the number of samples. However, this concentrated sampling target may not often be sampled adequately using an affordable number of model calls compared to an appropriately relaxed target. Consequently, the fitted GMM within the inverse importance sampling procedure may not accurately represent the target $h$. The accuracy of the estimated normalizing constant then slightly deteriorates in this 2-dimensional example. This decrease in accuracy becomes significantly more pronounced if the scenario in \cref{Effect_prob_fig}(c) is followed in high-dimensional problems, e.g., the high-dimensional examples discussed in \cref{Numerical_results}.

The deteriorated performance observed sometimes in certain importance sampling-based methods, particularly for high-dimensional and significantly nonlinear performance functions, can be attributed to the scenario outlined in \cref{Effect_prob_fig}(c). Specifically, some methods, see \citep{au1999new,tabandeh2022review} for example, aim to sample the unnormalized optimal distribution $(I_{\mathcal{F}}\, \pi_{\bm{X}})$, and then fit a normalized PDF based on these samples. Another sample set from the fitted PDF is then employed to compute the normalizing constant of $(I_{\mathcal{F}} \pi_{\bm{X}})$, which is the sought probability $p_\mathcal{F}$. This investigation here, therefore, underscores the sampling benefits of appropriately relaxing (smoothing) the indicator function, often leading to enhanced performance.

\section{Sampling the Approximate Target (\texorpdfstring{$\tilde{h}$}{h}) in ASTPA}\label{Sampling_target}

\noindent In this section, we discuss sampling the approximate target $\tilde{h}$ in ASTPA using gradient-based Markov Chain Monte Carlo (MCMC) methods in this work. Specifically, we focus on the state-of-the-art Hamiltonian MCMC (HMCMC) methods, including our developed Quasi-Newton mass preconditioned HMCMC (QNp-HMCMC), aimed at improving sampling efficiency for intricate probability distributions. We first present the standard HMCMC algorithm, followed by our developed QNp-HMCMC one, emphasizing its theoretical foundations and practical implementation techniques for complex non-Gaussian spaces. Connections with the preconditioned Metropolis-adjusted Langevin algorithm (MALA) are illustrated, highlighting the uniqueness of QNp-HMCMC, even with a single-step integration.
\subsection{Hamiltonian Markov Chain Monte Carlo (HMCMC)}\label{HMCMC}
\noindent HMCMC leverages Hamiltonian dynamics to generate distant Markov chain samples, thereby avoiding the slow exploration of high-dimensional random variable spaces encountered in random-walk samplers. This Hamiltonian approach was first introduced to molecular simulations by Alder and Wainwright in \citep{alder1959studies}, in which the motion of the molecules was deterministic. Subsequently, Duane et al. \citep{duane1987hybrid} integrated MCMC methods with molecular dynamics approaches. The enhanced sampling efficiency of HMCMC is mainly attributed to the augmentation of the space with momentum variables and the utilization of the target distribution's gradients, facilitating informed sampling steps.

Given $d$-dimensional variables of interest $\bm{X}$ with unnormalized density $\tilde{h}$(.), HMCMC introduces $d$-dimensional auxiliary momentum variables $\bm{Z}$ and then samples from the joint distribution characterized by:
\begin{equation}\label{HMC_joint}
    \pi(\bm{X},\textbf{Z}) \propto  \exp\big(-H(
    \bm{X},\bm{Z})\big)
\end{equation}
where $H(\bm{X},\bm{Z})=U(\bm{X})+K(\bm{X},\bm{Z})$ is termed the Hamiltonian function $H$. The functions $U(\boldsymbol{\bm{X}}) = -\log\tilde{h}(\bm{X})$ and $K(\bm{X},\bm{Z})=-\log\pi_{\bm{Z} | \bm{X}}(\bm{Z} | \bm{X})$ are introduced as the potential energy and kinetic energy, owing to the concept of the canonical distribution \citep{neal2011mcmc} and the physical laws which motivate the Hamiltonian Markov Chain Monte Carlo algorithm. The kinetic energy function is unconstrained and can be formed in various ways according to the implementation. In most typical cases, the momentum is sampled by a zero-mean normal distribution \citep{neal2011mcmc,betancourt2017conceptual}, and accordingly, the kinetic energy can be written as $K(\bm{X},\bm{Z})=-\log\pi_{\bm{Z} |\bm{X}}(\bm{Z} | \bm{X}) = \frac{1}{2}\bm{Z}^{T} \mathbf{M}^{-1} \bm{Z}$, where $\mathbf{M}$ is a symmetric, positive-definite matrix, termed mass matrix. The joint distribution $\pi(\bm{X},\bm{Z})$ can then be written as:
\begin{equation}\label{HMC_joint_expand}
\begin{aligned}
    \pi(\bm{X},\textbf{Z}) & \propto \exp\big(-\big(U(\bm{X})+K(\bm{X},\bm{Z})\big)\big) \\ &=
    \tilde{h}(\bm{X})\, \pi_{\bm{Z} |\bm{X}}(\bm{Z} | \bm{X})=\tilde{h}(\bm{X})\, \mathcal{N}(\bm{Z} |\textbf{\textit{0}}, \textbf{M})
    \end{aligned}
\end{equation} 
where $\mathcal{N}(\cdot)$ denotes a Gaussian distribution. HMCMC generates a Metropolis proposal on the joint state-space $(\bm{X},\textbf{Z})$ by sampling the momentum and simulating trajectories of Hamiltonian dynamics in which the time evolution of the state $(\bm{X},\textbf{Z})$ is governed by Hamilton's equations, expressed typically by: 
\begin{equation}
\begin{aligned}
&\frac{d\boldsymbol{\bm{X}}}{dt} = \frac{\partial H}{\partial \bm{Z}} =\frac{\partial K}{\partial \bm{Z}}= \mathbf{M}^{-1} \bm{Z} ,\ \ \  \ 
\\ &\frac{d \bm{Z}}{dt} = -\frac{\partial H}{\partial \bm{X}} = -\frac{\partial U}{\partial \bm{X}}= \nabla_{\bm{X}} \mathcal{L}(\bm{X}) \label{Hamilton_eq}
\end{aligned}
\end{equation}  
where $\mathcal{L}(\bm{X})=\log(\tilde{h}(\bm{X}))$ denotes here the log-density of the target distribution. Hamiltonian dynamics prove to be an effective proposal generation mechanism because the distribution 
$\pi(\bm{X},\textbf{Z})$ is invariant under the dynamics of \cref{Hamilton_eq}. These dynamics enable a proposal, triggered by an approximate solution of \cref{Hamilton_eq}, to be distant from the current state, yet with high probability acceptance. The solution to \cref{Hamilton_eq} is analytically intractable in general and thus the Hamiltonian equations need to be numerically solved by discretizing time using some small step size, $\varepsilon$. A symplectic integrator that can be used for the numerical solution is the \textit{leapfrog} one and works as follows:
\begin{equation} 
\begin{aligned} \label{leapfrog}
&\bm{z}\textsubscript{t+$\varepsilon$/2} = \bm{z}\textsubscript{t} - \dfrac{\varepsilon}{2}\dfrac{\partial U}{\partial \boldsymbol{\bm{X}}} (\bm{x}\textsubscript{t})  ,\ \ \  \
\bm{x}\textsubscript{t+$\varepsilon$} = \bm{x}\textsubscript{t} + \varepsilon \dfrac{\partial K}{\partial \bm{Z}} (\bm{z}\textsubscript{t+$\varepsilon$/2}) ,\ \ \  \ \\&
\bm{z}\textsubscript{t+$\varepsilon$} = \bm{z}\textsubscript{t+$\varepsilon$/2} - \dfrac{\varepsilon}{2}\dfrac{\partial U}{\partial \bm{X}} (\bm{x}\textsubscript{t+$\varepsilon$}) 
\end{aligned}
\end{equation}
where $\dfrac{\partial U}{\partial \bm{X}}(\bm{x})=-\nabla_{\bm{X}} \mathcal{L}(\bm{x})$ and $\dfrac{\partial K}{\partial \bm{Z}} (\bm{z})= \mathbf{M}^{-1} \bm{z} $. The main advantage of using the leapfrog integrator is its simplicity, that is volume-preserving, and that it is reversible, due to its symmetry, by simply negating $\bm{z}$, in order to generate a valid Metropolis proposal. See Neal \citep{neal2011mcmc} and Betancourt \citep{betancourt2017conceptual} for more details on energy-conservation, reversibility and volume-preserving integrators and their connections to HMCMC. It is noted that in the above leapfrog integration algorithm, the computationally expensive part is the one model call per step to acquire the $\dfrac{\partial U}{\partial \bm{X}}$ term. With $\tau$ the trajectory or else path length, taking $\textit{L}=\tau/\varepsilon$ leapfrog steps approximates the evolution $(\bm{x}(0),\bm{z}(0)) \longrightarrow (\bm{x}(\tau),\bm{z}(\tau))$, providing the exact solution in the limit $\varepsilon \longrightarrow 0$. \cref{HMCMC_par} discusses techniques for tuning these parameters.\par 

A generic procedure for drawing \textit{$N_{Iter}$} samples via HMCMC is described in \Cref{alg:stdHMCMC}, where again $\mathcal{L}(\bm{X})$ is the log-density of the target distribution of interest, $\bm{x}^{0}$ are initial values, and \textit{L} is the number of leapfrog steps, as explained before. For each HMCMC step, the momentum is first resampled, and then the \textit{L} leapfrog updates are performed, as seen in \cref{leapfrog}, before a typical accept/reject Metropolis step takes place. In \Cref{alg:stdHMCMC}, the mass matrix $\textbf{M}$ is set to the identity matrix, \textbf{I}, as often followed in the standard implementation of HMCMC. However, the efficiency of the HMCMC can be significantly enhanced by appropriately selecting the mass matrix, particularly when dealing with complex, high-dimensional distributions. In the subsequent section, we introduce our developed technique for adapting the mass matrix, which has shown substantial improvement in the sampling efficiency.

\begin{algorithm}[t!]
\caption{Hamiltonian Markov Chain Monte Carlo}\label{alg:stdHMCMC}
\begin{algorithmic}[1]
\Procedure{HMCMC}{$\bm{x}^{0}$, $\varepsilon$, \textit{L}, $\mathcal{L}(\bm{X})=\log(\tilde{h}(\bm{X}))$, \textit{$N_{Iter}$}} 
\For{\texttt{$m=1$ $to$ $N_{Iter}$}}
\State $\bm{z}^{0}$$\sim$$\mathcal{N}(\textbf{\textit{0}},\mathbf{I})$\\
\hspace{\algorithmicindent}\hspace{\algorithmicindent}\Comment{ sampling from independent standard}
\Statex\hspace{\algorithmicindent}\hspace{\algorithmicindent}\hspace{\algorithmicindent}\hspace{0.38in}
Gaussian distribution, $\mathbf{M}=\mathbf{I}$
\State $\bm{x}^{m}$ $\gets$ $\bm{x}^{m-1}$, $\tilde{\bm{x}}$ $\gets$ $\bm{x}^{m-1}$, $\tilde{\bm{z}}$ $\gets$ $\bm{z}^{0}$
\For{\texttt{$i=1$ $to$ $L$}}
\State $\tilde{\bm{x}}$, $\tilde{\bm{z}}$ $\gets$ Leapfrog($\tilde{\bm{x}}$, $\tilde{\bm{z}}$, $\varepsilon$) \\
\hspace{\algorithmicindent}\hspace{\algorithmicindent}\Comment{leapfrog integration in \cref{leapfrog}}
\EndFor\label{HMCMCfor2}
\State $with$ $probability$:\\ 
       \hspace{1cm} 
       $\alpha$ = min$\bigg\{$1,$\dfrac{\exp(\mathcal{L}(\tilde{\bm{x}})-\dfrac{1}{2} \tilde{\bm{z}}^{T}\tilde{\bm{z}})}{\exp(\mathcal{L}(\bm{x}^{m-1})-\dfrac{1}{2}{\bm{z}^{0}}^{T}\bm{z}^{0})}$
       $\bigg\}$ \\
       \hspace{\algorithmicindent}\hspace{\algorithmicindent}\Comment{Metropolis step}\\
       \hspace{1cm} $\bm{x}^{m}$ $\gets$ $\tilde{\bm{x}}$, $\bm{z}^{m}$ $\gets$ -$\tilde{\bm{z}}$
\EndFor\label{HMCMCfor}
\EndProcedure
\end{algorithmic}
\end{algorithm}

\subsection{Quasi-Newton mass preconditioned HMCMC (QNp-HMCMC)}\label{QNp_HMCMC}
\noindent In complex, high-dimensional problems, the performance of the standard HMCMC sampler, presented as \Cref{alg:stdHMCMC}, may deteriorate, necessitating a prohibitive number of model calls. Various strategies have been proposed to overcome this limitation, including the adaptation of the mass matrix $\mathbf{M}$, as highlighted in \citep{girolami2011riemann,zhang2011quasi,fu2016quasi,hirt2021entropy}. We have uniquely employed this mass matrix adaptation technique in our Quasi-Newton mass preconditioned HMCMC (QNp-HMCMC) sampler, as presented in \Cref{QNHMCMC} and discussed in this section. The role of the mass matrix in HMCMC is analogous to performing a linear transformation for the involved random variables $\bm{X}$, potentially simplifying the geometric complexity of the underlying sampling distribution. Such a transformation may thus make the transformed variables less correlated and to have a similar scale, facilitating a more efficient sampling process. Specifically, operating HMCMC in a transformed space \(\bm{X}' = \mathbf{A}\bm{X}\), where \(\mathbf{A}\) is a non-singular matrix, and using a standard mass matrix \(\mathbf{M}_1 = \mathbf{I}\) is equivalent to using a transformed mass matrix \(\mathbf{M}_2 = \mathbf{A}^T\mathbf{A}\) in the original space $\bm{X}$ \citep{neal2011mcmc,zhang2011quasi}.
\begin{algorithm}[htbp]
\caption{Quasi-Newton mass preconditioned Hamiltonian Markov Chain Monte Carlo (QNp-HMCMC)}\label{QNHMCMC}
\begin{algorithmic}[1]
\Procedure{QNp-HMCMC}{$\bm{x}^{0}$, $\varepsilon$, \textit{L}, $\mathcal{L}(\bm{X})$, $N_\textit{BurnIn}$, \textit{$N_{Iter}$}}\\
 \hspace{0.5cm}$\mathbf{W}^{0} = \mathbf{I}$
\For{\texttt{$m=1$ $to$ $N_{Iter}$}}
\If {$m$ $\leq$ $N_{BurnIn}$}
\State $\bm{z}^{0}$$\sim$$\mathcal{N}(\textbf{\textit{0}},\mathbf{M})$\Comment{where $\mathbf{M}=\mathbf{I}$}
\State $\bm{x}^{m}$ $\gets$ $\bm{x}^{m-1}$, $\tilde{\bm{x}}$ $\gets$ $\bm{x}^{m-1}$, $\tilde{\bm{z}}$ $\gets$ $\bm{z}^{0}$, 
\State $\mathbf{B}$ $\gets$ $\mathbf{W}^{m-1}$, $\mathbf{C}^{0}$ $\gets$ $\mathbf{W}^{m-1}$
\For{\texttt{$i=1$ $to$ $L$}}
\State $\tilde{\bm{x}}$, $\tilde{\bm{z}}$ $\gets$ Leapfrog-BurnIn($\tilde{\bm{x}}$, $\tilde{\bm{z}}$, $\varepsilon$, $\mathbf{B}$)\\
\vspace{.08cm}
\hspace{2cm} Compute $\mathbf{C}^{i}$ using $\mathbf{C}^{i-1}$ in \cref{BFGS}
\EndFor
\State   $\mathbf{W}^{m}$ $\gets$ $\mathbf{C}^{L}$
\State $with$ $probability$:\\ 
       \hspace{1.5cm}$\alpha = \min\big\{$1,$\dfrac{\exp(\mathcal{L}(\tilde{\bm{x}})-\dfrac{1}{2} \tilde{\bm{z}}^{T}\tilde{\bm{z}})}{\exp(\mathcal{L}(\bm{x}^{m-1})-\dfrac{1}{2}{\bm{z}^{0}}^{T}\bm{z}^{0})}$
       $\big\}$\\
       \hspace{1.5cm} $\bm{x}^{m}$ $\gets$ $\tilde{\bm{x}}$, $\bm{z}^{m}$ $\gets$ -$\tilde{\bm{z}}$
\Else \Comment{If $m$ $>$ $N_{BurnIn}$}     
\vspace{0.09cm}
\State $\bm{z}^{0}$$\sim$$\mathcal{N}(\textbf{\textit{0}},\mathbf{M})$\Comment{where $\mathbf{M}=(\mathbf{W}^{N_{BurnIn}})^{-1}$}
\State $\bm{x}^{m}$ $\gets$ $\bm{x}^{m-1}$, $\tilde{\bm{x}}$ $\gets$ $\bm{x}^{m-1}$, $\tilde{\bm{z}}$ $\gets$ $\bm{z}^{0}$
\For{\texttt{$i=1$ $to$ $L$}}
\State $\tilde{\bm{x}}$, $\tilde{\bm{z}}$ $\gets$ Leapfrog($\tilde{\bm{x}}$, $\tilde{\bm{z}}$, $\varepsilon$, $\mathbf{M}$)
\EndFor
\State $with$ $probability$:\\ 

       \hspace{1.5cm}$\alpha=\min\big\{$1,$\dfrac{\exp(\mathcal{L}(\tilde{\bm{x}})-\dfrac{1}{2} \tilde{\bm{z}}^{T}\ \mathbf{M}^{-1}\tilde{\bm{z}})}{\exp(\mathcal{L}(\bm{x}^{m-1})-\dfrac{1}{2}{\bm{z}^{0}}^{T}\ \mathbf{M}^{-1}\bm{z}^{0})}$
       $\big\}$\\
       \hspace{1.5cm} $\bm{x}^{m}$ $\gets$ $\tilde{\bm{x}}$, $\bm{z}^{m}$ $\gets$ -$\tilde{\bm{z}}$
\EndIf  
\EndFor
\EndProcedure
\\
\\
\\
\\
\Function {Leapfrog-BurnIn}{$\bm{x}, \bm{z}, \varepsilon, \mathbf{B}$}
\State $\tilde{\bm{z}} \gets \bm{z}+(\varepsilon/2)\mathbf{B}\nabla_{\bm{X}}\mathcal{L}(\bm{x})$
\State $\tilde{\bm{x}} \gets \bm{x}+\varepsilon\mathbf{B}\tilde{\bm{z}}$
\State $\tilde{\bm{z}} \gets \tilde{\bm{z}}+(\varepsilon/2)\mathbf{B}\nabla_{\bm{X}}\mathcal{L}(\tilde{\bm{x}})$\\
\Return $\tilde{\bm{x}}$, $\tilde{\bm{z}}$ 
\EndFunction
\\
\\
\Function {Leapfrog}{$\bm{x}, \bm{z}, \varepsilon, \mathbf{M}$}
\State $\tilde{\bm{z}} \gets \bm{z}+(\varepsilon/2)\nabla_{\bm{X}}\mathcal{L}(\bm{x})$
\State $\tilde{\bm{x}} \gets \bm{x}+\varepsilon\mathbf{M}^{-1}\tilde{\bm{z}}$
\State $\tilde{\bm{z}} \gets \tilde{\bm{z}}+(\varepsilon/2)\nabla_{\bm{X}}\mathcal{L}(\tilde{\bm{x}})$\\
\Return $\tilde{\bm{x}}$, $\tilde{\bm{z}}$ 
\EndFunction
\end{algorithmic}
\end{algorithm}

The mass matrix is commonly adapted to better describe the local geometry of the target distribution, enabling more efficient exploration of the space of random variables. For instance, the Riemannian Manifold Hamiltonian Monte Carlo (RMHMC) approach, suggested in \citep{girolami2011riemann}, leverages the manifold structure of the variable space through a position-specific Riemannian metric tensor. This tensor functions as a position-dependent mass matrix, leading to a non-separable Hamiltonian. Therefore, solving the Hamiltonian equations within RMHMC requires a generalized leapfrog scheme, demanding higher-order derivatives of distributions and additional model calls per leapfrog step. Conversely, within the scope of our research, a more computationally feasible approach is to introduce a position-independent mass matrix $\mathbf{M}(\bm{x})=\mathbf{M}$, as in \cref{alg:stdHMCMC}, describing the general local structure of the target distribution. Such description can be generally obtained in a Newton-type context, utilizing the Hessian information of the target distribution, as demonstrated in various MCMC methods \citep{qi2002hessian,martin2012stochastic,simsekli2016stochastic,leimkuhler2018ensemble}. In the context of HMCMC, Zhang and Sutton \citep{zhang2011quasi} employed a quasi-Newton-type approach to provide a position-independent, yet continuously adapted, mass matrix using a variant of the BFGS formula \citep{brodlie1973rank} to approximate the Hessian matrix. They maintained position independence by adopting the ensemble chain adaption (ECA) approach \citep{gilks1994adaptive}, updating the mass matrix for each chain based on the states of others. In QNp-HMCMC, we utilize an alternative approach to ensure position independence by approximating the Hessian matrix during a burn-in phase and subsequently utilizing it as a constant mass matrix.  Importantly, we enhance the burn-in stage's efficiency through a skew-symmetric preconditioning approach, while setting the mass matrix to identity. \par

In this work, we approximate the Hessian information of the potential energy function,  $U(\boldsymbol{\bm{X}}) =-\mathcal{L}(\bm{X})= -\log\tilde{h}(\bm{X})$, using the well-known BFGS approximation \citep{nocedal2006numerical}, solely based on the gradient information.
 Let $\bm{X}\in\mathbb{R}^{d}$, consistent with the previous sections. Given the $k^{th}$ estimate $\mathbf{W}_{k}$, where $\mathbf{W}_{k}$ is an approximation to the inverse Hessian of the potential energy function at $\bm{x}_{k}$, the BFGS update $\mathbf{W}_{k+1}$ can be expressed as:
\begin{align}
\mathbf{W}_{k+1} = (\mathbf{I}-\dfrac{\bm{s}_{k} \bm{y}_{k}^{T}}{\bm{y}_{k}^{T} \bm{s}_{k}})\mathbf{W}_{k}(\mathbf{I}-\dfrac{\bm{y}_{k} \bm{s}_{k}^{T}}{\bm{y}_{k}^{T} \bm{s}_{k}})+\dfrac{\bm{s}_{k} \bm{s}_{k}^{T}}{ \bm{y}_{k}^{T}  \bm{s}_{k}} \label{BFGS}
\end{align}
where $\mathbf{I}$ is the identity matrix, $\bm{s}_{k}=\bm{x}_{k+1}-\bm{x}_{k}$, and $\bm{y}_{k}= -\nabla \mathcal{L}(\bm{x}_{k+1})+\nabla \mathcal{L}(\bm{x}_{k})$ where $\mathcal{L}:\mathbb{R}^{d} \longrightarrow\mathbb{R}$ denotes the log density of the target distribution. There is a long history of efficient BFGS updates for very large systems and several numerical techniques can be used, including sparse and limited-memory approaches \citep{liu1989limited}. \par

To achieve a good approximation of the inverse Hessian matrix during the burn-in stage, we enhance the sampling efficiency by resorting to a skew-symmetric preconditioning approach for Hamiltonian dynamics. Such preconditioning approaches are originally introduced in \citep{ma2015complete} and utilized within HMCMC in \citep{fu2016quasi}, albeit in completely different settings compared to our QNp-HMCMC. Ma et al. in \citep{ma2015complete} devised a unified framework for continuous-dynamics samplers, such as HMCMC and Langevin Monte Carlo. Let $\boldsymbol{\theta}=(\bm{X}, \bm{Z}) \in \mathbb{R}^{2d}$ be an augmented state. The suggested dynamics in \citep{ma2015complete} can then be written as a stochastic differential equation (SDE) of the form:
\begin{equation}\label{unified_framework}
     d\boldsymbol{\theta}= \mathbf{f}(\boldsymbol{\theta})\,dt+\sqrt{2\,\mathbf{D}(\boldsymbol{\theta})}\,d\mathbf{W}(t)
\end{equation}
where $\mathbf{f}(\boldsymbol{\theta})$ denoted the deterministic drift and often related to the gradient of the Hamiltonian $H(\boldsymbol{\theta})$, $\mathbf{W}(t)$ is a $2d$-dimensional Wiener process, and $\mathbf{D}(\boldsymbol{\theta}) \in \mathbb{R}^{2d \times 2d}$ is a positive semidefinite diffusion matrix. The deterministic drift is further proposed as:
\begin{equation}\label{drift}
\begin{aligned}
     &\mathbf{f}(\boldsymbol{\theta})=-[\mathbf{D}(\boldsymbol{\theta})+\mathbf{S}(\boldsymbol{\theta})]\nabla H(\boldsymbol{\theta}) + \boldsymbol{\Gamma}(\boldsymbol{\theta}),\quad\\ & \boldsymbol{\Gamma}_i(\boldsymbol{\theta})=\sum_{j=1}^d \dfrac{\partial}{\partial\boldsymbol{\theta}_j} \big(\mathbf{D}_{ij}(\boldsymbol{\theta})+\mathbf{S}_{ij}(\boldsymbol{\theta})\big)
     \end{aligned}
\end{equation}
where $\mathbf{S}(\boldsymbol{\theta})\in \mathbb{R}^{2d \times 2d}$ is a skew-symmetric curl matrix representing the deterministic traversing effects seen in the HMCMC procedure. Ma et al. \citep{ma2015complete} proved that for any choice of positive semidefinite $\mathbf{D}(\boldsymbol{\theta})$ and skew-symmetric
$\mathbf{S}(\boldsymbol{\theta})$ parameterizing $\mathbf{f}(\boldsymbol{\theta})$, simulating from \cref{unified_framework} with $\mathbf{f}(\boldsymbol{\theta})$ as in \cref{drift} leads to the desired stationary distribution: $\pi(\boldsymbol{\theta}) \propto  \exp\big(-H(\boldsymbol{\theta})\big)$ in \cref{HMC_joint}. Deterministic dynamics, such as Hamiltonian dynamics, can be obtained under this unified framework in \cref{unified_framework} by setting the diffusion matrix $\mathbf{D}(\boldsymbol{\theta})=0$ as: 
\begin{equation}
     d\boldsymbol{\theta}=-\mathbf{S}(\boldsymbol{\theta})\nabla H(\boldsymbol{\theta}) + \boldsymbol{\Gamma}(\boldsymbol{\theta}),\quad  \boldsymbol{\Gamma}_i(\boldsymbol{\theta})=\sum_{j=1}^d \dfrac{\partial}{\partial\boldsymbol{\theta}_j} \big(\mathbf{S}_{ij}(\boldsymbol{\theta})\big)
\end{equation}
Hamiltonian dynamics in \cref{Hamilton_eq} can then be retrieved by setting $\mathbf{S}(\boldsymbol{\theta})=\begin{bmatrix}
0 &-\mathbf{I}_d\\ \mathbf{I}_d &0\\\end{bmatrix}$, with $\mathbf{I}_d \in  \mathbb{R}^{d \times d}$ being an identity matrix,  as:
\begin{equation}
\begin{aligned}
     \dfrac{d\boldsymbol{\theta}}{dt} &=\dfrac{d}{dt}\begin{bmatrix}
\bm{X} \\[6pt] 
\bm{Z} \\
\end{bmatrix}=-\begin{bmatrix}
0 &-\mathbf{I}_d\\[6pt]  \mathbf{I}_d &0\\\end{bmatrix} \begin{bmatrix}
\nabla_{\bm{X}} H(\bm{X},\bm{Z}) \\[6pt] 
\nabla_{\bm{Z}} H(\bm{X},\bm{Z}) \\
\end{bmatrix}\\&=-\begin{bmatrix}
0 &-\mathbf{I}_d\\[6pt]  \mathbf{I}_d &0\\\end{bmatrix} \begin{bmatrix}
-\nabla_{\bm{X}} \mathcal{L}(\bm{X}) \\[6pt] 
 \mathbf{M}^{-1} \bm{Z} \\
\end{bmatrix}= \begin{bmatrix}
\mathbf{M}^{-1} \bm{Z} \\[6pt] 
\nabla_{\bm{X}} \mathcal{L}(\bm{X}) 
\end{bmatrix}
\end{aligned}
\end{equation}
It is worth mentioning that if the mass matrix is position-dependent, such as in RMHMC, a different treatment for the unified dynamics in \cref{unified_framework} should be performed, to retrieve the corresponding dynamics; see \citep{ma2015complete} for more details.\par

In our suggested QNp-HMCMC, we utilize a skew-symmetric preconditioning matrix $\mathbf{S}(\boldsymbol{\theta})=\begin{bmatrix}
0 &-\mathbf{W}\\ \mathbf{W} &0\\\end{bmatrix}$, with $\mathbf{W}$ being the inverse Hessian of the potential energy function. Assuming $\mathbf{W}$ is position-independent, the correction term can be ignored, i.e., $\boldsymbol{\Gamma}_i(\boldsymbol{\theta})=0$, and therefore the corresponding dynamics can then be written as:
\begin{equation}
\begin{aligned}
     \dfrac{d\boldsymbol{\theta}}{dt} &=\dfrac{d}{dt}\begin{bmatrix}
\bm{X} \\[6pt] 
\bm{Z} \\
\end{bmatrix}=-\begin{bmatrix}
0 &-\mathbf{W}\\[6pt]  \mathbf{W} &0\\\end{bmatrix} \begin{bmatrix}
-\nabla_{\bm{X}} \mathcal{L}(\bm{X}) \\[6pt] 
 \mathbf{M}^{-1} \bm{Z} \\
\end{bmatrix}\\&= \begin{bmatrix}
\mathbf{W}\,\mathbf{M}^{-1} \bm{Z} \\[6pt] 
\mathbf{W}\,\nabla_{\bm{X}} \mathcal{L}(\bm{X}) 
\end{bmatrix}
\end{aligned}
\end{equation}
Simulating these dynamics keeps the target distribution invariant as it is cast under the discussed unified framework of continuous-dynamics samplers \citep{ma2015complete}. Accordingly, the leapfrog integrator is then reformulated as:
\begin{equation} \label{leapfrog_QNp}
\begin{aligned}
&\bm{z}\textsubscript{t+$\varepsilon$/2} = \bm{z}\textsubscript{t} + \dfrac{\varepsilon}{2} \mathbf{W}\textsubscript{t}\nabla_{\bm{X}}\mathcal{L}(\bm{x}\textsubscript{t}) ,\ \   
\bm{x}\textsubscript{t+$\varepsilon$} = \bm{x}\textsubscript{t} +\varepsilon\mathbf{W}\textsubscript{t}\,\mathbf{M}^{-1}\bm{z}\textsubscript{t+$\varepsilon$/2} ,\ \ \  \ \\&
\bm{z}\textsubscript{t+$\varepsilon$} = \bm{z}\textsubscript{t+$\varepsilon$/2} + \dfrac{\varepsilon}{2}\mathbf{W}\textsubscript{t}\nabla_{\bm{X}}\mathcal{L}(\bm{x}\textsubscript{t+$\varepsilon$}) 
\end{aligned}
\end{equation} 
Since the last update of the inverse Hessian, $\mathbf{W}\textsubscript{t}$, is employed to simulate the current step, the position-independence of $\mathbf{S}$ may not be achieved, and thus, a Metropolis correction should be devised to lead to the correct stationary distribution, as we follow in this work. Alternatively, an older update of the inverse Hessian, say $\mathbf{W}\textsubscript{t-5}$, can be employed to ensure position independence. A similar approach can be seen in \citep{simsekli2016stochastic} in the context of Langevin Monte Carlo. Additionally, we use here the same inverse Hessian for all steps of a simulated trajectory: $(\bm{x}(0),\bm{z}(0)) \longrightarrow (\bm{x}(\tau),\bm{z}(\tau))$. Following this preconditioning scheme thus implies that the new dynamics efficiently and compatibly adjust both the $\bm{Z}$ and $\bm{X}$ evolutions based on the local structure of the target distribution, also featuring and also features a Quasi-Newton direction for the momentum variables, allowing large jumps across the state space. In the burn-in stage of QNp-HMCMC, as presented in \Cref{QNHMCMC}, we set the mass matrix to identity $\big(\bm{Z} \sim \mathcal{N}(\textbf{\textit{0}},\mathbf{I})\big)$, while the skew-symmetric preconditioning is in effect. 


After this adaptive burn-in phase, we leverage the adapted inverse Hessian matrix to provide an informed, preconditioned mass matrix for the subsequent non-adaptive sampling phase of the algorithm. This is performed by inverting the final estimate of the inverse Hessian and utilizing this Hessian matrix itself as a constant mass matrix, i.e., $\mathbf{M}=\mathbf{W}^{-1}$. As such, in this non-adaptive phase of QNp-HMCMC, we follow the original dynamics in \cref{Hamilton_eq} and the leapfrog integrator in \cref{leapfrog}. However, we now utilize the properly constructed mass matrix that largely takes into account the scale and correlations of the position variables, leading to significant efficiency gains, particularly in complex, high-dimensional problems.

Utilizing the Hessian matrix as a mass matrix to simulate the momentum variables, $\bm{Z} \sim \mathcal{N}(\textbf{\textit{0}},\mathbf{M})$,  necessitates its positive definiteness. The BFGS procedure in \cref{BFGS} normally provides a symmetric, positive-definite $\mathbf{W}$ matrix in an optimization context. However, in our case, we are using BFGS under different settings that may not satisfy the curvature condition $\bm{s}_{k}^{T} \bm{y}_{k} > 0$, resulting in occasional deviations from positive-definiteness. Several standard techniques can be then implemented to ensure positive-definiteness, such as a damped BFGS updating \citep{nocedal2006numerical} or the simple addition $\mathbf{W}_{new} = \mathbf{W}_{old} + \delta \mathbf{I}$, where $\delta\geq 0$ is some appropriate number. Alternatively, $\mathbf{W}$ can be updated within QNp-HMCMC only when the curvature condition is satisfied, which directly guarantees positive definiteness. To further ensure the stability of the sampler, a positive threshold can be introduced to the curvature condition, e.g., $\bm{s}_{k}^{T} \bm{y}_{k} > 10$. This latter approach has been used and has worked well in this work. As can be seen in \Cref{QNHMCMC}, approximating the inverse Hessian is independent of the accept/reject Metropolis step. That is, we always use the simulated sample to adapt $\mathbf{W}$, even if the proposed sample is rejected later, since each accepted/rejected point provides insights about the space geometry. As a safeguard against faraway proposed samples, we may also return to the previous approximation of $\mathbf{W}$ when the acceptance probability is too low. 

To summarize, our developed \textit{Quasi-Newton mass preconditioned Hamiltonian Markov Chain Monte Carlo} (QNp-HMCMC) consists of two integrated coupled phases, an adaptive and a non-adaptive, as concisely presented in \Cref{QNHMCMC}. In the first adaptive burn-in phase, we approximate the inverse Hessian matrix, $\mathbf{W}$, using a BFGS formula, while the sampler dynamics are enhanced using a skew-symmetric scheme. The final estimate of the inverse Hessian matrix from this phase is then utilized to provide a constant, preconditioned mass matrix $(\mathbf{M}=\mathbf{W}^{-1})$ to be utilized during the non-adaptive sampling phase. Overall, QNp-HMCMC is a practical, efficient approach that only requires gradient information and provides important insight about the geometry of the target distribution, eventually improving computational performance and enabling faster mixing.\par
\subsection{(QNp-)HMCMC parameters}\label{HMCMC_par}
\noindent The performance of HMCMC algorithms relies on carefully selecting suitable values for the step size $\varepsilon$ and length $\tau$ of the simulated trajectory. The stepsize $\varepsilon$ can be efficiently and independently tuned to achieve some target average acceptance rate. In this work, we set this target acceptance rate to $65\%$, as values between $60\%$ and $80\%$ are typically assumed optimal \citep{neal2011mcmc,beskos2013optimal,hoffman2014no}. To this end, we adopt the dual averaging algorithm of Hoffman and Gelman \citep{hoffman2014no,nesterov2009primal} using its default parameters provided in \citep{hoffman2014no}, however, with a given appropriate initial step size. Therefore, in QNp-HMCMC, the step size $\varepsilon$ is automatically tuned, and we do that in the first $2N_{BurnIn}$ iterations, after which it is set at the estimated optimal value. This adaptation is extended beyond the burn-in phase, given that different dynamics are used in each phase of QNp-HMCMC, probably requiring different optimal step sizes.

In contrast, tuning the trajectory length $\tau$, or equivalently the number of steps, $\textit{L}=\tau/\varepsilon$, is more challenging and is often achieved by maximizing variants of the Expected Squared Jumping Distance (ESJD) \citep{pasarica2010adaptively}, ${\mathbb{E}}_{L}\lVert\bm{x}^{(m+1)} - \bm{x}^{(m)}\rVert^{2}$, such as normalized ESJD \citep{wang2013adaptive} and Change in the Estimator of the Expected Square (ChEES) \cite{hoffman2021adaptive}. Motivated by the idea of ESJD, Hoffman and Gelman \citep{hoffman2014no} proposed a reliable, robust HMCMC variant known as the No-U-Turn Sampler (NUTS), which basically relies on the Leapfrog integrator, building a tree procedure, until the simulation makes a “U-turn”. About half of these leapfrog steps are thus wasted to satisfy detailed balance \citep{hoffman2021adaptive, wu2018faster}. Obviously, such a computationally expensive approach is not suitable in this context of rare event estimation, as we aim to minimize the number of model calls. However, we observed that our QNp-HMCMC in non-Gaussian domains can perform well even when one leapfrog step is utilized. Since this aligns with our goal of minimizing the model calls, we adopted a single-step QNp-HMCMC implementation in this work, which worked very well in practice.  It is, however, understood that using multiple steps can generally achieve faster mixing rates for HMCMC \citep{chen2020fast}. Yet, the QNp-HMCMC has demonstrated good performance with a single-step implementation due to the described utilized preconditioned dynamics and derived mass matrix. Utilizing a single leapfrog step has also been adopted in the recent variant of the Generalized Hamiltonian Monte Carlo (GHMC), introduced in \citep{hoffman2022tuning}. 


\subsection{Connections with the preconditioned Metropolis adjusted Langevin algorithm (MALA)}\label{Connection_MALA}
\noindent In this section, we explore connections between QNp-HMCMC with a single-step implementation and preconditioned Langevin Monte Carlo (LMC) samplers, including the Metropolis adjusted Langevin algorithm (MALA). Such study provides insights into LMC preconditioning, leading to a novel, efficient preconditioning scheme. The preconditioned Langevin Monte Carlo proposal can be generally written as \citep{roberts2002langevin,girolami2011riemann}:  
\begin{equation}
\tilde{\bm{x}} = \bm{x}\textsubscript{t} +\dfrac{\varepsilon^2}{2}\,\mathbf{A}\nabla_{\bm{X}}\mathcal{L}(\bm{x}\textsubscript{t})+\varepsilon\mathbf{A}^{1/2}\bm{z}^{\prime}_{\text{t}} 
\label{LMC}
\end{equation}
where $\mathbf{A}$ is a preconditioning positive definite matrix, $\mathbf{A}^{1/2}$ can be obtained using Cholesky decomposition, as $\mathbf{A}=\mathbf{A}^{1/2}(\mathbf{A}^{1/2})^{T}$,  the log density $\mathcal{L}(\bm{X})= \log\tilde{h}(\bm{X})$, with $\tilde{h}$ being the target distribution, and  $\bm{Z}^\prime \sim \mathcal{N}(\textbf{\textit{0}},\mathbf{I})$. 
\begin{lemma}\label{MALA_proposal_equivalence}
    QNp-HMCMC with a single-step implementation is equivalent to a preconditioned Langevin Monte Carlo (LMC) with an adaptive preconditioning matrix $\mathbf{A}_\text{t}=\mathbf{W}_\text{t}^2$ in the burn-in stage and a fixed preconditioning matrix  $\mathbf{A}=\mathbf{M}^{-1}=\mathbf{W}_{N_{BurnIn}}$ afterwards, where $\mathbf{W}_\text{t}$ is the inverse Hessian matrix approximated at iteration $t$ of the burn-in stage using the BFGS formula, and $\mathbf{W}_{N_{BurnIn}}$ is the final approximation of the inverse Hessian matrix. 
\end{lemma}

\begin{lemma}\label{MALA_acceptance_equivalence}
    Considering the accept/reject Metropolis step, QNp-HMCMC with a single-step implementation is equivalent to a preconditioned Metropolis adjusted Langevin algorithm (MALA), with preconditioning matrices described in \Cref{MALA_proposal_equivalence}.
\end{lemma}
Proofs of \Cref{MALA_proposal_equivalence,MALA_acceptance_equivalence} are provided in Appendix A. Despite many existing LMC variants, the proposed preconditioning approach, $\mathbf{A}_\text{t}=\mathbf{W}_\text{t}^2$, is unique, to the best of our knowledge. Based on \Cref{MALA_proposal_equivalence,MALA_acceptance_equivalence}, the preconditioned MALA scheme corresponding to QNp-HMCMC with a single-step integration is a novel, efficient algorithm, as shown in this work, and can be easily retrieved from \Cref{QNHMCMC} by setting $L=1$. Further, the computational cost of QNp-HMCMC can be reduced by following the decomposition approach in \cref{LMC} and utilizing a BFGS variant directly providing the decomposed (inverse) Hessian, such as the formula in \citep{brodlie1973rank}, for example.

\section{Rare Event Domain Discovery}\label{Discovery}
\noindent As is well known, good initial points can significantly enhance the efficiency of any MCMC sampler by facilitating faster convergence to the stationary target distribution. This is particularly crucial in the context of estimating probabilities of rare events, where identifying such points poses challenges due to the arbitrary existence of rare event domains within the tail regions of the original distribution; see \cref{Corr_Gumbel_fig}, for example. Therefore, randomly starting at a sample from the original distribution of the involved random variables most often initializes the sampler far from the regions of interest. In our previous work \citep{Papakon2023HMCMC}, we adopted an annealing approach for the approximate sampling target in ASTPA, $\tilde{h}$, in which the sampling target is continuously moving toward the rare event domain in the burn-in phase, facilitating rare event discovery. This approach, accompanied by the powerful HMCMC samplers, has demonstrated excellent performance when used in Gaussian spaces.

In this work, we alternatively use an optimization approach to initialize the sampler, given our focus on complex non-Gaussian spaces. Specifically, the Adam optimizer \citep{kingma2014adam} is employed to minimize the negative logarithm of the target distribution $(-\log\tilde{h})$, given its ability to automatically tune its step size, while also utilizing momentum for accelerated convergence. We initialize Adam at the mean of the original distribution, $\boldsymbol{\mu}_{\pi_{\bm{X}}}$, and set its learning rate to $0.1$ and its other parameters to their default values, as recommended in \citep{kingma2014adam}. This optimization is performed over a maximum of 500 iterations or else until convergence is reached, indicated by the L2 norm of the optimizer's update dropping below $10^{-7}$.

This discussed optimization approach can also be used to verify the appropriateness of the ASTPA parameters, $g_c$ and $\sigma$. Convergence of the Adam optimizer to a state, $\bm{x}_{Adam}$, within the rare event domain, or to a state with a limit-state value relatively near zero, suggests that the sampling target is well-placed within the regions of interest. Otherwise, we may reduce the $\sigma$ value and/or change the scaling constant $g_c$ to shift the sampling target closer to these regions of interest. In this case, running Adam on the new target, starting from $\bm{x}_{Adam}$, often requires a few iterations to converge. Our observations, however, indicate that the recommended parameter ranges for $\sigma$ and  $g_c$ in \cref{Targ_form} are generally effective in practice.\par

\section{ASTPA in Bounded Spaces}\label{bounded}
\noindent To obviate the need to deal with variable bounds in non-Gaussian spaces while simulating the Hamiltonian dynamics during sampling, a transformation to an unconstrained space can be performed \cite{PapakonICOSSAR2022}. A new random variable $\bm{Y} =\bm{R}(\bm{X}) $ can be characterized by transforming a $d$-dimensional random variable $\bm{X}$, with PDF $\pi_{\bm{X}}(\bm{X})$, using a properly well-defined function $\bm{R}$. If $\bm{R}$ is a one-to-one functional mapping and its inverse $\bm{R}^{-1}$ is available, with an explicit Jacobian, then the density of $\bm{Y}$ is expressed as: 
\begin{equation}
    \pi_{\bm{Y}}(\bm{Y}) = \pi_{\bm{X}}\big (\bm{R}^{-1}(\bm{Y})\big ) \,  \mathopen \bigg| \det J_{\bm{R}^{-1}}(\bm{Y})  \mathclose \bigg|
\end{equation}    
where "det" denotes the determinant operation of the Jacobian matrix, $J_{\bm{R}^{-1}}(\bm{Y})$. We are interested in constructing a map $R_i: X_i \rightarrow Y_i$, with $i=1,2,...,d$, where $Y_i$ are unbounded random variables. In this scenario, the Jacobian matrix will be diagonal and thus easy to compute. Subsequently, we present some commonly utilized mappings based on distinct relevant categories of bounds.\par
\textit{Lower bounded variables}:
A one-dimensional random variable $X_i$, having a lower bound $\alpha_i$, can be transformed to an unconstrained variable $Y_i=R_i(X_i)$ using a logarithmic transform, as $Y_i = \log(X_i-\alpha_i)$, and thus  $X_i = R_i^{-1}(Y_i) =\exp(Y_i) + \alpha_i$. Therefore, the corresponding $i^{th}$ diagonal element in $J_{R^{-1}}(\bm{Y})$ will be $\exp(Y_i)$.\par
	
\textit{Upper bounded variables}: The logarithmic transformation $Y_i = \log(\beta_i - X_i)$
can be utilized to transform a one-dimensional random variable $X_i$, with an upper bound $\beta_i$, to an unconstrained random variable $Y_i$. $X_i$ can be then also retrieved by $ X_i=R_i^{-1}(Y_i)= \beta_i - \exp(Y_i)$. The corresponding $i^{th}$ diagonal element in the Jacobian matrix will accordingly be $\big(-\exp(Y_i)\big)$.

\textit{Lower and upper bounded variables}:  A random variable $X_i$ with a lower bound $\alpha_i$ and an upper bound $\beta_i$ can be transformed to an unconstrained random variable $Y_i$, as $Y_i = \text{logit}(\largesfrac{X_i-\alpha_i}{\beta_i-\alpha_i})$, where the logit function is defined for $w \in (0,1)$ as $\text{logit}(w)=\log(\largesfrac{w}{1-w})$. The inverse map and its derivative are then expressed as $X_i =  R_i^{-1}(Y_i)=\alpha_i + (\beta_i - \alpha_i) \,\text{logit}^{-1}(Y_i)$ and $\largesfrac{dX_i}{dY_i}=(\beta_i - \alpha_i)  \,\text{logit}^{-1}(Y_i) \, (1 - \text{logit}^{-1}(Y_i))$, respectively, where $\text{logit}^{-1}(\kappa) = \largesfrac{1}{\big(1 + \exp(-\kappa)\big)}$ for $\kappa \in (-\infty,+\infty)$.\par

\section{Summary of (QNp-)HMCMC-based ASTPA}\label{ASTPA_summary}
\noindent The proposed (Quasi-Newton mass preconditioned) HMCMC based ASTPA framework can be directly applied in non-Gaussian spaces as follows:
\begin{enumerate}
    \item Constructing the approximate sampling target $\tilde{h}$ in \cref{Tar_ASTPA}, where its two parameters $\sigma$ and $g_c$ are determined as recommended in \cref{Targ_form}.

    \item Running Adam optimizer to get a good initial state for the sampler, as described in \cref{Discovery}, with a number of iterations, $N_{Adam}$, typically less than 500.
    
    \item Utilizing the (QNp-)HMCMC with single-step implementation to draw a sample set $\{\bm{x}_{i}\}_{i=1}^{N} \sim h$. We consider a burn-in phase with a length $N_{BurnIn} \approx 10$-$15\%\, \text{of}\, N$, particularly needed to approximate the inverse Hessian in QNp-HMCMC. The leapfrog step size $\varepsilon$ is automatically tuned during the first $2N_{BurnIn}$ iterations using the dual averaging algorithm of Hoffman and Gelman \citep{hoffman2014no}. 
    
    \item Computing the expected value of the weighted indicator function, $\hat{\tilde{p}}_\mathcal{F}$, using \cref{p_f_tilde_ASTPA}, based on $\{\bm{x}_{i}\}_{i=1}^{N}$.

    \item Applying Inverse Importance Sampling (IIS) to compute the normalizing constant of $\hat{C}_h$. This process initiates by fitting a GMM $Q(\boldsymbol{\bm{X}})$ based on $\{\bm{x}_{i}\}_{i=1}^{N}$, with the $Q(\boldsymbol{\bm{X}})$ structured as recommended in \cref{IIS}. An i.i.d. sample set $\{\bm{x}_i\}_{i=1}^{M}$ is then drawn from $Q(\boldsymbol{\bm{X}})$ to compute $\hat{C}_h$ using \cref{IIS_est_hat}. $M$ can be generally around $30\%$ of $N$.
    
    \item Computing the sought rare event probability as $\hat{p}_\mathcal{F}= \hat{\tilde{p}}_\mathcal{F} \,\, \hat{C}_h $.
\end{enumerate}
Naturally, the required total number of model calls $(N_{Total}=N_{Adam}+N_{BurnIn}+N+M)$ is case dependent, and convergence of the estimator can be checked through \cref{var_IS_p_f_tilde,var_IIS_c_h_tilde}. However, $N_{total}$ can roughly be $5,000$-$10,000$ modal calls for high-dimensional problems and target probabilities lower than $10^{-4}$.

\section{Numerical Results}\label{Numerical_results}
\noindent Several numerical examples are studied in this section to examine the performance and efficiency of the proposed methods. In all examples, (QNp-)HMCMC-based ASTPA is implemented as summarized in \cref{ASTPA_summary}. Results are compared with the Component-Wise Metropolis-Hastings based Subset Simulation (CWMH-SuS) \cite{au2001estimation}, with a uniform proposal distribution of width $2$, a samples percentile of $0.1$ to determine the subsets, and an appropriate number of samples in each subset level. In Example 3, our results are also compared against the state-of-the-art Riemannian Manifold Hamiltonian Monte Carlo-based subset simulation (RMHMC-SuS), using the available results reported in \citep{chen2022riemannian}. As a related note,  we could not identify any existing advanced, principled importance sampling-based method that can consistently, competitively and directly operate in the studied complex non-Gaussian problems. Comparisons are provided in terms of accuracy and computational cost based on $100$ independent simulations. Specifically, for each simulation $j$, we record the target probability estimate $(\hat{p}_\mathcal{F})^j$, the total number of model calls $(N_{Total})^j$, and the analytical Coefficient of Variation of ASTPA, $(\text{C.o.V-Anal})^j$, computed by \cref{cov_ASTPA}. $N_{Total}$ is defined for ASTPA in \cref{ASTPA_summary}. We then report the mean of the rare event probability estimates, $\mathop{\mathbb{E}}[\hat{p}_\mathcal{F}]$, the mean of the total number of model calls $\mathop{\mathbb{E}}[N_{Total}]$, and the sampling C.o.V, computed as:
\begin{equation}
\begin{aligned}
    &\textrm{C.o.V}=\dfrac{\sqrt{\widehat{\text{Var}}(\hat{p}_\mathcal{F})}}{\mathop{\mathbb{E}}[\hat{p}_\mathcal{F}]},\quad \\& \widehat{\text{Var}}(\hat{p}_{\mathcal{F}})=\dfrac{1}{100-1}\,\sum_{j=1}^{100}\big((\hat{p}_\mathcal{F})^j-\mathop{\mathbb{E}}[\hat{p}_\mathcal{F}]\big)^2
    \end{aligned}
\end{equation}
The mean of the analytical C.o.V, $\mathbb{E}[\text{C.o.V-Anal}]$, is also reported in parentheses in the relevant tables. The total number of limit-state function evaluations  $N_{Total}$ for ASTPA has been determined based on achieving C.o.V values $ \in [0.1,\, 0.35]$. The reference rare event probabilities are computed based on the standard Monte Carlo estimator, described in \cref{rare_event_est}, using $10^{6}$-$10^{8}$ samples, as appropriate. The problem dimensions are denoted by $d$, and all ASTPA parameters are carefully chosen for all examples, based on our general recommendations reported in \Cref{Targ_form},  but are not optimized. Analytical gradients are provided in all examples; hence, one limit-state/model call can provide both the relevant function value and all pertinent gradients.

\begin{figure*}[t!]
\centerline{\subfigure[]{\includegraphics[trim=1cm 3.5cm 0cm 3cm,width=0.3\textwidth]{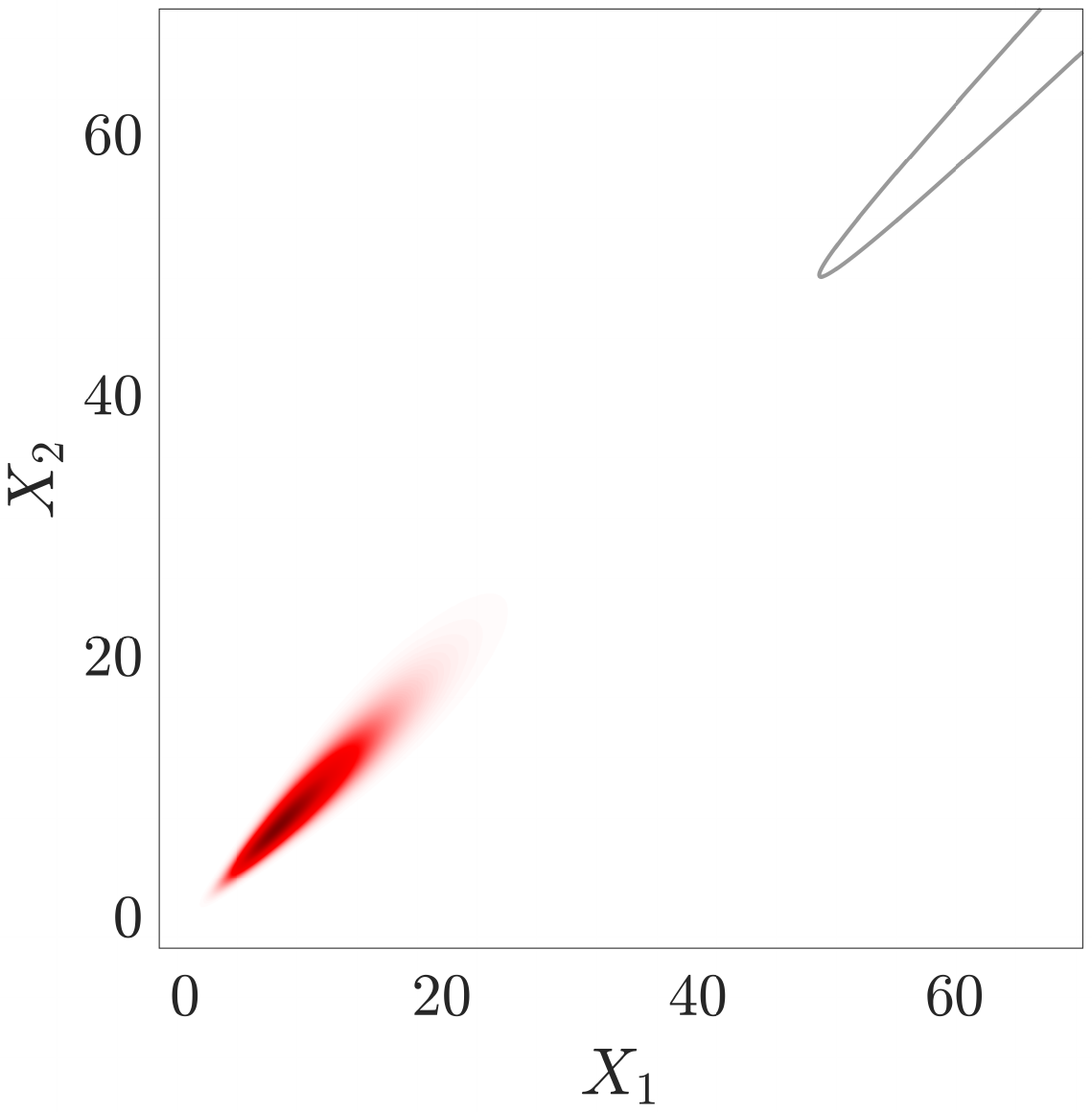}}
\subfigure[]{\includegraphics[trim=1cm 3.5cm 0cm 3cm,width=0.3\textwidth]{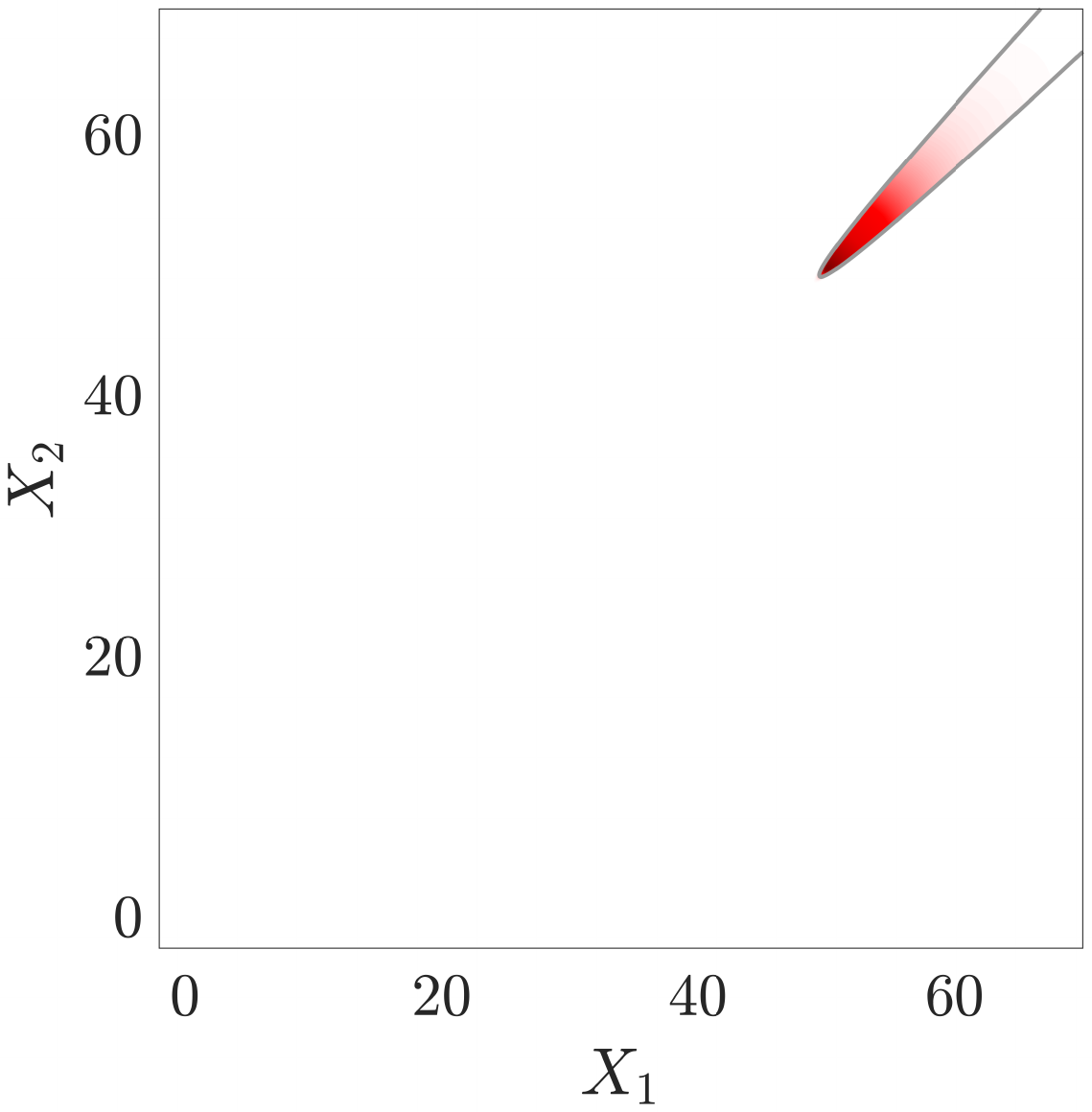}}
\subfigure[]{\includegraphics[trim=1cm 3.5cm 0cm 3cm,width=0.3\textwidth]{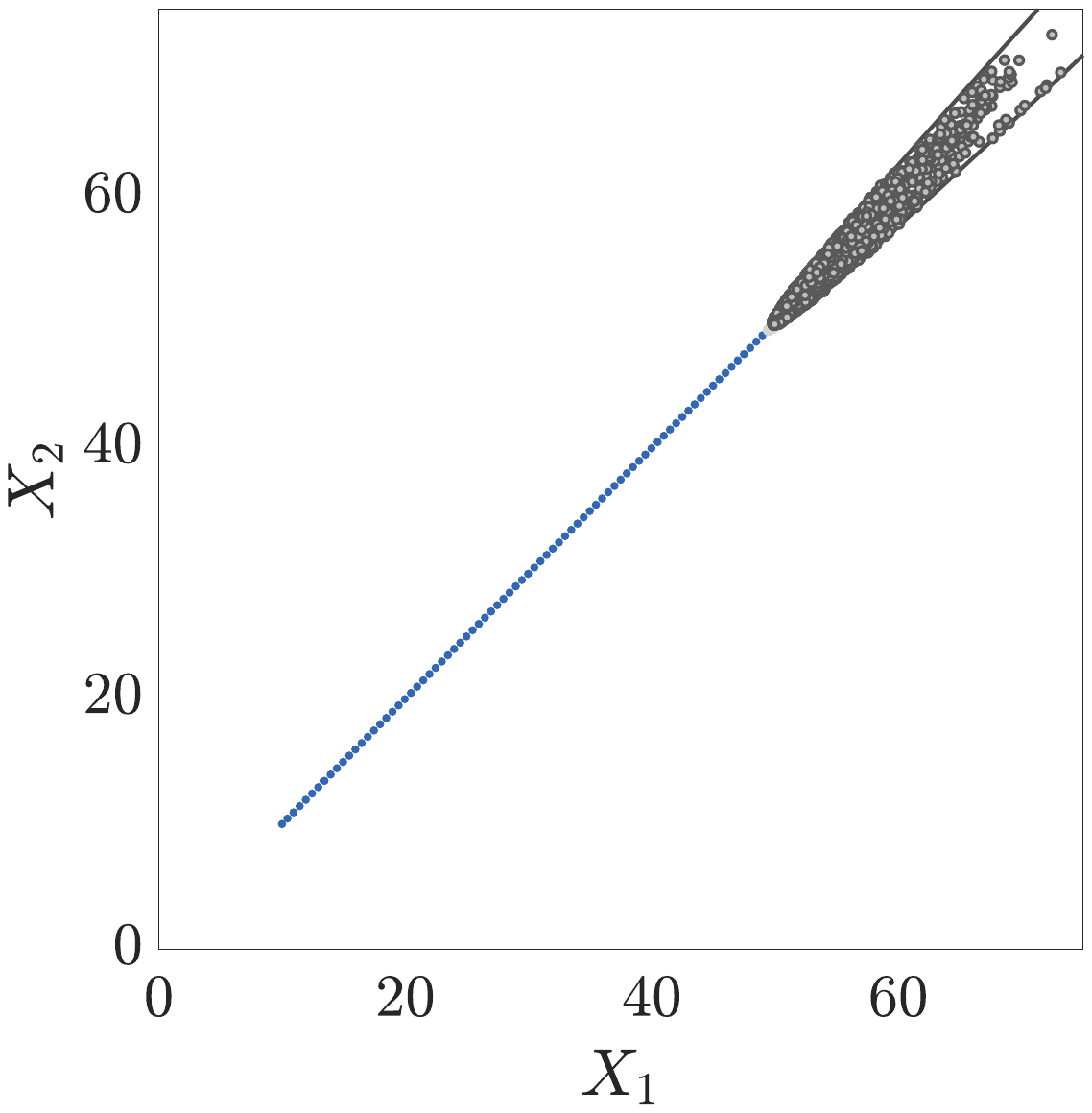}}}
\caption{Example 1: (a) Correlated Gumbel distribution plotted in red; the gray line represents the limit-state surface at $g(\bm{X})=0$, (b) The approximate sampling target $\tilde{h}$ in ASTPA, (c) Indicative Adam steps plotted in blue; gray samples drawn from $h$ using QNp-HMCMC.\vspace{-0.1in}}\label{Corr_Gumbel_fig}
\end{figure*}

\newcommand{\head}[1]{\textnormal{\textbf{#1}}}

\begin{table*}[t!]
\caption{Example 1: Performance of various methods for a quadratic limit-state function with highly correlated Gumbel distributions.}
\centering
\footnotesize
\setlength\tabcolsep{4pt}
\begin{tabular}{p{1.5cm}C{2cm}C{2cm}C{2cm}C{2cm}C{2cm}}
  \toprule[1.5pt]
  \multirow{2}{*}{\cref{Gumbel_limit}}& \textbf{100 Independent} & \multirow{2}{*}{\textbf{MCS}}& \multicolumn{2}{c}{\textbf{ASTPA}  ($\sigma = 0.1, \, q = 20$)}& \multirow{2}{*}{\textbf{CWMH-SuS}}\\ 
  
  \cline{4-5}
    \addlinespace[2pt]
  & \textbf{Simulations} && \multicolumn{1}{c}{\head{HMCMC}} & \multicolumn{1}{c}{\head{QNp-HMCMC}}&  \\ 
 \cmidrule(lr){1-6}
         \multirow{4}{*}{\shortstack[l]{\vspace{-0.14in}\\$d=2$\\$\lambda=70$\\$\gamma=2$}}\rule{0pt}{2.5ex}    &$\mathop{\mathbb{E}}[N_{Total}]$ &1.00E9 & 5,348 & 4,048  & 44,690\\
      &C.o.V &0.06& 0.09$\color{ForestGreen}($0.05$\color{ForestGreen})$  &0.09$\color{ForestGreen}($0.06$\color{ForestGreen})$&6.78 \\
       &$\mathop{\mathbb{E}}[\hat{p}_{\mathcal{F}}]$    &2.51E-7& 2.37E-7& 2.43E-7  &4.55E-8  \\
      \cmidrule(lr){1-6}

    \multirow{4}{*}{\shortstack[l]{\vspace{-0.14in}\\$d=3$\\$\lambda=5$\\$\gamma=3$}}\rule{0pt}{2.5ex}  &$\mathop{\mathbb{E}}[N_{Total}]$ &1.00E9& 8,598&4,598 & 31,730 \\
       &C.o.V&0.05 &  0.23$\color{ForestGreen}($0.14$\color{ForestGreen})$  &0.16$\color{ForestGreen}($0.13$\color{ForestGreen})$& 0.82\\
       &$\mathop{\mathbb{E}}[\hat{p}_{\mathcal{F}}]$   &4.17E-7 &4.18E-7  &4.19E-7 &5.19E-7  \\
      \cmidrule(lr){1-6}

    \multirow{4}{*}{\shortstack[l]{\vspace{-0.14in}\\$d=40$\\$\lambda=-200$\\$\gamma=20$}}\rule{0pt}{2.5ex}    &$\mathop{\mathbb{E}}[N_{Total}]$&1.00E8 & 13,298&5,298  &40,685  \\
       &C.o.V &0.04&   0.19$\color{ForestGreen}($0.07$\color{ForestGreen})$  &0.08$\color{ForestGreen}($0.08$\color{ForestGreen})$&4.63\\
       &$\mathop{\mathbb{E}}[\hat{p}_{\mathcal{F}}]$    &4.60E-6& 4.56E-6&4.60E-6 &6.32E-6  \\
      \bottomrule[1.5pt]

\end{tabular}\label{Cor_Gumbel_results}
\end{table*}

\subsection{Example 1: Multi-dimensional quadratic limit-state function with correlated Gumbel marginal distributions}\label{Corr_Gumbel}
\noindent Our proposed framework is first tested based on a joint probability distribution formulated using a Gaussian copula and Gumbel marginal distributions, as:
\begin{equation}
\begin{aligned}
\pi_{\bm{X}}(\bm{X})&=\phi_d \big( [\Phi^{-1}(F_{X_1}(X_1)), \Phi^{-1}(F_{X_2}(X_2)),...,\\ &\Phi^{-1}(F_{X_d}(X_d)) ];\mathbf{R} \big)  \frac{ \prod_{i=1}^{d} f_{X_i}(X_i)}{\prod_{i=1}^{d} \phi\big(\Phi^{-1}(F_{X_i}(X_i))\big)}
\end{aligned}
\end{equation}
where $\phi_d(\cdot)$ is the $d$-dimensional standard Gaussian PDF, with a correlation matrix $\mathbf{R}$, having a pair-wise Gaussian correlation $\rho_{ij}=0.9528$, corresponding to a $0.95$ Gumbel correlation; $\phi(\cdot)$ and $\Phi(\cdot)$ are the univariate standard normal PDF and CDF, respectively; $f_{X_i}$ and $F_{X_i}$ are the Gumbel distribution PDF and CDF of $X_i$, respectively. All $X_i$'s have a mean value $\mu_i=10$ and a Coefficient of Variation of 0.40. 
The model is represented here by the following quadratic limit-state function:
\begin{equation}
\begin{aligned}
g(\bm{X}) = \lambda - \frac{1}{\sqrt{d}}\ \sum_{i=1}^{d} X_{i} + 2.5\ \bigg (X_{1} - \sum_{j=2}^{\gamma} X_{j} \bigg )^{2}
\end{aligned}\label{Gumbel_limit}
\end{equation}
where the threshold $\lambda$ determines the rarity  of the considered event, and $\gamma$ defines the number of involved nonlinear variables. \cref{Corr_Gumbel_fig}(a) showcases the joint probability distribution and the limit-state surface for $d=2$, $\lambda=70$, and $\gamma=2$.\par

\cref{Cor_Gumbel_results} compares the performance of ASTPA variants with CWMH-SuS for dimensions $d=2,3,\text{and } 40$. The approximate sampling target, $\tilde{h}$, in ASTPA is constructed for all dimensions using a likelihood dispersion factor, $\sigma$, of 0.1, and a scaling constant $g_c=\sfrac{g(\boldsymbol{\mu}_{\pi_{\bm{X}}})}{q}$, with $q=20$, and $\boldsymbol{\mu}_{\pi_{\bm{X}}}=[\mu_i, \mu_i, ..., \mu_i] \in \mathbb{R}^d$, as explained in \cref{g_c}. The Adam optimizer is then run, starting at $\boldsymbol{\mu}_{\pi_{\bm{X}}}$, to provide an initial state for the HMCMC samplers. In \cref{Cor_Gumbel_results}, it is clear that our ASTPA $\mathbb{E}[\hat{p}_{\mathcal{F}}]$ estimates have very good agreement with the reference probability given by the standard Monte Carlo simulation (MCS) for all studied cases. Samples for MCS and the first set in Subset Simulation are generated using Nataf transformation \citep{nataf1962determination,lebrun2009innovating}. Compared to CWMH-SuS, ASTPA results are improved, as demonstrated by achieving low C.o.V using a significantly lower total number of model calls $N_{Total}$. Comparing HMCMC and QNp-HMCMC within ASTPA, it is obvious that QNp-HMCMC exhibits enhanced performance, considering the comparatively lower number of model calls, i.e., $(N+N_{BurnIn})$ (see \cref{ASTPA_summary}), required to achieve similar or improved accuracy. The generally good agreement between the sampling C.o.V and analytical C.o.V, as reported in parentheses, also shows the effectiveness and accuracy of our analytical C.o.V expression, particularly when utilizing the QNp-HMCMC sampler. \cref{Corr_Gumbel_fig} outlines the ASTPA framework in the 2D case here, with the joint distribution $\pi_{\bm{X}}$ and the limit-state surface at $g(\bm{X})=0$ shown in \cref{Corr_Gumbel_fig}(a). The approximate sampling target is visualized in \cref{Corr_Gumbel_fig}(b), with its samples plotted in gray in \cref{Corr_Gumbel_fig}(c). The blue points in \cref{Corr_Gumbel_fig}(c) are the steps of the Adam optimizer, efficiently discovering the rare event domain. As illustrated in \cref{Effect_prob_fig}, the developed framework can also provide better estimates for the 2-dimensional case by employing alternative parameter values, $\sigma$ and $q$, to construct the sampling target. Yet, since we have not optimized these parameters for any of the examples and for consistency purposes herein, we use the same parameter values across all considered dimensionalities within this example.

\begin{figure*}[t!]
\centerline{\subfigure[]{\includegraphics[trim=1cm 3.5cm 0cm 3cm,width=0.248\textwidth]{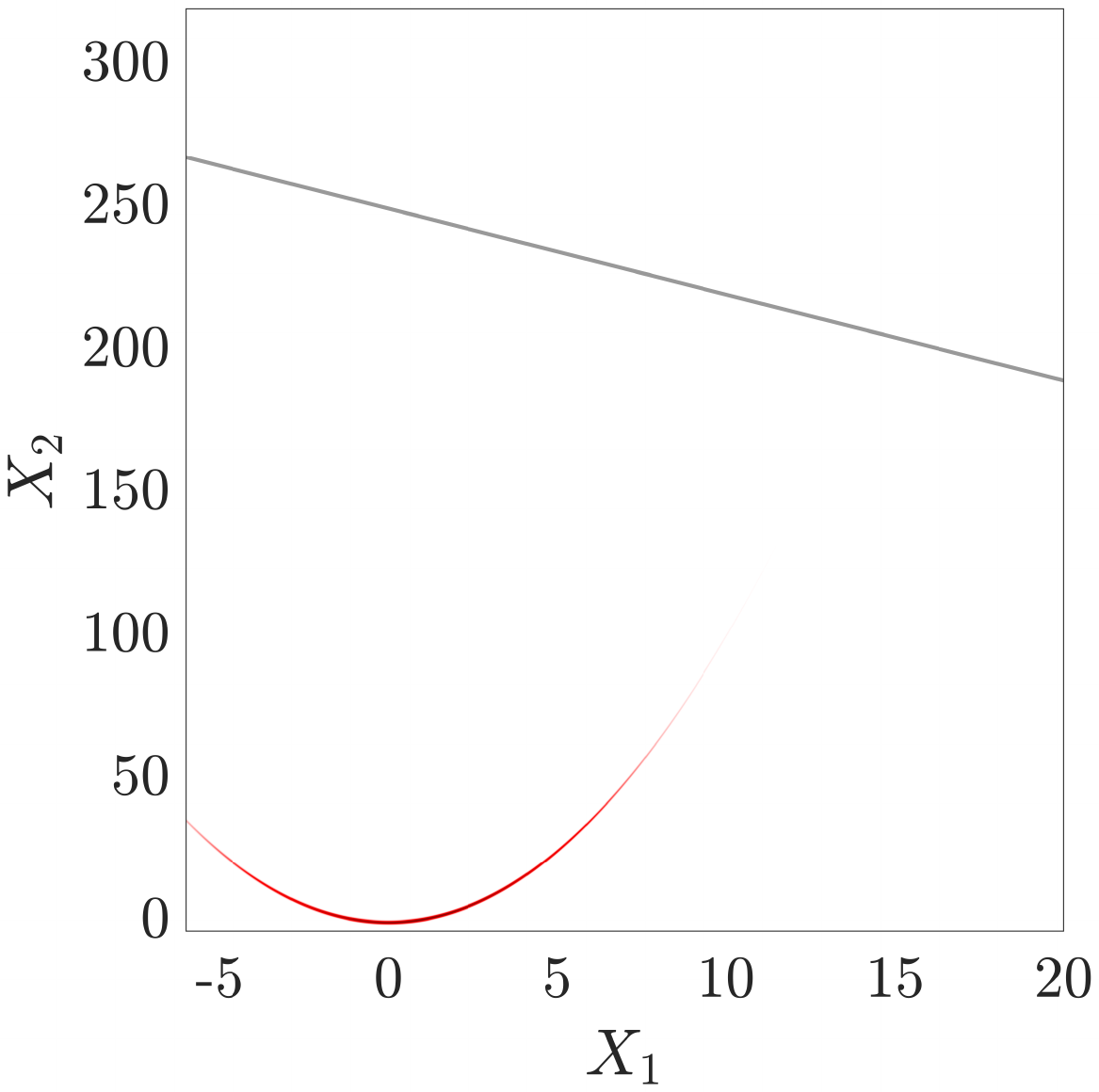}}
\subfigure[]{\includegraphics[trim=1cm 3.5cm 0cm 3cm,width=0.248\textwidth]{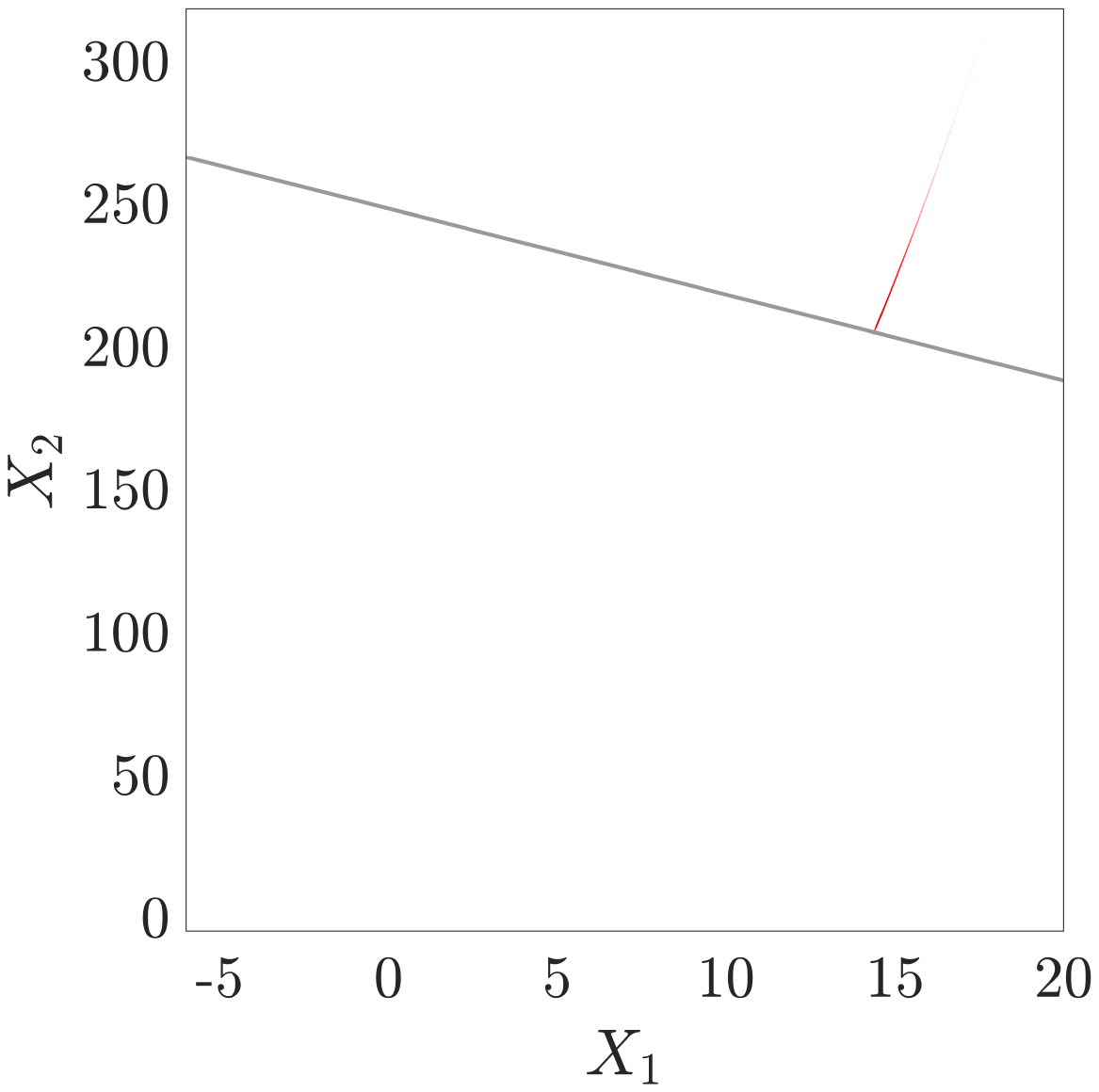}}
\subfigure[]{\includegraphics[trim=1cm 3.5cm 0cm 3cm,width=0.248\textwidth]{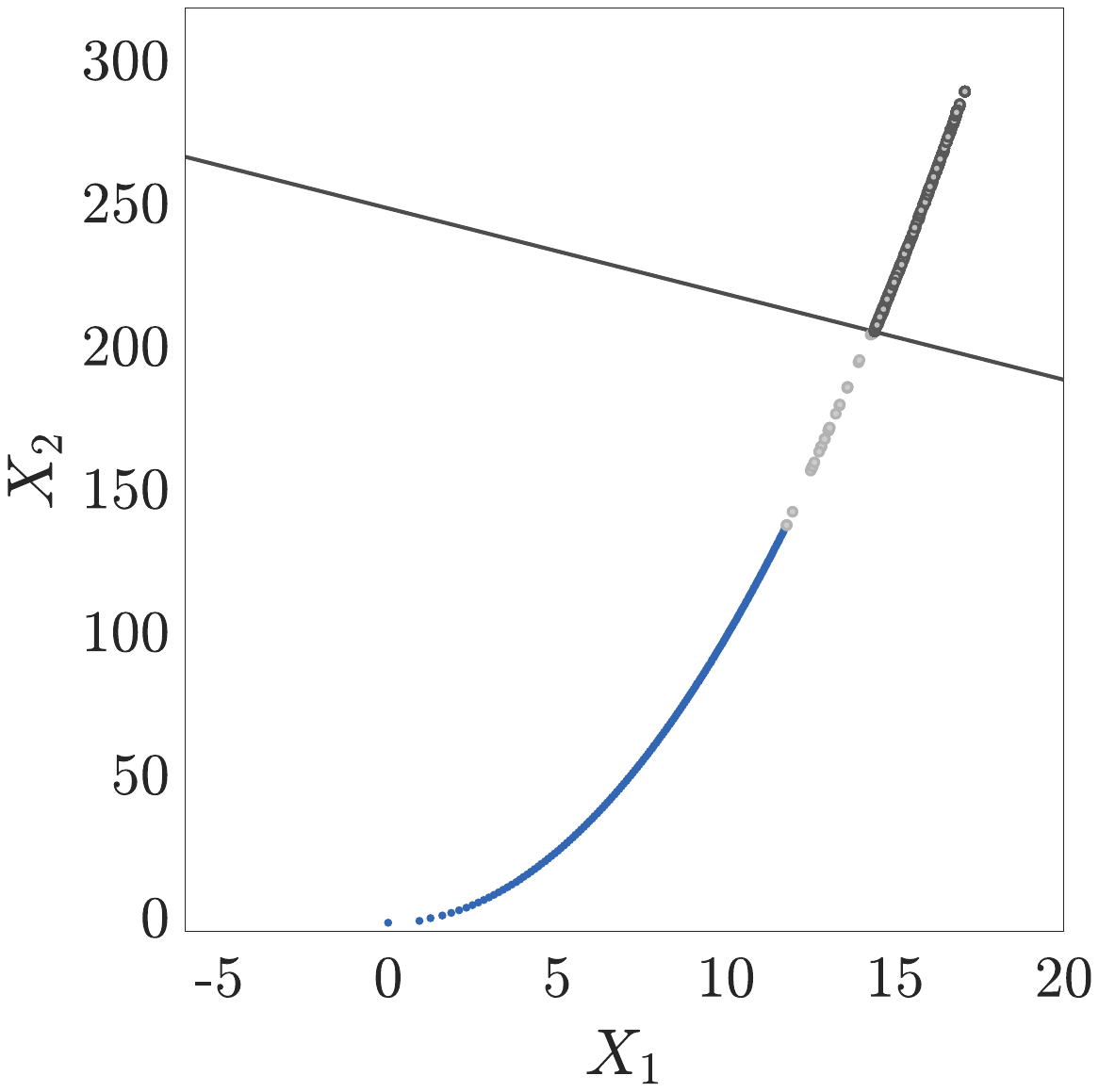}}
\subfigure[]{\includegraphics[trim=1cm 3.5cm 0cm 3cm,width=0.248\textwidth]{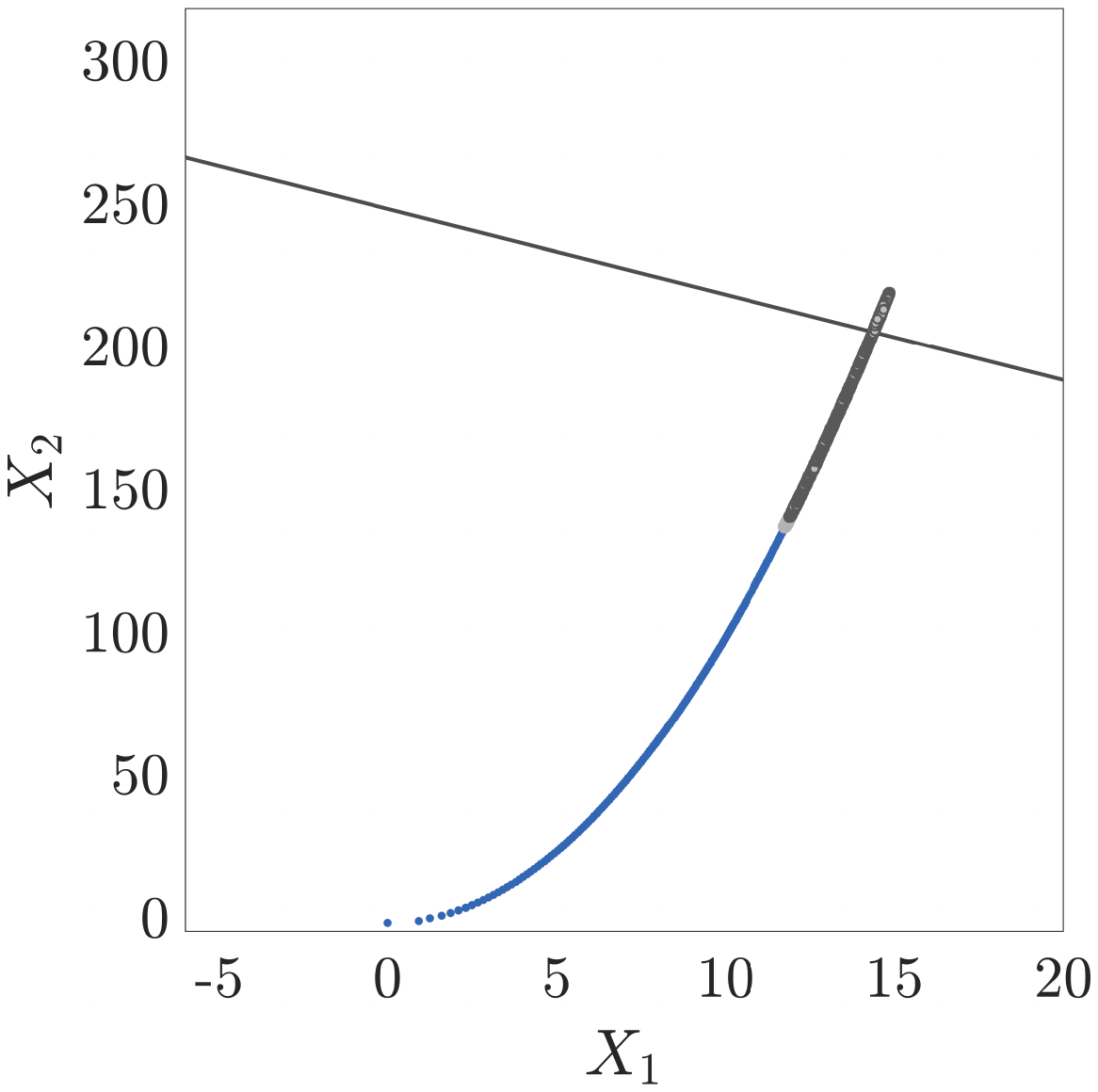}}}
  \captionsetup{labelfont={color=Black}}
\caption{Example 2: (a) 2D Rosenbrock distribution plotted in red; the gray line represents the limit-state surface at $g(\bm{X})=0$, (b) The approximate sampling target $\tilde{h}$ in ASTPA, (c) Indicative Adam steps plotted in blue; 1,650 gray samples drawn from $h$ using QNp-HMCMC, including burn-in samples shown in light gray, (d) Around $10^6$ gray samples generated using the standard HMCMC; Comparing (c) and (d) demonstrates the efficiency of QNp-HMCMC in adequately exploring the sampling target.\vspace{-0.1in}}\label{Rosenbrock_fig}
\end{figure*}

\begin{table*}[t!]
\caption{Example 2: Performance of various methods for 2D and 3D Rosenbrock distributions.}
\centering
\footnotesize
\setlength\tabcolsep{4pt}
\begin{tabular}{p{1.5cm}C{2cm}C{2cm}C{2cm}C{2cm}C{2cm}}
  \toprule[1.5pt]
  \multirow{2}{*}{\cref{Rosen_limit_3D}}& \textbf{100 Independent}   & \multirow{2}{*}{\textbf{MCS}}& \multicolumn{2}{c}{\textbf{ASTPA}  ($\sigma = 0.1, \, q = 20$)}& \multirow{2}{*}{\textbf{CWMH-SuS}}\\ 
  
  \cline{4-5}
  \addlinespace[2pt]
  &\textbf{Simulations} & &\multicolumn{1}{c}{\head{HMCMC}} & \multicolumn{1}{c}{\head{QNp-HMCMC}}&  \\ 
 \cmidrule(lr){1-6}
         \multirow{4}{*}{\shortstack[l]{\vspace{-0.18in}\\$d=2$\\$\gamma=1$\\$a=0.05$\\$b=5$}}\rule{0pt}{2.5ex}    
         &$\mathop{\mathbb{E}}[N_{Total}]$  &1.00E8&1.02E6 & 3,848  &633,700   \\
      &C.o.V&0.03 & 0.20$\color{ForestGreen}($0.03$\color{ForestGreen})$  &0.12$\color{ForestGreen}($0.11$\color{ForestGreen})$&1.09 \\
       &$\mathop{\mathbb{E}}[\hat{p}_{\mathcal{F}}]$   &1.15E-5& 0.67E-5& 1.10E-5  &1.15E-5  \\
      
 \cmidrule(lr){1-6}
         \multirow{4}{*}{\shortstack[l]{\vspace{-0.18in}\\$d=3$\\$\gamma=0.5$\\$a=1$\\$b=5$}}\rule{0pt}{2.5ex}    
         &$\mathop{\mathbb{E}}[N_{Total}]$  &1.00E8&1.02E6 & 4,948  & 848,800  \\
      &C.o.V &0.10& 0.41$\color{ForestGreen}($0.07$\color{ForestGreen})$  &0.16$\color{ForestGreen}($0.13$\color{ForestGreen})$& 2.22 \\
       &$\mathop{\mathbb{E}}[\hat{p}_{\mathcal{F}}]$    &1.00E-6 &0.20E-6 & 0.87E-6  &1.70E-6\\
      \bottomrule[1.5pt]

\end{tabular}\label{Rosen_table}
\end{table*}

\subsection{Example 2: 2D and 3D Rosenbrock distributions}
\noindent In this example, we consider the challenging Rosenbrock distribution, defined for multi-dimensional random variables as \citep{goodman2010ensemble,pagani2019n}:
\begin{equation}
\begin{aligned}
\pi_{\bm{X}}(\bm{X}) = &\dfrac{\sqrt{a} \prod_{i=2}^d\sqrt{b_i}}{\pi^{d/2}}\\ & \cdot \exp\bigg( -a(X_1-\gamma)^2-\sum_{i=2}^d b_i(X_i-X_{i-1}^2)^2\bigg)
\end{aligned}\label{Rosen_limit_3D}
\end{equation}
where the parameters $a$, $b_i$, and $\gamma$ define the geometry and complexity of the distribution. An interesting discussion on the sampling complexity of the two-dimensional Rosenbrock distribution can be found in \citep{shields2021subset, thaler2024reliability}. In this example, we include two- and three-dimensional cases of \cref{Rosen_limit_3D}, where the relevant parameter sets are given in \cref{Rosen_table}. The chosen parameters of \cref{Rosen_limit_3D} here result in significantly challenging sampling targets, as showcased for the two-dimensional case in \cref{Rosenbrock_fig}(a).
To further complicate this testing benchmark, a low probability level ($\sim 10^{-5} - 10^{-6}$) is considered by defining the limit-state function as: 
\begin{equation}
\begin{aligned}
g(\bm{X}) = 250-3X_1-\sum_{i=2}^d X_i
\end{aligned}
\end{equation}
The approximate sampling target in ASTPA is depicted in \cref{Rosenbrock_fig}(b). The results, accompanied by the ASTPA parameters, are reported in \cref{Rosen_table}. Samples in MCS and the first set in Subset Simulation (CWMH-SuS) are generated directly based on the conditional sampling scheme provided in \citep{pagani2019n}. The Adam optimizer is run in this example for 1,500 iterations because of the complexity of the problem. As can be seen in \cref{Rosen_table}, QNp-HMCMC-based ASTPA significantly outperforms the other methods. The performance difference between the two HMCMC schemes can be explained by observing the gray samples in \cref{Rosenbrock_fig}, where \cref{Rosenbrock_fig}(c) visualizes 1,650 QNp-HMCMC samples, and \cref{Rosenbrock_fig}(d) showcases around one million HMCMC samples. Comparing these two plots demonstrates the incomparable ability of QNp-HMCMC to sample complex target distributions representatively and efficiently. As discussed in \cref{QNp_HMCMC}, this sampling efficiency is mainly attributed to the preconditioned dynamics and mass matrix incorporated in QNp-HMCMC. In QNp-HMCMC-based ASTPA, the estimates of the developed analytical C.o.V expression exhibit very good agreement with the sampling ones. However, this is not the case for the HMCMC-based ones, considering its highly correlated samples, requiring a thining scheme different from the general one in \cref{thining}.

\subsection{Example 3: High-dimensional hyperspherical limit-state function with multivariate funnel distribution}\label{funnel}
\begin{figure*}[t!]
\centerline{\subfigure[]{\includegraphics[trim=1cm 3.5cm 0cm 3cm,width=0.3\textwidth]{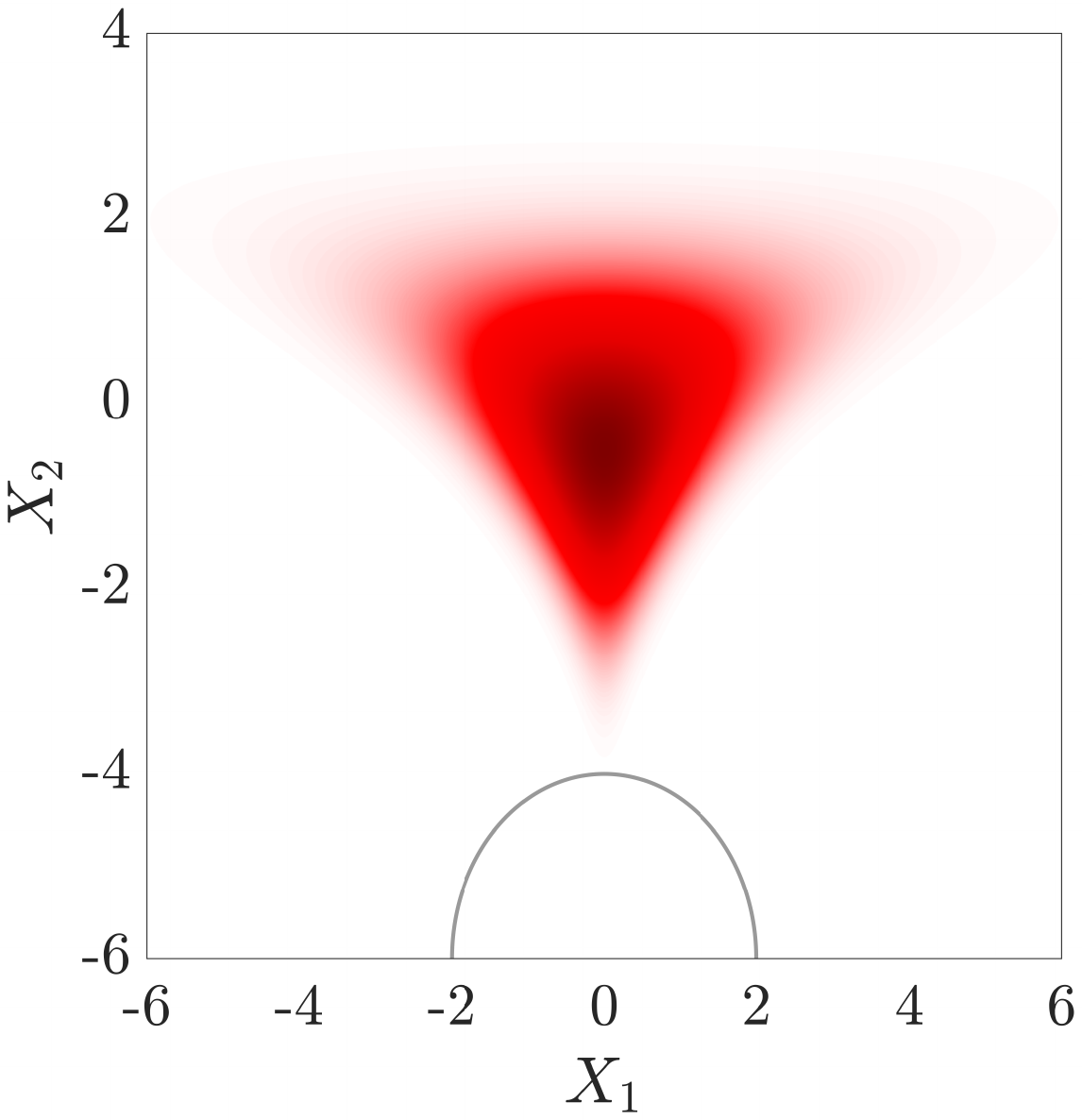}}
\subfigure[]{\includegraphics[trim=1cm 3.5cm 0cm 3cm,width=0.3\textwidth]{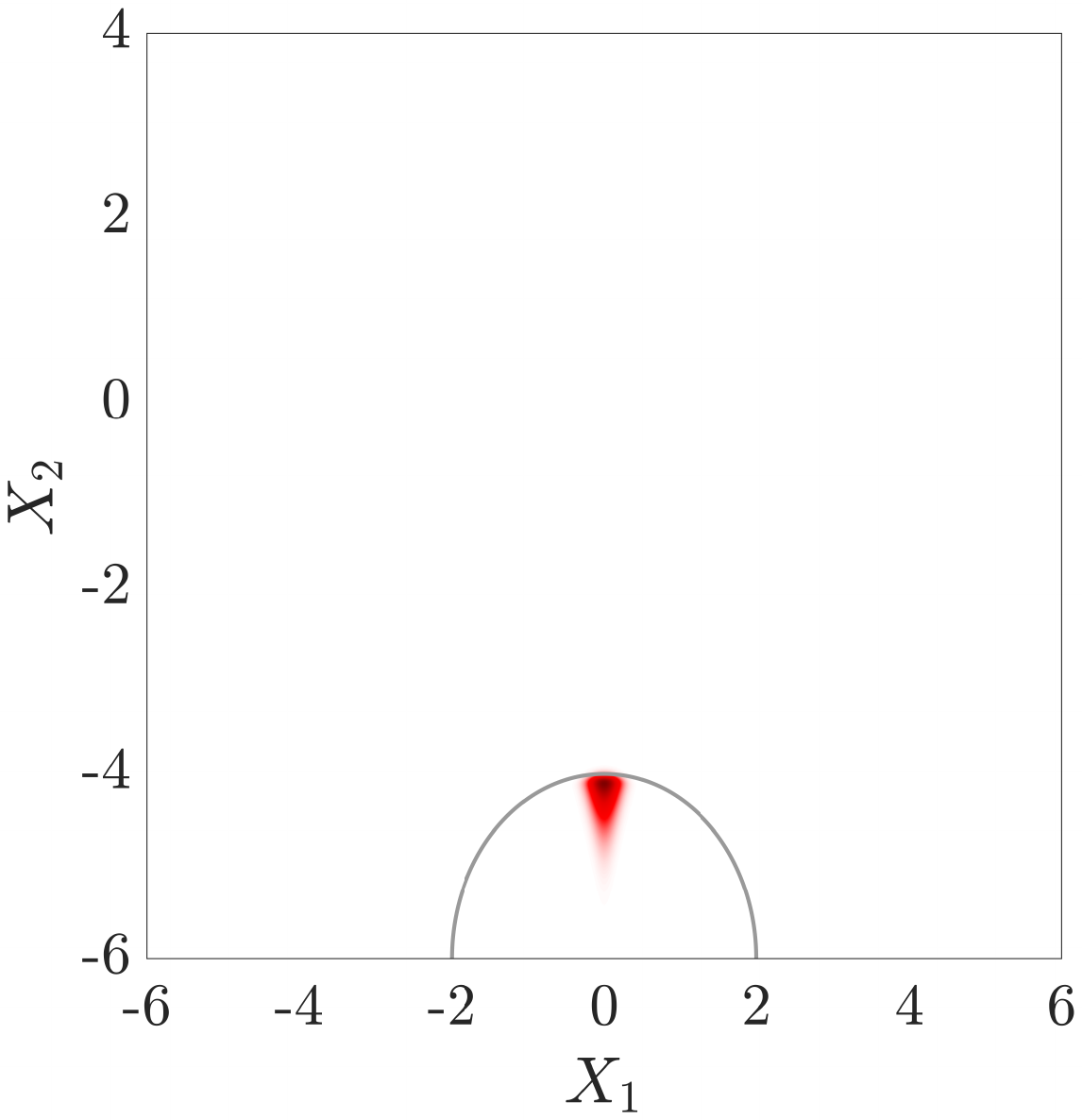}}
\subfigure[]{\includegraphics[trim=1cm 3.5cm 0cm 3cm,width=0.3\textwidth]{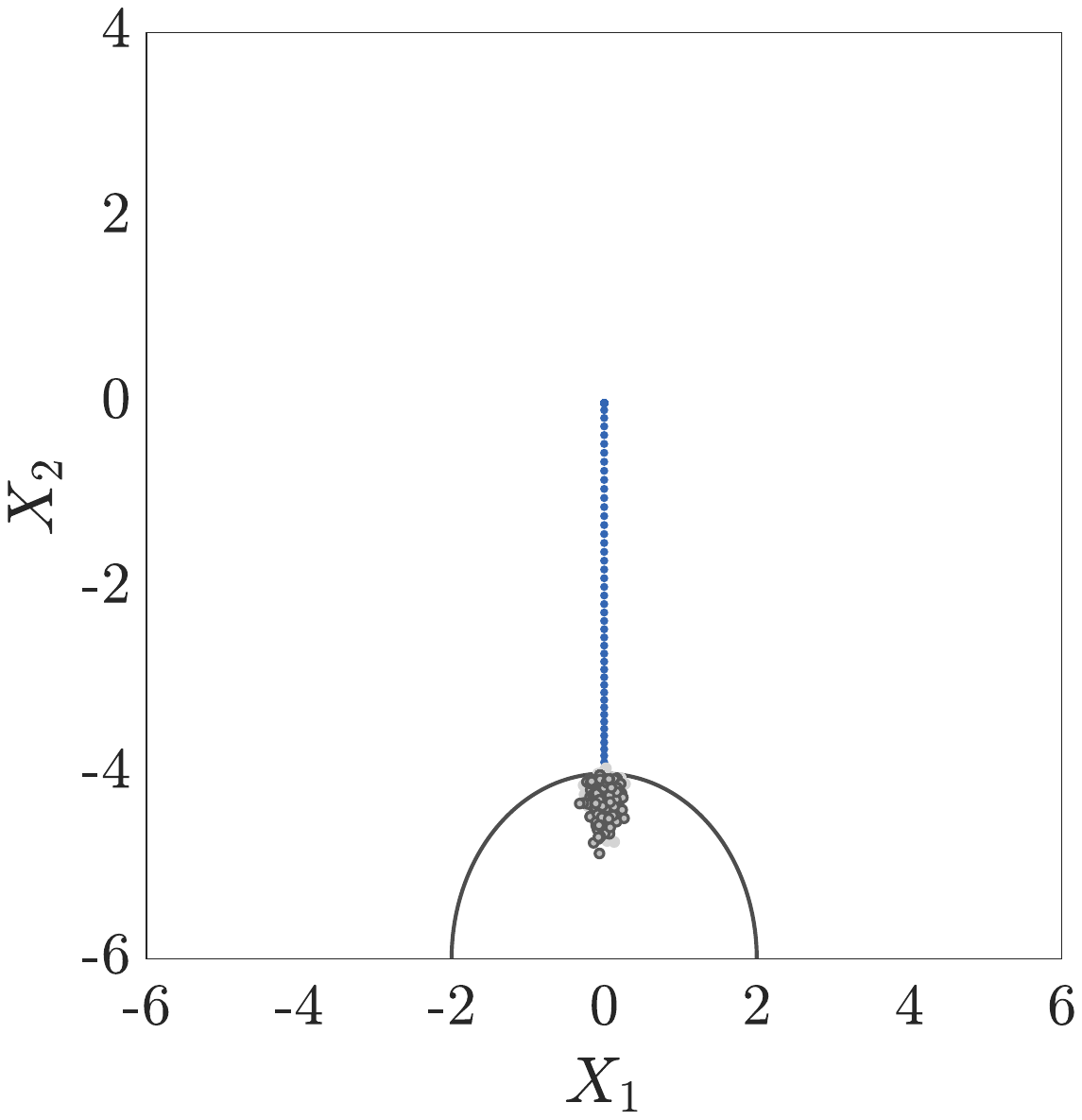}}}
\caption{Example 3: (a) Neal's funnel distribution plotted in red; the gray line represents the limit-state surface at $g(\bm{X})=0$, (b) The approximate sampling target $\tilde{h}$ in ASTPA, (c) Adam steps plotted in blue; gray samples drawn from $h$ using QNp-HMCMC.\vspace{-0.1in}}\label{Neal_fig}
\end{figure*}

\noindent Our proposed framework is here evaluated on a challenging non-Gaussian sampling distribution considering a $d$-dimensional random variable $\bm{X}=[X_{1}, X_{2},...,X_{d}]$ following a $d$-dimensional Neal's funnel distribution, defined as \citep{neal2003slice,chen2022riemannian}: 
\begin{equation}
\pi_{\bm{X}}(\bm{X})=\prod_{i=1}^{d-1} \mathcal{N}(X_{i}\, \vert 0,\,\exp(X_{d}))\cdot\,\mathcal{N}(X_{d} \,\vert 0,\,1)
\label{Neal_target}
 \end{equation}
A hyperspherical limit-state function is utilized in this example, expressed as:
\begin{equation}
g(\bm{X}) = \sum_{i=1}^{d-1} X_{i}^2 + (X_{d} + 6)^{2}-r^2\\
\label{spherical}
\end{equation}
where $r$ is a threshold parameter determining the level of the rare event probability. \cref{Neal_fig}(a) depicts Neal's funnel distribution $\pi_{\bm{X}}(\bm{X})$ and the limit-state function for $d=2$, and $r=2$. The approximate sampling target, its samples, and the pertinent Adam steps are also visualized in \cref{Neal_fig}. The results, accompanied by ASTPA parameters, are presented in \cref{Neal_tabel}, for dimensions up to $101$. Samples in MCS and the first set in Subset Simulation (CWMH-SuS) are generated directly based on the joint distribution in \cref{Neal_target}. Notably, ASTPA is successfully compared with the state-of-the-art RMHMC-SuS, using results as reported for the studied examples up to $51$ dimensions in \citep{chen2022riemannian}. Overall, the ASTPA results in this set of examples further emphasize the superior performance of ASTPA in challenging scenarios. HMCMC and QNp-HMCMC exhibit similar performance in these examples, except for the $101$-dimensional case, wherein QNp-HMCMC-based ASTPA is significantly more efficient.

\begin{table*}[t!]
\caption{Example 3: Performance of various methods for a high-dimensional hyperspherical limit-state function with Neal's funnel distribution.}
\centering
\footnotesize
\setlength\tabcolsep{4pt}
\begin{tabular}{p{1.5cm}C{2cm}C{2cm}C{2cm}C{2cm}C{2.5cm}C{2cm}}
  \toprule[1.5pt]
  \multirow{2}{*}{\cref{spherical}}& \textbf{100 Independent}& \multirow{2}{*}{\textbf{MCS}} & \multicolumn{2}{c}{\textbf{ASTPA}  ($\sigma = 0.1, \, q = 20$)}& \multirow{2}{*}{\textbf{RMHMC-SuS} \citep{chen2022riemannian}}& \multirow{1}{*}{\textbf{CWMH-SuS}} \\ 
  
  \cline{4-5} 
    \addlinespace[2pt]
  &\textbf{Simulations}& &\multicolumn{1}{c}{\head{HMCMC}} & \multicolumn{1}{c}{\head{QNp-HMCMC}}&&\\ 
 \cmidrule(lr){1-7}
         \multirow{4}{*}{\shortstack[l]{\vspace{-0.14in}\\$d=2$\\$r=2$}}\rule{0pt}{2.5ex}    &$\mathop{\mathbb{E}}[N_{Total}]$  & 1.00E7& 1,213  & 1,213&  4,761& 4,618 \\
      &C.o.V &0.06& 0.09$\color{ForestGreen}($0.08$\color{ForestGreen})$  &0.10$\color{ForestGreen}($0.09$\color{ForestGreen})$& 0.35& 0.57 \\
       &$\mathop{\mathbb{E}}[\hat{p}_{\mathcal{F}}]$  &3.11E-5& 3.09E-5 & 3.09E-5 & 3.11E-5&3.13E-5  \\
     \cmidrule(lr){1-7}

    \multirow{4}{*}{\shortstack[l]{\vspace{-0.14in}\\$d=31$\\$r=2$}}\rule{0pt}{2.5ex}  &$\mathop{\mathbb{E}}[N_{Total}]$& 1.00E7 &3,213 & 3,213 &4,905&20,884  \\
       &C.o.V &0.07& 0.10$\color{ForestGreen}($0.11$\color{ForestGreen})$  &0.12$\color{ForestGreen}($0.11$\color{ForestGreen})$& 0.41 & 1.95\\
       &$\mathop{\mathbb{E}}[\hat{p}_{\mathcal{F}}]$ &1.87E-5& 1.83E-5  &1.84E-5 & 1.99E-5&1.52E-5 \\
      \cmidrule(lr){1-7}
      
    \multirow{4}{*}{\shortstack[l]{\vspace{-0.14in}\\$d=51$\\$r=2$}}\rule{0pt}{2.5ex}  &$\mathop{\mathbb{E}}[N_{Total}]$& 1.00E7&4,313 & 4,313&5,152& 22,072\\
       &C.o.V &0.08& 0.15$\color{ForestGreen}($0.14$\color{ForestGreen})$  &0.13$\color{ForestGreen}($0.11$\color{ForestGreen})$&0.45 & 2.63\\
       &$\mathop{\mathbb{E}}[\hat{p}_{\mathcal{F}}]$  &1.37E-5 &1.34E-5  & 1.33E-5 & 1.37E-5&1.32E-5  \\
     \cmidrule(lr){1-7}
      
    \multirow{4}{*}{\shortstack[l]{\vspace{-0.14in}\\$d=51$\\$r=1$}}\rule{0pt}{2.5ex}  &$\mathop{\mathbb{E}}[N_{Total}]$ & 1.00E8&4,840 & 4,840&7,065& 32,260\\
       &C.o.V  &0.29& 0.15$\color{ForestGreen}($0.11$\color{ForestGreen})$  &0.17$\color{ForestGreen}($0.11$\color{ForestGreen})$&0.55 & 5.21\\
       &$\mathop{\mathbb{E}}[\hat{p}_{\mathcal{F}}]$&1.28E-7&1.26E-7  & 1.28E-7 & 1.36E-7&0.77E-7  \\
      \cmidrule(lr){1-7}
            
          \multirow{4}{*}{\shortstack[l]{\vspace{-0.14in}\\$d=101$\\$r=2$}}\rule{0pt}{2.5ex}    &$\mathop{\mathbb{E}}[N_{Total}]$ & 1.00E8& 14,313& 7,813& & 25,564\\
       &C.o.V &0.04& 0.16$\color{ForestGreen}($0.14$\color{ForestGreen})$  &0.16$\color{ForestGreen}($0.11$\color{ForestGreen})$ & &4.04\\
       &$\mathop{\mathbb{E}}[\hat{p}_{\mathcal{F}}]$ &6.82E-6& 6.55E-6& 6.55E-6&&5.68E-6  \\
      \bottomrule[1.5pt]

\end{tabular}\label{Neal_tabel}
\end{table*}

\subsection{Example 4: High-dimensional octic problem}
\noindent To further investigate the ASTPA performance in high-dimensional non-Gaussian spaces with significant nonlinearities, the limit-state function in this example is expressed in the lognormal space as: 
\begin{equation}
\begin{aligned}
g(\bm{X}) = &Y_{0} - \frac{1}{\sqrt{200}}\ \sum_{i=1}^{200} X_{i} + 2.5\ \bigg (X_{1} - \sum_{j=2}^{10} X_{j} \bigg )^{2} \\ & + \bigg (X_{11} - \sum_{k=12}^{14} X_{k} \bigg )^{4} + \bigg (X_{15} - \sum_{l=16}^{17} X_{l} \bigg )^{8}\label{eq:31}
\end{aligned} 
\end{equation}
where $\bm{X}$ follows an independent lognormal distribution with a mean $\mu_{X_i}=1.0$ and a standard deviation $\sigma_{X_i}=1.0$. Two values are considered for the threshold $Y_0$, $15$ and $16$, to examine different probability levels. Prior to applying ASTPA, these positive lognormal variables are transformed to an unconstrained non-Gaussian space by following the approach discussed in \cref{bounded}. The results, accompanied by the ASTPA parameters, are presented in \cref{lognormal_tabel}, wherein ASTPA has significantly outperformed CWMH-SuS, considering all metrics, while giving an accurate mean value for the sought probability. Once more, QNp-HMCMC, which is implemented here with diagonal inverse Hessian and mass matrices, has significantly outperformed the original HMCMC in this highly nonlinear scenario, reinforcing our previous findings.

\begin{table*}[t!]
\caption{Example 4: Performance of various methods for a high-dimensional, octic limit-state function with lognormal distribution.}
\centering
\footnotesize
\setlength\tabcolsep{4pt}
\begin{tabular}{p{1.5cm}C{2cm}C{2cm}C{2cm}C{2cm}C{2cm}}

  \toprule[1.5pt]
  \multirow{2}{*}{\cref{eq:31}}& \textbf{100 Independent}  & \multirow{2}{*}{\textbf{MCS}}& \multicolumn{2}{c}{\textbf{ASTPA}  ($\sigma = 0.2, \, q = 10$)}& \multirow{2}{*}{\textbf{CWMH-SuS}}\\ 
  \cline{4-5}
  \addlinespace[2pt]
  &\textbf{Simulations}& & \multicolumn{1}{c}{\head{HMCMC}} & \multicolumn{1}{c}{\head{QNp-HMCMC}}&  \\ 
 \cmidrule(lr){1-6}
         \multirow{4}{*}{\shortstack[l]{\vspace{-0.14in}\\$d=200$\\$Y_0=15$}}\rule{0pt}{2.5ex}    
         &$\mathop{\mathbb{E}}[N_{Total}]$ &1.00E7&30,312& 8,812& 13,719  \\
      &C.o.V &0.06& 0.24$\color{ForestGreen}($0.18$\color{ForestGreen})$  &0.22$\color{ForestGreen}($0.26$\color{ForestGreen})$&0.41 \\
       &$\mathop{\mathbb{E}}[\hat{p}_{\mathcal{F}}]$ \ \ \  &2.22E-5&2.22E-5 & 2.20E-5  & 5.62E-5\\
      
 \cmidrule(lr){1-6}
          \multirow{4}{*}{\shortstack[l]{\vspace{-0.14in}\\$d=200$\\$Y_0=16$}}\rule{0pt}{2.5ex}     
         &$\mathop{\mathbb{E}}[N_{Total}]$&1.00E7& 30,333& 11,833 & 23,855\\
      &C.o.V &0.16&0.26$\color{ForestGreen}($0.26$\color{ForestGreen})$   &0.24$\color{ForestGreen}($0.39$\color{ForestGreen})$& 0.46\\
       &$\mathop{\mathbb{E}}[\hat{p}_{\mathcal{F}}]$ \ \ \ &3.54E-6 & 3.69E-6& 3.62E-6  &  1.66E-5\\
      \bottomrule[1.5pt]

\end{tabular}\label{lognormal_tabel}
\end{table*}

\subsection{Example 5: High-dimensional ring-shaped conditional distribution in a Bayesian context}\label{sec:Ring}
\begin{figure*}[t!]
\centerline{\subfigure[]{\includegraphics[trim=1cm 3.5cm 0cm 3cm,width=0.248\textwidth]{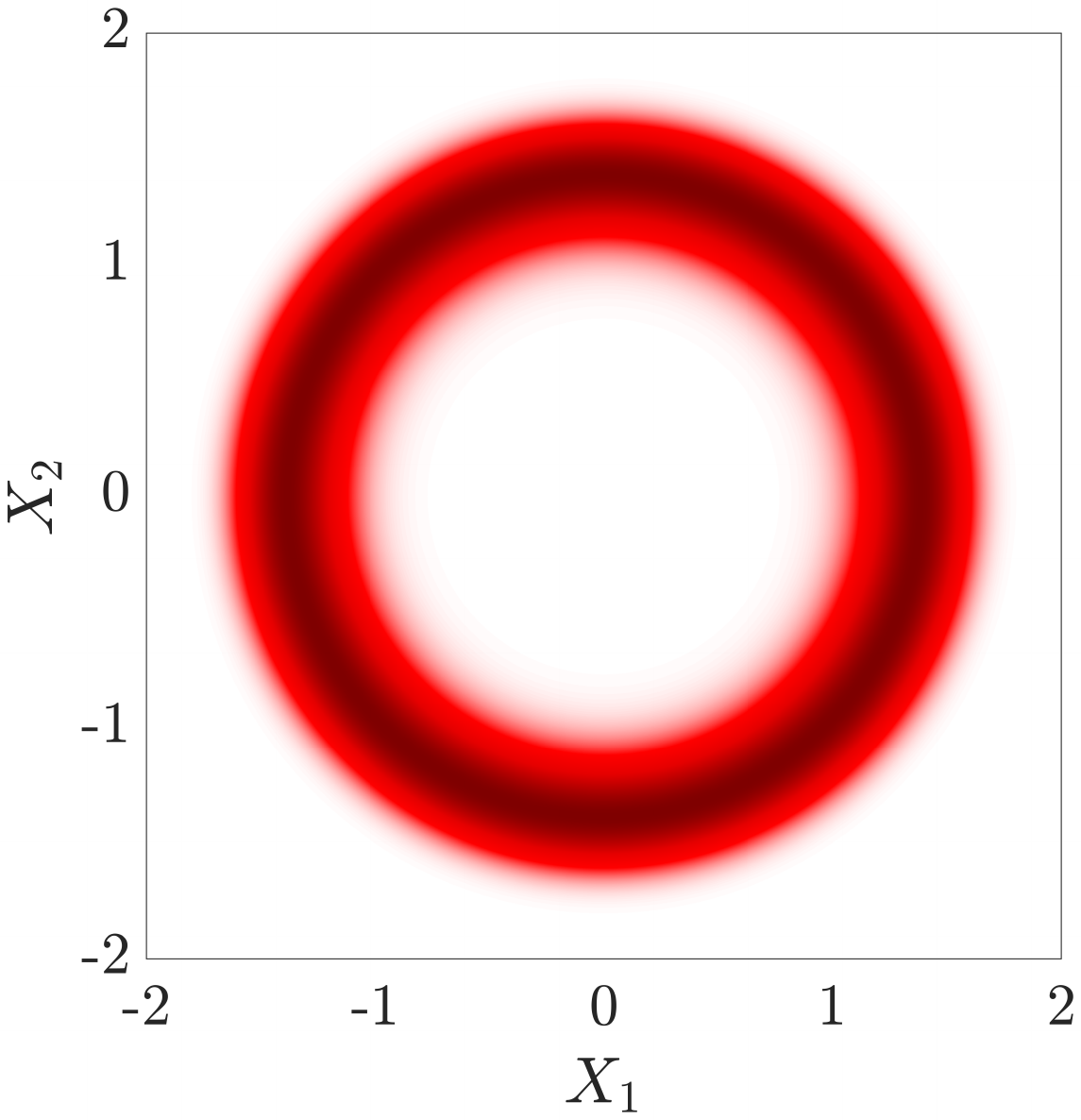}}
\subfigure[]{\includegraphics[trim=1cm 3.5cm 0cm 3cm,width=0.248\textwidth]{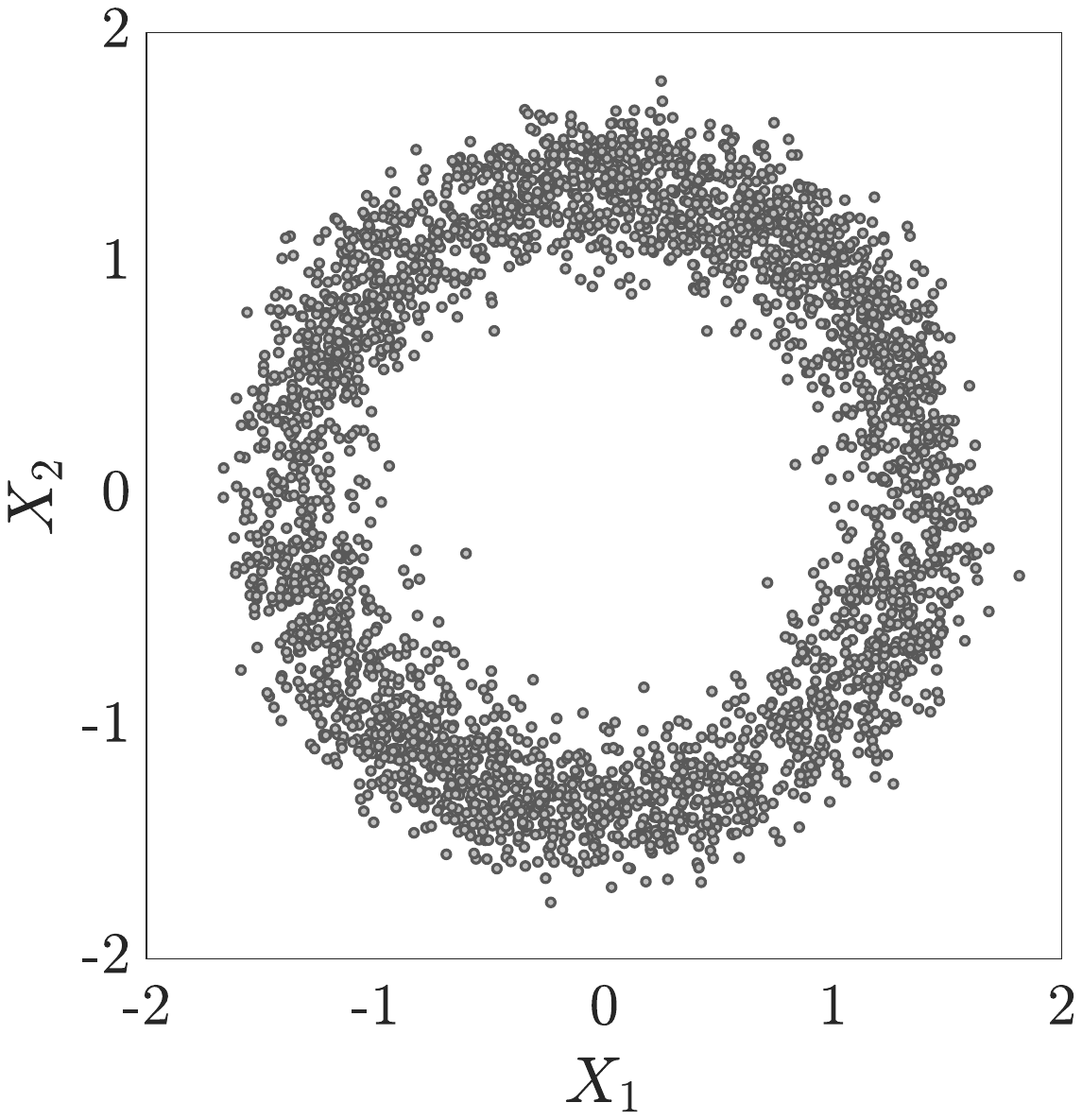}}
\subfigure[]{\includegraphics[trim=1cm 3.5cm 0cm 3cm,width=0.248\textwidth]{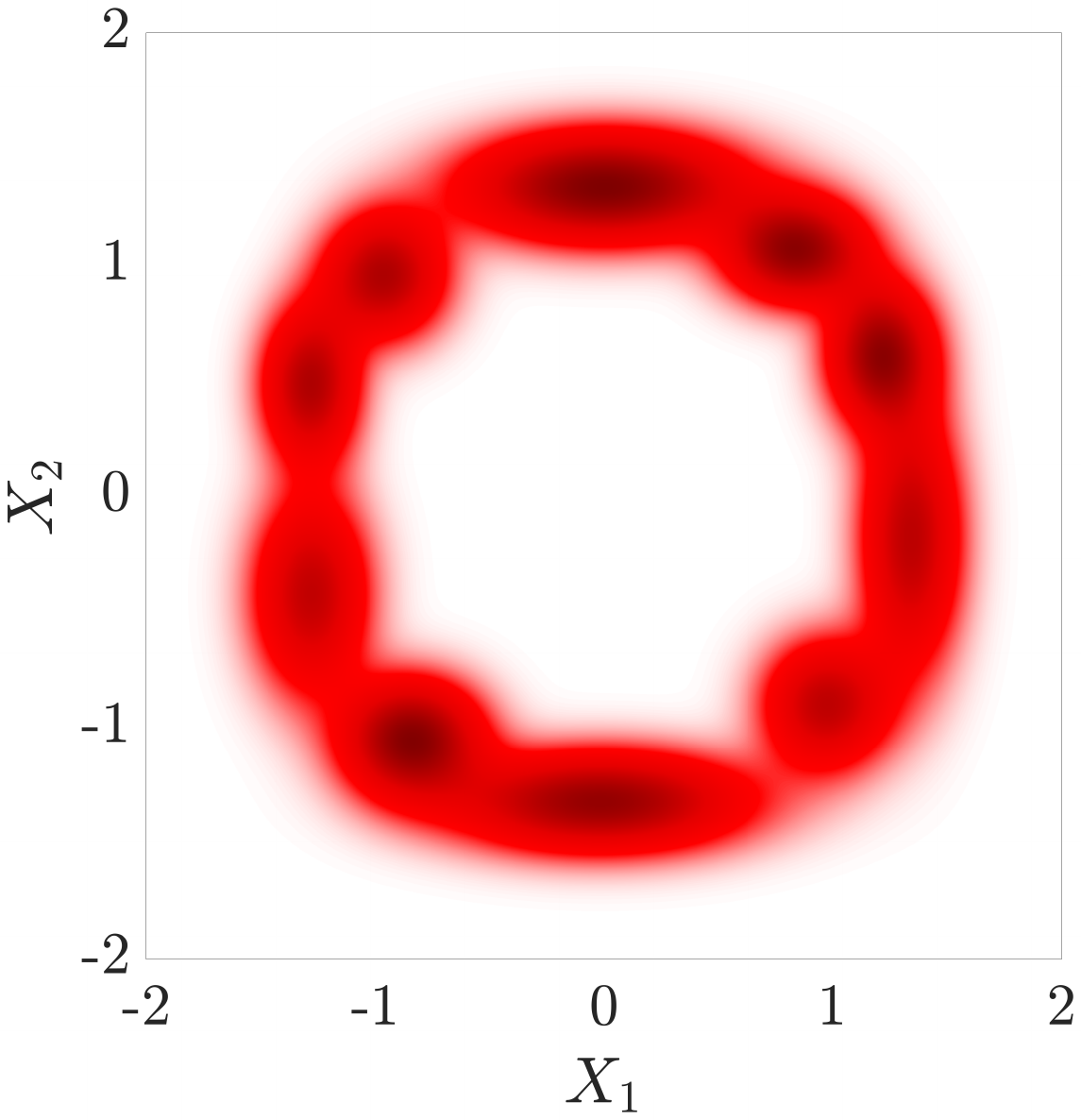}}
\subfigure[]{\includegraphics[trim=1cm 3.5cm 0cm 3cm,width=0.248\textwidth]{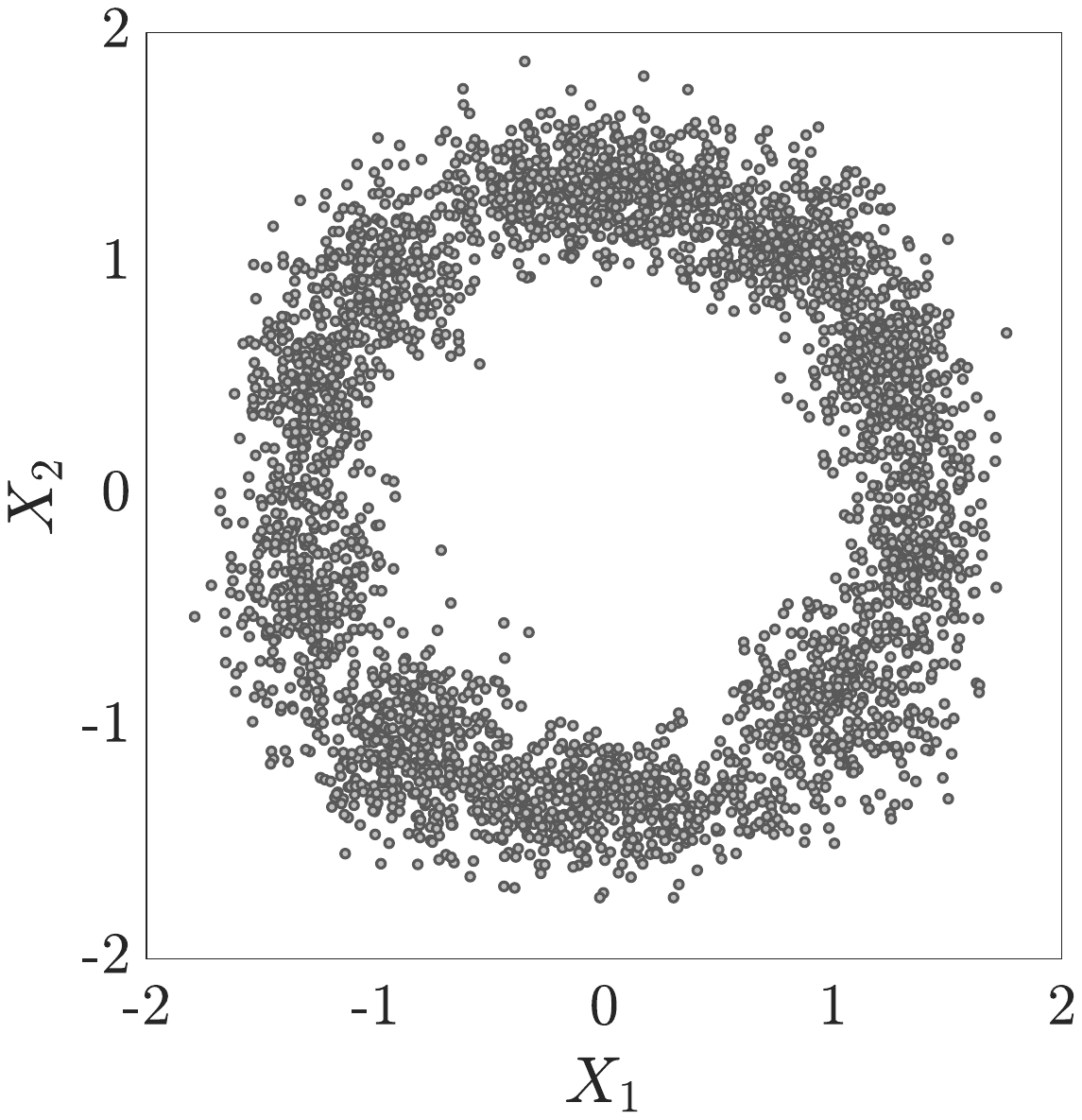}}}
\caption{Example 5: Outlining the normalizing constant, $C_\pi$, estimation in \cref{pi_norm_const}; (a) Ring-shaped distribution $\tilde{\pi}_{\bm{X}}$, (b) $N_\pi$ samples from $\tilde{\pi}_{\bm{X}}$, (c) a GMM $Q_\pi(\bm{X})$ fitted based on the $\tilde{\pi}_{\bm{X}}$ samples, and (d) $M_\pi$ GMM samples utilized to compute $C_\pi$ in \cref{c_pi_hat}.\vspace{-0.1in}}\label{Ring_normalizing_constant}
\end{figure*}

\noindent Another challenging, high dimensional, complex example, originally introduced in \citep{chen2022riemannian}, is considered in this section. The original joint distribution, $\pi_{\bm{X}}(\bm{X})=\pi_{\bm{X}}(\bm{X}\vert y)$, is constructed in a Bayesian context. The observed random variable $y$ is assumed to follow a Gaussian distribution with mean $\mu_Y=2$ and $\sigma_Y=4$, and 100 data points are then generated accordingly. A likelihood function $L(y\vert \bm{X})$ is then defined as:
\begin{equation}
L(y\vert \bm{X}) = \prod_{j=1}^{100}  \frac{1}{\sigma_Y \sqrt{2\pi}}  \,\exp{\bigg(-\dfrac{(y_j-s_{\bm{X}})^2}{2\,\sigma_Y^2}\bigg)}
\label{Likelihood_ring}
\end{equation}
where $s_{\bm{X}} \coloneqq \sum_{i=1}^{d} X_i^2$ is assumed hereafter to be an unknown mean parameter, intended to be estimated using the Bayes' rule. An independent standard Gaussian prior distribution, $f_{\bm{X}}(\bm{X})=\prod_{i=1}^{d} N(X_i \vert 0,\,1)$, is then assumed for $\bm{X}$. The posterior distribution can thus be expressed as: 
\begin{equation}\label{Ring_posterior}
\begin{aligned}
\pi_{\bm{X}}(&\bm{X}\vert y) =  \dfrac{L(y\vert \bm{x}) f_{\bm{X}}(\bm{x})}{\pi(y)} \\& = \dfrac{\exp\bigg(-\dfrac{s_{\bm{X}}}{2}-\dfrac{1}{2\sigma_Y^2} \sum_{j=1}^{100} (y_j-s_{\bm{X}})^2\bigg)}{C_\pi}=\dfrac{\tilde{\pi}_{\bm{X}}(\bm{X}\vert y)}{C_\pi}
\end{aligned}
\end{equation}
where $\pi(y)$ is the Bayes' evidence, subsequently absorbed in $C_\pi$, a normalizing constant of the unnormalized posterior $\tilde{\pi}_{\bm{X}}(\bm{X}\vert y)$. \cref{Ring_normalizing_constant}(a) depicts this unnormalized posterior distribution for $d=2$. Since ASTPA necessitates a normalized original distribution, the normalizing constant $C_\pi$ is first computed here appropriately.

\begin{table*}[t!]
\captionsetup{justification=centering}
\caption{Example 5: The estimated normalizing constant $C_\pi$ for the ring-shaped distribution.}
\centering
\footnotesize
\setlength\tabcolsep{4pt}
\begin{tabular}{C{3.5cm}C{2cm}C{2cm}C{2cm}C{2cm}}
  \toprule[1.5pt]
    \textbf{50 Independent Simulations}&  $d=2$ & $d=50$ & $d=150$&$d=500$\\ 
 \cmidrule(lr){1-5}
    $N_{\pi}^{Total}$ & 11,000 &11,000&11,000& 20,000\\
       C.o.V & 0.01&0.01&0.02&0.03\\
       $\mathop{\mathbb{E}}[\hat{C}_\pi]$    & 3.43E-19& 3.26E-21&1.05E-50& 6.41E-210 \\
      \bottomrule[1.5pt]
\end{tabular}\label{table_Ring_c_pi}
\end{table*}

\subsubsection{Normalizing constant of the posterior distribution}\label{pi_norm_const}
\noindent The normalizing constant $C_\pi$ can be expressed as:
\begin{equation}
   C_\pi=\int_{\mathcal{X}} \tilde{\pi}_{\bm{X}}(\bm{x}\vert y) d\bm{x}
\end{equation}
We originally employ our developed inverse importance sampling (IIS) technique to evaluate this integral. First, we draw a sample set $\{\bm{x}_i\}_{i=1}^{N_\pi}$ from the posterior distribution $\tilde{\pi}_{\bm{X}}(\bm{X}\vert y)$ using the HMCMC sampler. A GMM $Q_\pi(\bm{X})$ structured as recommended in \cref{IIS} is subsequently fitted based on $\{\bm{x}_i\}_{i=1}^{N_\pi}$, which is then used as ISD, as follows:
\begin{equation}\label{c_pi}
   C_\pi=\int_{\mathcal{X}} \dfrac{\tilde{\pi}_{\bm{X}}(\bm{x}\vert y)}{Q_\pi(\bm{x})}Q_\pi(\bm{x}) d\bm{x}=\mathbb{E}_{Q_\pi} \big[ \dfrac{\tilde{\pi}_{\bm{X}}(\bm{x}\vert y)}{Q_\pi(\bm{X})}\big] 
\end{equation}
By drawing an i.i.d sample set $\{\bm{x}_i^\prime\}_{i=1}^{M_{\pi}}$ from $Q_\pi$, $\hat{C}_\pi$ can be eventually computed as: 
\begin{equation} \label{c_pi_hat}
\hat{C}_\pi = \dfrac{1}{M_\pi}\sum_{i=1}^{M_\pi} \dfrac{\tilde{\pi}_{\bm{X}}(\bm{x}_i^\prime \vert y)}{Q_\pi(\bm{x}_i^\prime)}
\end{equation}

\begin{table*}[t!]
\caption{Example 5: Performance of various methods for the high dimensional problem with the ring-shaped distribution}
\centering
\footnotesize
\setlength\tabcolsep{4pt}
\begin{tabular}{p{1.5cm}C{2cm}C{2cm}C{2cm}C{2cm}C{2cm}}
  \toprule[1.5pt]
  \multirow{2}{*}{\cref{Ring2}}& \textbf{100 Independent} & \multirow{2}{*}{\textbf{MCMC}}& \multicolumn{2}{c}{\textbf{ASTPA}  ($\sigma = 0.3, \, q = 10$)}& \multirow{2}{*}{\textbf{CWMH-SuS}} \\ 
  
  \cline{4-5}
    \addlinespace[2pt]
  & \textbf{Simulations}&&\multicolumn{1}{c}{\head{HMCMC}} & \multicolumn{1}{c}{\head{QNp-HMCMC}}&\\ 
 \cmidrule(lr){1-6}

         \multirow{4}{*}{\shortstack[l]{\vspace{-0.14in}\\$d=2$\\$r=3.8$}}\rule{0pt}{2.5ex}    &$\mathop{\mathbb{E}}[N_{Total}]$ & 1.00E7& 1,639 & 1,639 &9,812    \\
      &C.o.V &0.20& 0.07$\color{ForestGreen}($0.16$\color{ForestGreen})$  &0.09$\color{ForestGreen}($0.10$\color{ForestGreen})$& 2.05 \\
       &$\mathop{\mathbb{E}}[\hat{p}_{\mathcal{F}}]$ \ \ \ &3.38E-5 & 3.48E-5 & 3.45E-5 & 3.34E-5 \\
      \cmidrule(lr){1-6}
           
    \multirow{4}{*}{\shortstack[l]{\vspace{-0.14in}\\$d=50$\\$r=3.4$}}\rule{0pt}{2.5ex}  &$\mathop{\mathbb{E}}[N_{Total}]$ & 1.00E7&3,156 &3,156  &18,724  \\
       &C.o.V &0.09&  0.20$\color{ForestGreen}($0.22$\color{ForestGreen})$  &0.15$\color{ForestGreen}($0.17$\color{ForestGreen})$& 0.73  \\
       &$\mathop{\mathbb{E}}[\hat{p}_{\mathcal{F}}]$ \ \ \  &2.32E-5& 2.36E-5   &2.29E-5 & 3.53E-5 \\
     \cmidrule(lr){1-6}
      
    \multirow{4}{*}{\shortstack[l]{\vspace{-0.14in}\\$d=150$\\$r=3.4$}}\rule{0pt}{2.5ex}  &$\mathop{\mathbb{E}}[N_{Total}]$ &1.00E6& 5,056& 5,056 & 17,330  \\
       &C.o.V &0.22& 0.14$\color{ForestGreen}($0.16$\color{ForestGreen})$  &0.17$\color{ForestGreen}($0.17$\color{ForestGreen})$&0.72  \\
       &$\mathop{\mathbb{E}}[\hat{p}_{\mathcal{F}}]$ \ \ \    &6.78E-5&6.91E-5 & 6.85E-5 & 9.26E-5 \\
      \cmidrule(lr){1-6}
      
    \multirow{4}{*}{\shortstack[l]{\vspace{-0.14in}\\$d=150$\\$r=3.6$}}\rule{0pt}{2.5ex}  &$\mathop{\mathbb{E}}[N_{Total}]$ &&4,998 & 4,998 &32,476  \\
       &C.o.V & &0.10$\color{ForestGreen}($0.15$\color{ForestGreen})$  &0.12$\color{ForestGreen}($0.16$\color{ForestGreen})$&2.25\\
       &$\mathop{\mathbb{E}}[\hat{p}_{\mathcal{F}}]$ \ \ \   &&1.65E-8  & 1.69E-8 &1.54E-8 \\
      \cmidrule(lr){1-6} 
      
    \multirow{4}{*}{\shortstack[l]{\vspace{-0.14in}\\$d=500$\\$r=3.6$}}\rule{0pt}{2.5ex}  &$\mathop{\mathbb{E}}[N_{Total}]$ &1.00E5& 6,597& 6,597& 11,200 \\
       &C.o.V &0.10 &0.18$\color{ForestGreen}($0.17$\color{ForestGreen})$ &0.18$\color{ForestGreen}($0.16$\color{ForestGreen})$&0.35  \\
       &$\mathop{\mathbb{E}}[\hat{p}_{\mathcal{F}}]$ \ \ \   &2.90E-3&2.89E-3  & 2.88E-3 & 3.30E-3 \\
     \cmidrule(lr){1-6}
            
          \multirow{4}{*}{\shortstack[l]{\vspace{-0.14in}\\$d=500$\\$r=3.8$}}\rule{0pt}{2.5ex}    &$\mathop{\mathbb{E}}[N_{Total}]$ &&6,639 &6,639&   27,976\\
       &C.o.V& & 0.23$\color{ForestGreen}($0.35$\color{ForestGreen})$  &0.23$\color{ForestGreen}($0.35$\color{ForestGreen})$ & 5.49\\
       &$\mathop{\mathbb{E}}[\hat{p}_{\mathcal{F}}]$ \ \ \  & & 0.99E-7& 1.00E-7& 4.99E-7\\
      \bottomrule[1.5pt]
\end{tabular}\label{Ring_tabel}
\end{table*}

\cref{Ring_normalizing_constant} showcases this procedure for the two-dimensional case. \cref{table_Ring_c_pi} shows the estimated $C_\pi$ for dimensions $d=2, 50, 150,$ and $500$, based on $50$ independent simulations. We report the mean estimates for the normalizing constant,  
 $\mathop{\mathbb{E}}[\hat{C}_\pi]$, the Coefficient of Variation (C.o.V) of $\hat{C}_\pi$, and the number of total $\tilde{\pi}_{\bm{X}}(\bm{x}_i^\prime \vert y)$ evaluations, $N_{\pi}^{Total}=N_\pi+M_\pi$. Estimating $C_\pi$ does not require any (expensive) limit-state function evaluations, and thus, the focus is on obtaining accurate estimates characterized by lower C.o.V regardless of the number of samples. We did not compare our computed $\hat{C}_\pi$ estimator with other numerical methods for evaluating normalizing constants, as this is not the main scope of this paper, and its accuracy is instead verified through the computed probabilities in \Cref{Target probability estimation}. 

\subsubsection{Target probability estimation} \label{Target probability estimation}
\begin{figure*}[t!]
\centerline{\subfigure[]{\includegraphics[trim=1cm 3.5cm 0cm 3cm,width=0.3\textwidth]{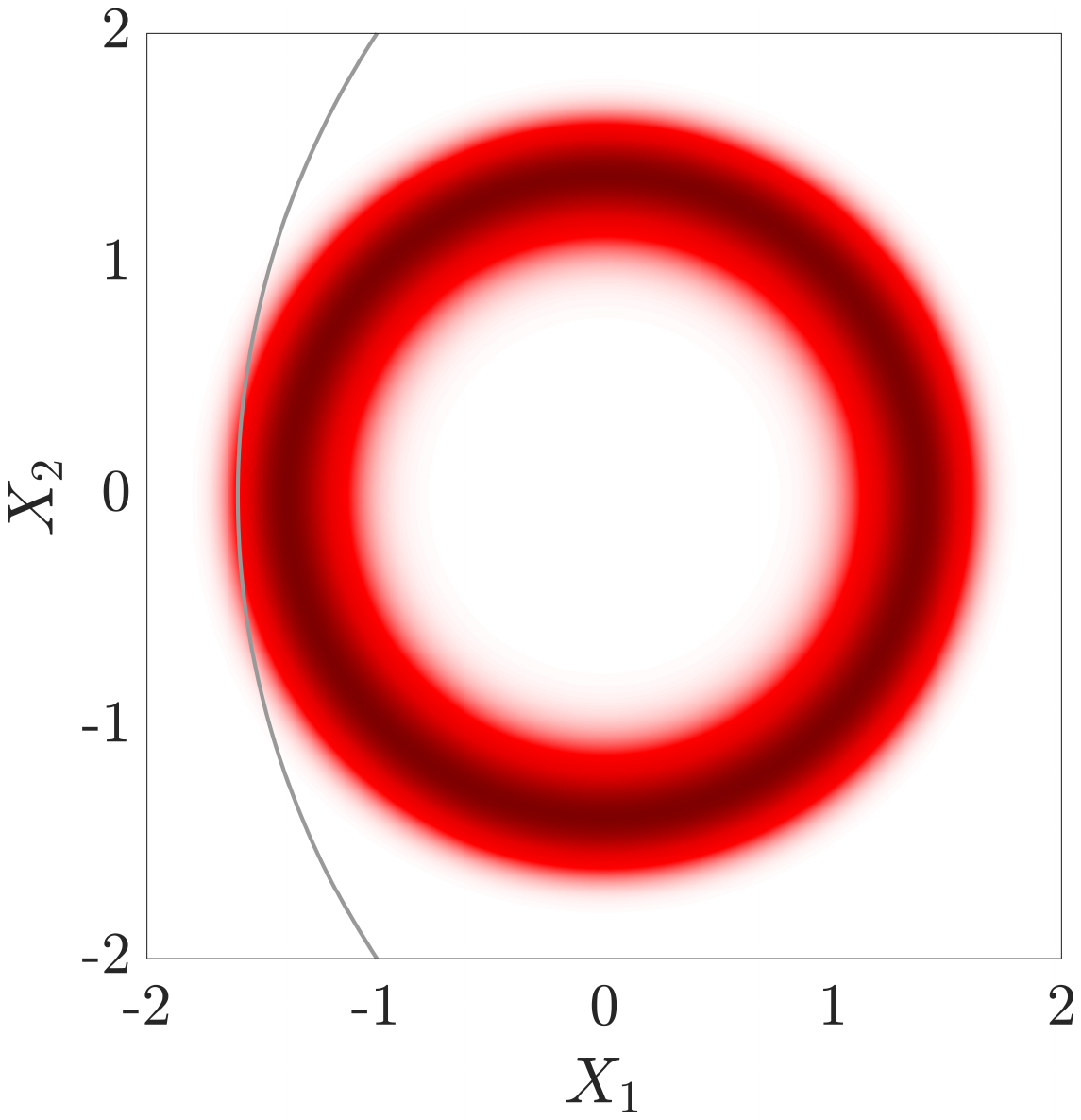}}
\subfigure[]{\includegraphics[trim=1cm 3.5cm 0cm 3cm,width=0.3\textwidth]{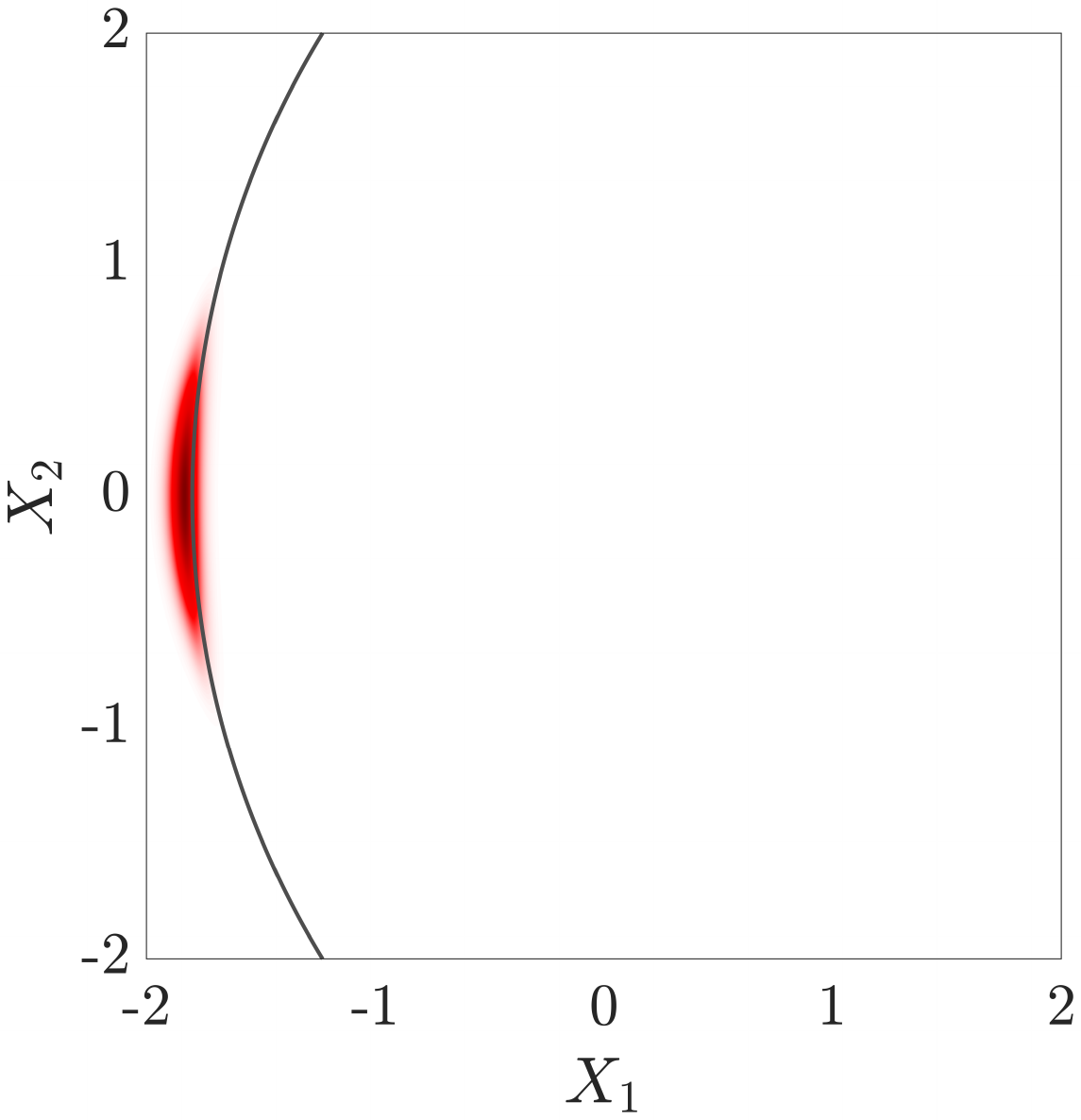}}
\subfigure[]{\includegraphics[trim=1cm 3.5cm 0cm 3cm,width=0.3\textwidth]{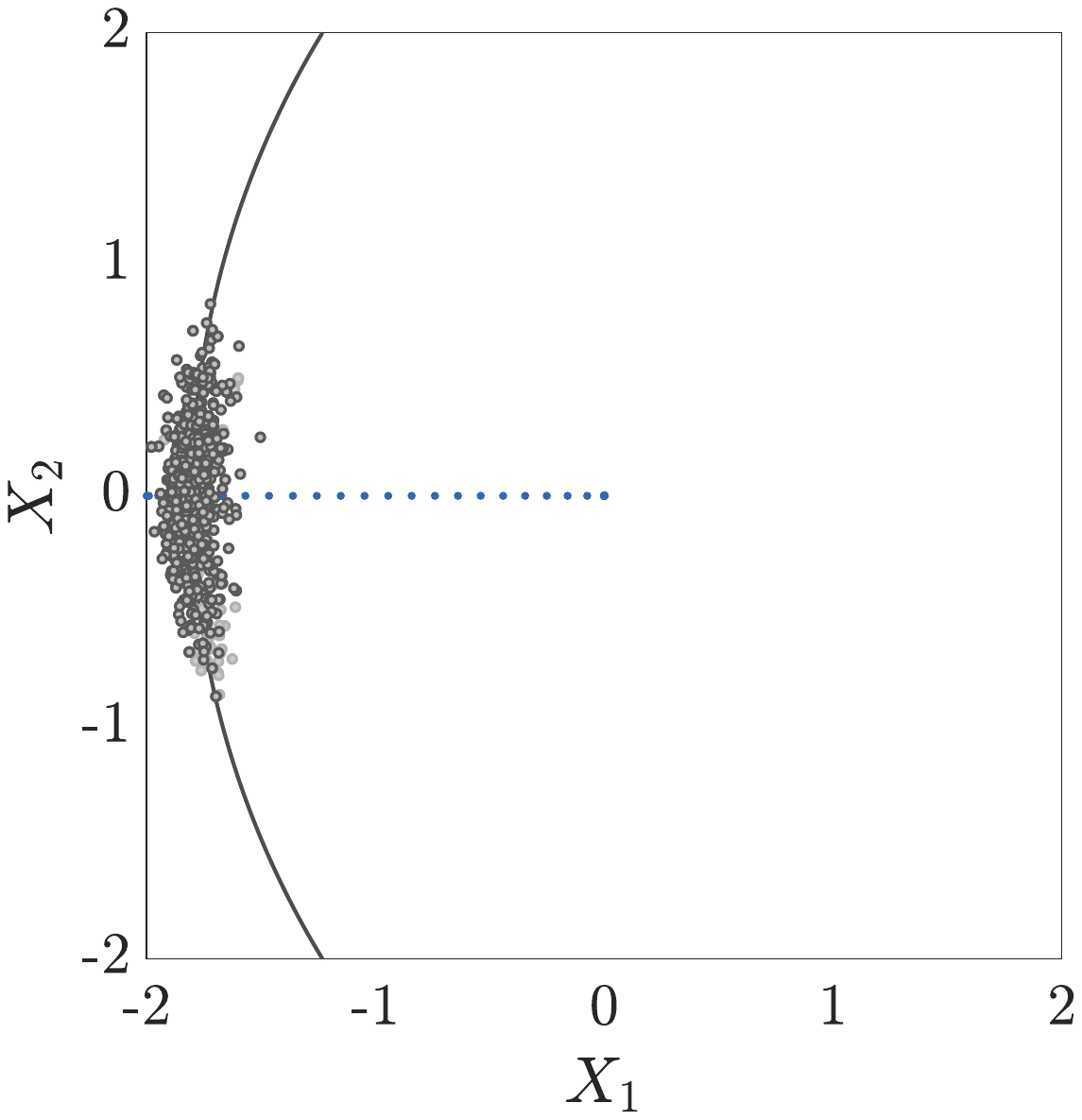}}}
  \captionsetup{labelfont={color=Black}}
\caption{Example 5: (a) Ring-shaped distribution plotted in red; the gray line represents the limit-state surface at $g(\bm{X})=0$, (b) The approximate sampling target $\tilde{h}$ in ASTPA, (c) Adam steps plotted in blue; gray samples drawn from $h$ using QNp-HMCMC.\vspace{-0.1in}}\label{Ring_fig}
\end{figure*}

\noindent The following quadratic limit-state function is now considered:
\begin{equation}
g(\bm{X}) = r^2-(X_1-2)^2-\sum_{i=2}^{d} X_{i}^2
\label{Ring2}
\end{equation} 
where the threshold $r$ determines the level of rarity. \cref{Ring_fig} depicts for $d=2$ the ringed-shaped distribution $\pi_{\bm{X}}(\bm{X}\vert y)$, the approximate sampling target $\tilde{h}(\bm{X})$, its samples, and the Adam steps. 

The presented results in \cref{Ring_tabel} demonstrate the superiority of the proposed framework, examined here for dimensions as high as $500$. The MCMC term in \cref{Ring_tabel} denotes utilizing the standard Monte Carlo approach, with samples generated using an MCMC algorithm (HMCMC), since direct sampling is not applicable for the considered posterior distribution $\pi_{\bm{X}}(\bm{X}\vert y)$. However, we could not computationally afford these standard MCMC computations for the two cases with $d=150$ and $500$, for which the target probability is extremely low ($\sim 10^{-8})$. Comparisons with MCMC results show the accuracy of the mean estimate $\mathop{\mathbb{E}}[\hat{p}_{\mathcal{F}}]$, implying also the accuracy of the computed normalizing constant $C_\pi$ in \cref{table_Ring_c_pi}. ASTPA has again significantly outperformed CWMH-SuS, which utilized HMCMC to generate the first set. Similar performance is observed in this example for HMCMC and QNp-HMCMC, which utilized diagonal inverse Hessian and mass matrices for $d=500$. Once more, the analytical C.o.V demonstrates good agreement with the sampling C.o.V. \par

The target probability in this example is referred to in the literature as a posteriori rare event probability \citep{cui2024deep}. The ASTPA estimator, in this case, can be expressed as:
\begin{equation} 
p_\mathcal{F} = \tilde{p}_\mathcal{F} \,\, C_h = \dfrac{C_h}{C_\pi}\,{\mathop{\mathbb{E}}}_{h} \big[ I_{\mathcal{F}} (\bm{X}) \dfrac{\tilde{\pi}_{\bm{X}}(\bm{X})}{\tilde{h}(\bm{X})}\big] = \dfrac{\tilde{\tilde{p}}_\mathcal{F}\,\,C_h}{C_\pi} 
\end{equation} 
This estimate is known to be theoretically biased due to the existence of the ratio estimator, even if $\tilde{\tilde{p}}_\mathcal{F}$, $C_\pi$, and $C_h$ are independent \citep{cui2024deep}. However, in \citep{cui2024deep}, where the ISD $h$ is normalized, i.e., $C_h$ is dropped, the authors proved theoretically that this biasedness is negligible if $h$ and $Q_\pi$ closely resemble $I_{\mathcal{F}} \pi_{\bm{X}}$ and $\pi_{\bm{X}}$, respectively. This is true in our proposed framework since, in fact, $h$ is a smoothed version of $I_{\mathcal{F}} \pi_{\bm{X}}$, and $Q_\pi$ is fitted on samples from $\pi_{\bm{X}}$, as demonstrated in \Cref{Ring_normalizing_constant}(a) and (c).

\section{Conclusions}\label{conclusion_sec}

\noindent A novel framework for accurately and efficiently estimating rare event probabilities directly in complex non-Gaussian spaces is presented in this paper, suitable for low- and high-dimensional problems, building upon our foundational Approximate Sampling Target with Post-processing Adjustment (ASTPA) approach \citep{Papakon2023HMCMC}. An unnormalized sampling target is first constructed and sampled, relaxing the optimal importance sampling distribution, and appropriately designed for non-Gaussian spaces.  The obtained samples are subsequently utilized not only to compute a shifted estimate of the sought probability but also to guide the second stage in ASTPA. Post-sampling, the sampling target's normalizing constant is estimated using a stable inverse importance sampling procedure, which employs an importance sampling density based on the already available samples. As shown and discussed in the paper, a very approximately fitted Gaussian Mixture Model (GMM), even using one component with a diagonal covariance matrix, is adequate at this step, thus avoiding the known scalability issues of GMMs, as also showcased by the high-dimensional numerical examples, with dimensions up to 500 presented in this work. The target probability is eventually computed based on the estimations in these two stages. We employed our developed Quasi-Newton mass preconditioned Hamiltonian MCMC (QNp-HMCMC) algorithm to sample the constructed target, achieving outstanding performance in sampling high-dimensional, complex distributions.

In comparison to our relevant framework for estimating rare event probabilities in Gaussian spaces in \citep{Papakon2023HMCMC}, this paper introduces significant enhancements and novel contributions to enable direct applicability in non-Gaussian spaces. We have first adjusted the recommended values for parameters of the ASTPA sampling target to suit non-Gaussian spaces. The inverse importance sampling technique has also advanced through a stable implementation approach, making it robust against possible anomalous samples. The proposed estimator is further theoretically analyzed in this work, proving its unbiasedness and deriving its analytical coefficient of variation (C.o.V). The analytical C.o.V expression, accompanied by an applied implementation technique based on the effective sample size, demonstrates accurate agreement with empirical results. Further, our developed Quasi-Newton mass preconditioned Hamiltonian MCMC (QNp-HMCMC) is theoretically reinforced in this work, proving that it converges to the correct stationary target distribution. Avoiding the challenging task of tuning the trajectory length in general, complex, non-Gaussian spaces, a variant of the QNp-HMCMC is effectively utilized in this manuscript based on a single-step integration, and we are showing its equivalence to an original and efficient preconditioned Metropolis adjusted Langevin algorithm (MALA). An optimization approach based on the Adam optimizer is also devised, to initiate the QNp-HMCMC effectively. The utilization of the proposed framework in bounded spaces is also discussed and examined. 

We have demonstrated the notable performance of our developed methodology through comparisons with the state-of-the-art Subset Simulation, and related variants, in a series of diverse, low- and high-dimensional non-Gaussian problems. Our methodology sets new efficiency benchmarks on existing and new, challenging non-Gaussian problems, including high-dimensional Neal's funnel distribution and two- and three-dimensional Rosenbrock distributions, as well as a high-dimensional, highly correlated Gumbel distribution. In addition, we have successfully applied our framework to scenarios involving unnormalized probability distributions, such as those often appearing in Bayesian contexts, where the target probability is related to a posteriori rare events. 

Since we are utilizing gradient-based sampling methods in this work, all of our results and conclusions are based on the fact that analytical gradients can be computed. Some of our ongoing and future works are further directed toward exploring various ASTPA variants. That includes devising effective discovery schemes for challenging multi-modal rare event domains involving distant modes, developing a related non-gradient-based sampling framework, and estimating high-dimensional first-passage problems under various settings \citep{PapakonICASP2023}.

\section*{Acknowledgements}
\noindent The authors would like to thank Prof. Dr. Daniel Straub, Dr. Iason Papaioannou, and Dr. Max Ehre at the Technical University of Munich, for very fruitful scientific discussions in relation to this manuscript. This material is based upon work partially supported by the U.S. National Science Foundation under CAREER Grant No. 1751941.

\bibliography{references}  

\begin{thebibliography}{93}
\expandafter\ifx\csname natexlab\endcsname\relax\def\natexlab#1{#1}\fi
\providecommand{\url}[1]{\texttt{#1}}
\providecommand{\href}[2]{#2}
\providecommand{\path}[1]{#1}
\providecommand{\DOIprefix}{doi:}
\providecommand{\ArXivprefix}{arXiv:}
\providecommand{\URLprefix}{URL: }
\providecommand{\Pubmedprefix}{pmid:}
\providecommand{\doi}[1]{\href{http://dx.doi.org/#1}{\path{#1}}}
\providecommand{\Pubmed}[1]{\href{pmid:#1}{\path{#1}}}
\providecommand{\bibinfo}[2]{#2}
\ifx\xfnm\relax \def\xfnm[#1]{\unskip,\space#1}\fi
\bibitem[{Ditlevsen and Madsen(2007)}]{ditlevsen1996structural}
\bibinfo{author}{O.~Ditlevsen}, \bibinfo{author}{H.~O. Madsen}, \bibinfo{title}{Structural {Reliability Methods}}, \bibinfo{publisher}{Department of Mechanical Engineering, Technical University of Denmark}, \bibinfo{year}{2007}.
\bibitem[{Nikolaidis et~al.(2005)Nikolaidis, Ghiocel, and Singhal}]{nikolaidisengineering}
\bibinfo{author}{E.~Nikolaidis}, \bibinfo{author}{D.~Ghiocel}, \bibinfo{author}{S.~Singhal}, \bibinfo{title}{{Engineering Design Reliability Handbook}}, \bibinfo{publisher}{CRC press}, \bibinfo{year}{2005}.
\bibitem[{Lemaire et~al.(2009)Lemaire, Chateauneuf, and Mitteau}]{lemaire2009structural}
\bibinfo{author}{M.~Lemaire}, \bibinfo{author}{A.~Chateauneuf}, \bibinfo{author}{J.~Mitteau}, \bibinfo{title}{Structural {Reliability}}, \bibinfo{publisher}{Wiley}, \bibinfo{year}{2009}.
\bibitem[{Au and Wang(2014)}]{au2014engineering}
\bibinfo{author}{S.-K. Au}, \bibinfo{author}{Y.~Wang}, \bibinfo{title}{Engineering {Risk Assessment} with {Subset Simulation}}, \bibinfo{publisher}{Wiley}, \bibinfo{year}{2014}.
\bibitem[{Melchers and Beck(2018)}]{melchers2018structural}
\bibinfo{author}{R.~E. Melchers}, \bibinfo{author}{A.~T. Beck}, \bibinfo{title}{Structural {Reliability Analysis and Prediction}}, \bibinfo{publisher}{Wiley}, \bibinfo{year}{2018}.
\bibitem[{Der~Kiureghian(2022)}]{der2022structural}
\bibinfo{author}{A.~Der~Kiureghian}, \bibinfo{title}{Structural and system reliability}, \bibinfo{publisher}{Cambridge University Press}, \bibinfo{year}{2022}.
\bibitem[{Rackwitz(2001)}]{rackwitz2001reliability}
\bibinfo{author}{R.~Rackwitz},
\newblock \bibinfo{title}{Reliability analysis—a review and some perspectives},
\newblock \bibinfo{journal}{Structural Safety} \bibinfo{volume}{23} (\bibinfo{year}{2001}) \bibinfo{pages}{365--395}.
\bibitem[{Breitung(2015)}]{breitung201540}
\bibinfo{author}{K.~Breitung},
\newblock \bibinfo{title}{40 years {FORM}: Some new aspects?},
\newblock \bibinfo{journal}{Probabilistic Engineering Mechanics} \bibinfo{volume}{42} (\bibinfo{year}{2015}) \bibinfo{pages}{71--77}.
\bibitem[{Breitung(1984)}]{breitung1984asymptotic}
\bibinfo{author}{K.~Breitung},
\newblock \bibinfo{title}{Asymptotic approximations for multinormal integrals},
\newblock \bibinfo{journal}{Journal of Engineering Mechanics} \bibinfo{volume}{110} (\bibinfo{year}{1984}) \bibinfo{pages}{357--366}.
\bibitem[{Asmussen and Glynn(2007)}]{asmussen2007stochastic}
\bibinfo{author}{S.~Asmussen}, \bibinfo{author}{P.~W. Glynn}, \bibinfo{title}{Stochastic simulation: algorithms and analysis}, volume~\bibinfo{volume}{57}, \bibinfo{publisher}{Springer}, \bibinfo{year}{2007}.
\bibitem[{Dembo and Zeitouni(2009)}]{dembo2009large}
\bibinfo{author}{A.~Dembo}, \bibinfo{author}{O.~Zeitouni}, \bibinfo{title}{Large deviations techniques and applications}, volume~\bibinfo{volume}{38}, \bibinfo{publisher}{Springer Science \& Business Media}, \bibinfo{year}{2009}.
\bibitem[{Grafke and Vanden-Eijnden(2019)}]{grafke2019numerical}
\bibinfo{author}{T.~Grafke}, \bibinfo{author}{E.~Vanden-Eijnden},
\newblock \bibinfo{title}{Numerical computation of rare events via large deviation theory},
\newblock \bibinfo{journal}{Chaos: An Interdisciplinary Journal of Nonlinear Science} \bibinfo{volume}{29} (\bibinfo{year}{2019}).
\bibitem[{Breitung(2006)}]{breitung2006asymptotic}
\bibinfo{author}{K.~W. Breitung}, \bibinfo{title}{Asymptotic approximations for probability integrals}, \bibinfo{publisher}{Springer}, \bibinfo{year}{2006}.
\bibitem[{Valdebenito et~al.(2010)Valdebenito, Pradlwarter, and Schu{\"e}ller}]{valdebenito2010role}
\bibinfo{author}{M.~Valdebenito}, \bibinfo{author}{H.~Pradlwarter}, \bibinfo{author}{G.~Schu{\"e}ller},
\newblock \bibinfo{title}{The role of the design point for calculating failure probabilities in view of dimensionality and structural nonlinearities},
\newblock \bibinfo{journal}{Structural Safety} \bibinfo{volume}{32} (\bibinfo{year}{2010}) \bibinfo{pages}{101--111}.
\bibitem[{Bai et~al.(2023)Bai, Dieker, and Lam}]{bai2023curse}
\bibinfo{author}{Y.~Bai}, \bibinfo{author}{A.~B. Dieker}, \bibinfo{author}{H.~Lam},
\newblock \bibinfo{title}{Curse of dimensionality in rare-event simulation},
\newblock in: \bibinfo{booktitle}{2023 Winter Simulation Conference (WSC)}, \bibinfo{organization}{IEEE}, \bibinfo{year}{2023}, pp. \bibinfo{pages}{375--384}.
\bibitem[{Au and Beck(2001)}]{au2001estimation}
\bibinfo{author}{S.-K. Au}, \bibinfo{author}{J.~L. Beck},
\newblock \bibinfo{title}{Estimation of small failure probabilities in high dimensions by {Subset Simulation}},
\newblock \bibinfo{journal}{Probabilistic {Engineering Mechanics}} \bibinfo{volume}{16} (\bibinfo{year}{2001}) \bibinfo{pages}{263--277}.
\bibitem[{Zuev and Katafygiotis(2011)}]{zuev2011modified}
\bibinfo{author}{K.~M. Zuev}, \bibinfo{author}{L.~S. Katafygiotis},
\newblock \bibinfo{title}{Modified {Metropolis--Hastings} algorithm with delayed rejection},
\newblock \bibinfo{journal}{Probabilistic Engineering Mechanics} \bibinfo{volume}{26} (\bibinfo{year}{2011}) \bibinfo{pages}{405--412}.
\bibitem[{Zuev et~al.(2012)Zuev, Beck, Au, and Katafygiotis}]{zuev2012bayesian}
\bibinfo{author}{K.~M. Zuev}, \bibinfo{author}{J.~L. Beck}, \bibinfo{author}{S.-K. Au}, \bibinfo{author}{L.~S. Katafygiotis},
\newblock \bibinfo{title}{Bayesian post-processor and other enhancements of {Subset Simulation} for estimating failure probabilities in high dimensions},
\newblock \bibinfo{journal}{Computers \& {Structures}} \bibinfo{volume}{92} (\bibinfo{year}{2012}) \bibinfo{pages}{283--296}.
\bibitem[{Au and Patelli(2016)}]{au2016rare}
\bibinfo{author}{S.-K. Au}, \bibinfo{author}{E.~Patelli},
\newblock \bibinfo{title}{Rare event simulation in finite-infinite dimensional space},
\newblock \bibinfo{journal}{Reliability Engineering \& System Safety} \bibinfo{volume}{148} (\bibinfo{year}{2016}) \bibinfo{pages}{67--77}.
\bibitem[{Papaioannou et~al.(2015)Papaioannou, Betz, Zwirglmaier, and Straub}]{papaioannou2015mcmc}
\bibinfo{author}{I.~Papaioannou}, \bibinfo{author}{W.~Betz}, \bibinfo{author}{K.~Zwirglmaier}, \bibinfo{author}{D.~Straub},
\newblock \bibinfo{title}{{MCMC} algorithms for {Subset Simulation}},
\newblock \bibinfo{journal}{Probabilistic Engineering Mechanics} \bibinfo{volume}{41} (\bibinfo{year}{2015}) \bibinfo{pages}{89--103}.
\bibitem[{Breitung(2019)}]{breitung2019geometry}
\bibinfo{author}{K.~Breitung},
\newblock \bibinfo{title}{The geometry of limit state function graphs and subset simulation: Counterexamples},
\newblock \bibinfo{journal}{Reliability Engineering \& System Safety} \bibinfo{volume}{182} (\bibinfo{year}{2019}) \bibinfo{pages}{98--106}.
\bibitem[{Neal(2012)}]{neal2011mcmc}
\bibinfo{author}{R.~M. Neal},
\newblock \bibinfo{title}{{MCMC} using {Hamiltonian dynamics}.},
\newblock \bibinfo{journal}{arXiv preprint arXiv:1206.1901}  (\bibinfo{year}{2012}).
\bibitem[{Wang et~al.(2019)Wang, Broccardo, and Song}]{wang2019hamiltonian}
\bibinfo{author}{Z.~Wang}, \bibinfo{author}{M.~Broccardo}, \bibinfo{author}{J.~Song},
\newblock \bibinfo{title}{{Hamiltonian Monte Carlo} methods for {Subset Simulation} in reliability analysis},
\newblock \bibinfo{journal}{Structural Safety} \bibinfo{volume}{76} (\bibinfo{year}{2019}) \bibinfo{pages}{51--67}.
\bibitem[{Girolami and Calderhead(2011)}]{girolami2011riemann}
\bibinfo{author}{M.~Girolami}, \bibinfo{author}{B.~Calderhead},
\newblock \bibinfo{title}{{Riemann manifold Langevin} and {Hamiltonian Monte Carlo} methods},
\newblock \bibinfo{journal}{Journal of the Royal Statistical Society: Series B (Statistical Methodology)} \bibinfo{volume}{73} (\bibinfo{year}{2011}) \bibinfo{pages}{123--214}.
\bibitem[{Chen et~al.(2022)Chen, Wang, Broccardo, and Song}]{chen2022riemannian}
\bibinfo{author}{W.~Chen}, \bibinfo{author}{Z.~Wang}, \bibinfo{author}{M.~Broccardo}, \bibinfo{author}{J.~Song},
\newblock \bibinfo{title}{{Riemannian Manifold Hamiltonian Monte Carlo} based {Subset Simulation} for reliability analysis in non-{G}aussian space},
\newblock \bibinfo{journal}{Structural Safety} \bibinfo{volume}{94} (\bibinfo{year}{2022}) \bibinfo{pages}{102134}.
\bibitem[{Thaler et~al.(2024)Thaler, Dhulipala, Bamer, Markert, and Shields}]{thaler2024reliability}
\bibinfo{author}{D.~Thaler}, \bibinfo{author}{S.~L. Dhulipala}, \bibinfo{author}{F.~Bamer}, \bibinfo{author}{B.~Markert}, \bibinfo{author}{M.~D. Shields},
\newblock \bibinfo{title}{Reliability analysis of complex systems using {Subset Simulations} with {Hamiltonian Neural Networks}},
\newblock \bibinfo{journal}{arXiv preprint arXiv:2401.05244}  (\bibinfo{year}{2024}).
\bibitem[{Goodman and Weare(2010)}]{goodman2010ensemble}
\bibinfo{author}{J.~Goodman}, \bibinfo{author}{J.~Weare},
\newblock \bibinfo{title}{Ensemble samplers with affine invariance},
\newblock \bibinfo{journal}{Communications in Applied Mathematics and Computational Science} \bibinfo{volume}{5} (\bibinfo{year}{2010}) \bibinfo{pages}{65--80}.
\bibitem[{Shields et~al.(2021)Shields, Giovanis, and Sundar}]{shields2021subset}
\bibinfo{author}{M.~D. Shields}, \bibinfo{author}{D.~G. Giovanis}, \bibinfo{author}{V.~Sundar},
\newblock \bibinfo{title}{{Subset Simulation} for problems with strongly {non-Gaussian}, highly anisotropic, and degenerate distributions},
\newblock \bibinfo{journal}{Computers \& Structures} \bibinfo{volume}{245} (\bibinfo{year}{2021}) \bibinfo{pages}{106431}.
\bibitem[{Guyader et~al.(2011)Guyader, Hengartner, and Matzner-L{\o}ber}]{guyader2011simulation}
\bibinfo{author}{A.~Guyader}, \bibinfo{author}{N.~Hengartner}, \bibinfo{author}{E.~Matzner-L{\o}ber},
\newblock \bibinfo{title}{Simulation and estimation of extreme quantiles and extreme probabilities},
\newblock \bibinfo{journal}{Applied Mathematics \& Optimization} \bibinfo{volume}{64} (\bibinfo{year}{2011}) \bibinfo{pages}{171--196}.
\bibitem[{C{\'e}rou et~al.(2012)C{\'e}rou, Del~Moral, Furon, and Guyader}]{cerou2012sequential}
\bibinfo{author}{F.~C{\'e}rou}, \bibinfo{author}{P.~Del~Moral}, \bibinfo{author}{T.~Furon}, \bibinfo{author}{A.~Guyader},
\newblock \bibinfo{title}{Sequential {Monte Carlo} for rare event estimation},
\newblock \bibinfo{journal}{Statistics and Computing} \bibinfo{volume}{22} (\bibinfo{year}{2012}) \bibinfo{pages}{795--808}.
\bibitem[{Walter(2015)}]{walter2015moving}
\bibinfo{author}{C.~Walter},
\newblock \bibinfo{title}{Moving particles: A parallel optimal multilevel splitting method with application in quantiles estimation and meta-model based algorithms},
\newblock \bibinfo{journal}{Structural {Safety}} \bibinfo{volume}{55} (\bibinfo{year}{2015}) \bibinfo{pages}{10--25}.
\bibitem[{Huang et~al.(2016)Huang, Chen, and Zhu}]{huang2016assessing}
\bibinfo{author}{X.~Huang}, \bibinfo{author}{J.~Chen}, \bibinfo{author}{H.~Zhu},
\newblock \bibinfo{title}{Assessing small failure probabilities by {AK--SS}: An active learning method combining kriging and {Subset Simulation}},
\newblock \bibinfo{journal}{Structural Safety} \bibinfo{volume}{59} (\bibinfo{year}{2016}) \bibinfo{pages}{86--95}.
\bibitem[{Marelli and Sudret(2018)}]{marelli2018active}
\bibinfo{author}{S.~Marelli}, \bibinfo{author}{B.~Sudret},
\newblock \bibinfo{title}{An active-learning algorithm that combines sparse polynomial chaos expansions and bootstrap for structural reliability analysis},
\newblock \bibinfo{journal}{Structural Safety} \bibinfo{volume}{75} (\bibinfo{year}{2018}) \bibinfo{pages}{67--74}.
\bibitem[{Moustapha et~al.(2022)Moustapha, Marelli, and Sudret}]{moustapha2022active}
\bibinfo{author}{M.~Moustapha}, \bibinfo{author}{S.~Marelli}, \bibinfo{author}{B.~Sudret},
\newblock \bibinfo{title}{Active learning for structural reliability: Survey, general framework and benchmark},
\newblock \bibinfo{journal}{Structural Safety} \bibinfo{volume}{96} (\bibinfo{year}{2022}) \bibinfo{pages}{102174}.
\bibitem[{Papakonstantinou et~al.(2023)Papakonstantinou, Nikbakht, and Eshra}]{Papakon2023HMCMC}
\bibinfo{author}{K.~G. Papakonstantinou}, \bibinfo{author}{H.~Nikbakht}, \bibinfo{author}{E.~Eshra},
\newblock \bibinfo{title}{Hamiltonian {MCMC} methods for estimating rare events probabilities in high-dimensional problems},
\newblock \bibinfo{journal}{Probabilistic Engineering Mechanics} \bibinfo{volume}{74} (\bibinfo{year}{2023}) \bibinfo{pages}{103485}.
\bibitem[{Hoffman et~al.(2021)Hoffman, Radul, and Sountsov}]{hoffman2021adaptive}
\bibinfo{author}{M.~Hoffman}, \bibinfo{author}{A.~Radul}, \bibinfo{author}{P.~Sountsov},
\newblock \bibinfo{title}{An adaptive-{MCMC} scheme for setting trajectory lengths in {Hamiltonian Monte Carlo}},
\newblock in: \bibinfo{booktitle}{International Conference on Artificial Intelligence and Statistics}, \bibinfo{organization}{PMLR}, \bibinfo{year}{2021}, pp. \bibinfo{pages}{3907--3915}.
\bibitem[{Duane et~al.(1987)Duane, Kennedy, Pendleton, and Roweth}]{duane1987hybrid}
\bibinfo{author}{S.~Duane}, \bibinfo{author}{A.~D. Kennedy}, \bibinfo{author}{B.~J. Pendleton}, \bibinfo{author}{D.~Roweth},
\newblock \bibinfo{title}{Hybrid {Monte Carlo}},
\newblock \bibinfo{journal}{Physics {Letters} B} \bibinfo{volume}{195} (\bibinfo{year}{1987}) \bibinfo{pages}{216--222}.
\bibitem[{Zhang and Sutton(2011)}]{zhang2011quasi}
\bibinfo{author}{Y.~Zhang}, \bibinfo{author}{C.~A. Sutton},
\newblock \bibinfo{title}{{Quasi-Newton} methods for {Markov Chain Monte Carlo}},
\newblock in: \bibinfo{booktitle}{Advances in Neural Information Processing Systems}, \bibinfo{year}{2011}, pp. \bibinfo{pages}{2393--2401}.
\bibitem[{Fu et~al.(2016)Fu, Luo, and Zhang}]{fu2016quasi}
\bibinfo{author}{T.~Fu}, \bibinfo{author}{L.~Luo}, \bibinfo{author}{Z.~Zhang},
\newblock \bibinfo{title}{{Quasi-Newton} {Hamiltonian Monte Carlo}},
\newblock in: \bibinfo{booktitle}{{Uncertainty in Artificial Intelligence}}, \bibinfo{year}{2016}.
\bibitem[{Nikbakht and Papakonstantinou(2019)}]{nikbakht2019HMCMC}
\bibinfo{author}{H.~Nikbakht}, \bibinfo{author}{K.~G. Papakonstantinou},
\newblock \bibinfo{title}{A direct {Hamiltonian MCMC} approach for reliability estimation},
\newblock in: \bibinfo{booktitle}{3rd {International Conference on Uncertainty Quantification in Computational Sciences and Engineering}}, \bibinfo{year}{2019}.
\bibitem[{Ma et~al.(2015)Ma, Chen, and Fox}]{ma2015complete}
\bibinfo{author}{Y.-A. Ma}, \bibinfo{author}{T.~Chen}, \bibinfo{author}{E.~Fox},
\newblock \bibinfo{title}{A complete recipe for stochastic gradient {MCMC}},
\newblock in: \bibinfo{booktitle}{Advances in Neural Information Processing Systems}, \bibinfo{year}{2015}, pp. \bibinfo{pages}{2917--2925}.
\bibitem[{Hoffman and Gelman(2014)}]{hoffman2014no}
\bibinfo{author}{M.~D. Hoffman}, \bibinfo{author}{A.~Gelman},
\newblock \bibinfo{title}{The {No-U-turn} sampler: {Adaptively} setting path lengths in {Hamiltonian Monte Carlo}},
\newblock \bibinfo{journal}{Journal of Machine Learning Research} \bibinfo{volume}{15} (\bibinfo{year}{2014}) \bibinfo{pages}{1593--1623}.
\bibitem[{Nesterov(2009)}]{nesterov2009primal}
\bibinfo{author}{Y.~Nesterov},
\newblock \bibinfo{title}{Primal-dual subgradient methods for convex problems},
\newblock \bibinfo{journal}{Mathematical Programming} \bibinfo{volume}{120} (\bibinfo{year}{2009}) \bibinfo{pages}{221--259}.
\bibitem[{Kingma and Ba(2014)}]{kingma2014adam}
\bibinfo{author}{D.~P. Kingma}, \bibinfo{author}{J.~Ba},
\newblock \bibinfo{title}{Adam: {A} method for stochastic optimization},
\newblock \bibinfo{journal}{arXiv preprint arXiv:1412.6980}  (\bibinfo{year}{2014}).
\bibitem[{Ang et~al.(1992)Ang, Ang, and Tang}]{ang1992optimal}
\bibinfo{author}{G.~L. Ang}, \bibinfo{author}{A.~H.-S. Ang}, \bibinfo{author}{W.~H. Tang},
\newblock \bibinfo{title}{Optimal importance-sampling density estimator},
\newblock \bibinfo{journal}{Journal of Engineering Mechanics} \bibinfo{volume}{118} (\bibinfo{year}{1992}) \bibinfo{pages}{1146--1163}.
\bibitem[{Schu{\"e}ller and Stix(1987)}]{schueller1987critical}
\bibinfo{author}{G.~I. Schu{\"e}ller}, \bibinfo{author}{R.~Stix},
\newblock \bibinfo{title}{A critical appraisal of methods to determine failure probabilities},
\newblock \bibinfo{journal}{Structural Safety} \bibinfo{volume}{4} (\bibinfo{year}{1987}) \bibinfo{pages}{293--309}.
\bibitem[{Kurtz and Song(2013)}]{kurtz2013cross}
\bibinfo{author}{N.~Kurtz}, \bibinfo{author}{J.~Song},
\newblock \bibinfo{title}{Cross-entropy-based adaptive importance sampling using {Gaussian} mixture},
\newblock \bibinfo{journal}{Structural Safety} \bibinfo{volume}{42} (\bibinfo{year}{2013}) \bibinfo{pages}{35--44}.
\bibitem[{Wang and Song(2016)}]{wang2016cross}
\bibinfo{author}{Z.~Wang}, \bibinfo{author}{J.~Song},
\newblock \bibinfo{title}{Cross-entropy-based adaptive importance sampling using von {Mises-Fisher} mixture for high dimensional reliability analysis},
\newblock \bibinfo{journal}{Structural Safety} \bibinfo{volume}{59} (\bibinfo{year}{2016}) \bibinfo{pages}{42--52}.
\bibitem[{Papaioannou et~al.(2019)Papaioannou, Geyer, and Straub}]{papaioannou2019improved}
\bibinfo{author}{I.~Papaioannou}, \bibinfo{author}{S.~Geyer}, \bibinfo{author}{D.~Straub},
\newblock \bibinfo{title}{Improved cross entropy-based importance sampling with a flexible mixture model},
\newblock \bibinfo{journal}{Reliability Engineering \& System Safety} \bibinfo{volume}{191} (\bibinfo{year}{2019}) \bibinfo{pages}{106564}.
\bibitem[{Uribe et~al.(2021)Uribe, Papaioannou, Marzouk, and Straub}]{uribe2021cross}
\bibinfo{author}{F.~Uribe}, \bibinfo{author}{I.~Papaioannou}, \bibinfo{author}{Y.~M. Marzouk}, \bibinfo{author}{D.~Straub},
\newblock \bibinfo{title}{Cross-entropy-based importance sampling with failure-informed dimension reduction for rare event simulation},
\newblock \bibinfo{journal}{SIAM/ASA Journal on Uncertainty Quantification} \bibinfo{volume}{9} (\bibinfo{year}{2021}) \bibinfo{pages}{818--847}.
\bibitem[{Demange-Chryst et~al.(2023)Demange-Chryst, Bachoc, Morio, and Krauth}]{demange2023variational}
\bibinfo{author}{J.~Demange-Chryst}, \bibinfo{author}{F.~Bachoc}, \bibinfo{author}{J.~Morio}, \bibinfo{author}{T.~Krauth},
\newblock \bibinfo{title}{Variational autoencoder with weighted samples for high-dimensional non-parametric adaptive importance sampling},
\newblock \bibinfo{journal}{arXiv preprint arXiv:2310.09194}  (\bibinfo{year}{2023}).
\bibitem[{Tong and Stadler(2023)}]{tong2023large}
\bibinfo{author}{S.~Tong}, \bibinfo{author}{G.~Stadler},
\newblock \bibinfo{title}{Large deviation theory-based adaptive importance sampling for rare events in high dimensions},
\newblock \bibinfo{journal}{SIAM/ASA Journal on Uncertainty Quantification} \bibinfo{volume}{11} (\bibinfo{year}{2023}) \bibinfo{pages}{788--813}.
\bibitem[{Chiron et~al.(2023)Chiron, Genest, Morio, and Dubreuil}]{chiron2023failure}
\bibinfo{author}{M.~Chiron}, \bibinfo{author}{C.~Genest}, \bibinfo{author}{J.~Morio}, \bibinfo{author}{S.~Dubreuil},
\newblock \bibinfo{title}{Failure probability estimation through high-dimensional elliptical distribution modeling with multiple importance sampling},
\newblock \bibinfo{journal}{Reliability Engineering \& System Safety} \bibinfo{volume}{235} (\bibinfo{year}{2023}) \bibinfo{pages}{109238}.
\bibitem[{Au and Beck(1999)}]{au1999new}
\bibinfo{author}{S.-K. Au}, \bibinfo{author}{J.~L. Beck},
\newblock \bibinfo{title}{A new adaptive importance sampling scheme for reliability calculations},
\newblock \bibinfo{journal}{Structural {Safety}} \bibinfo{volume}{21} (\bibinfo{year}{1999}) \bibinfo{pages}{135--158}.
\bibitem[{Au and Beck(2003)}]{au2003important}
\bibinfo{author}{S.-K. Au}, \bibinfo{author}{J.~L. Beck},
\newblock \bibinfo{title}{Important sampling in high dimensions},
\newblock \bibinfo{journal}{Structural {Safety}} \bibinfo{volume}{25} (\bibinfo{year}{2003}) \bibinfo{pages}{139--163}.
\bibitem[{Tabandeh et~al.(2022)Tabandeh, Jia, and Gardoni}]{tabandeh2022review}
\bibinfo{author}{A.~Tabandeh}, \bibinfo{author}{G.~Jia}, \bibinfo{author}{P.~Gardoni},
\newblock \bibinfo{title}{A review and assessment of importance sampling methods for reliability analysis},
\newblock \bibinfo{journal}{Structural Safety} \bibinfo{volume}{97} (\bibinfo{year}{2022}) \bibinfo{pages}{102216}.
\bibitem[{Ehre et~al.(2023)Ehre, Papaioannou, and Straub}]{ehre2023stein}
\bibinfo{author}{M.~Ehre}, \bibinfo{author}{I.~Papaioannou}, \bibinfo{author}{D.~Straub},
\newblock \bibinfo{title}{Stein variational rare event simulation},
\newblock \bibinfo{journal}{arXiv preprint arXiv:2308.04971}  (\bibinfo{year}{2023}).
\bibitem[{Cui et~al.(2024)Cui, Dolgov, and Scheichl}]{cui2024deep}
\bibinfo{author}{T.~Cui}, \bibinfo{author}{S.~Dolgov}, \bibinfo{author}{R.~Scheichl},
\newblock \bibinfo{title}{Deep importance sampling using tensor trains with application to a priori and a posteriori rare events},
\newblock \bibinfo{journal}{SIAM Journal on Scientific Computing} \bibinfo{volume}{46} (\bibinfo{year}{2024}) \bibinfo{pages}{C1--C29}.
\bibitem[{Del~Moral et~al.(2006)Del~Moral, Doucet, and Jasra}]{del2006sequential}
\bibinfo{author}{P.~Del~Moral}, \bibinfo{author}{A.~Doucet}, \bibinfo{author}{A.~Jasra},
\newblock \bibinfo{title}{Sequential {Monte Carlo} samplers},
\newblock \bibinfo{journal}{Journal of the {Royal Statistical Society: Series B} (Statistical Methodology)} \bibinfo{volume}{68} (\bibinfo{year}{2006}) \bibinfo{pages}{411--436}.
\bibitem[{Papaioannou et~al.(2016)Papaioannou, Papadimitriou, and Straub}]{papaioannou2016sequential}
\bibinfo{author}{I.~Papaioannou}, \bibinfo{author}{C.~Papadimitriou}, \bibinfo{author}{D.~Straub},
\newblock \bibinfo{title}{Sequential importance sampling for structural reliability analysis},
\newblock \bibinfo{journal}{Structural Safety} \bibinfo{volume}{62} (\bibinfo{year}{2016}) \bibinfo{pages}{66--75}.
\bibitem[{Bennett(1976)}]{bennett1976efficient}
\bibinfo{author}{C.~H. Bennett},
\newblock \bibinfo{title}{Efficient estimation of free energy differences from {Monte Carlo} data},
\newblock \bibinfo{journal}{Journal of Computational Physics} \bibinfo{volume}{22} (\bibinfo{year}{1976}) \bibinfo{pages}{245--268}.
\bibitem[{Meng and Wong(1996)}]{meng1996simulating}
\bibinfo{author}{X.-L. Meng}, \bibinfo{author}{W.~H. Wong},
\newblock \bibinfo{title}{Simulating ratios of normalizing constants via a simple identity: a theoretical exploration},
\newblock \bibinfo{journal}{Statistica Sinica}  (\bibinfo{year}{1996}) \bibinfo{pages}{831--860}.
\bibitem[{Sinha et~al.(2020)Sinha, O'Kelly, Tedrake, and Duchi}]{sinha2020neural}
\bibinfo{author}{A.~Sinha}, \bibinfo{author}{M.~O'Kelly}, \bibinfo{author}{R.~Tedrake}, \bibinfo{author}{J.~C. Duchi},
\newblock \bibinfo{title}{Neural bridge sampling for evaluating safety-critical autonomous systems},
\newblock \bibinfo{journal}{Advances in Neural Information Processing Systems} \bibinfo{volume}{33} (\bibinfo{year}{2020}) \bibinfo{pages}{6402--6416}.
\bibitem[{Gelman and Meng(1998)}]{gelman1998simulating}
\bibinfo{author}{A.~Gelman}, \bibinfo{author}{X.-L. Meng},
\newblock \bibinfo{title}{Simulating normalizing constants: {From} importance sampling to bridge sampling to path sampling},
\newblock \bibinfo{journal}{Statistical Science}  (\bibinfo{year}{1998}) \bibinfo{pages}{163--185}.
\bibitem[{Johansen et~al.(2006)Johansen, Del~Moral, and Doucet}]{johansen2005sequential}
\bibinfo{author}{A.~M. Johansen}, \bibinfo{author}{P.~Del~Moral}, \bibinfo{author}{A.~Doucet},
\newblock \bibinfo{title}{Sequential {Monte Carlo} samplers for rare events},
\newblock in: \bibinfo{booktitle}{6th International Workshop on Rare Event Simulation}, \bibinfo{year}{2006}, pp. \bibinfo{pages}{256--267}.
\bibitem[{Xian and Wang(2024)}]{xian2024relaxation}
\bibinfo{author}{J.~Xian}, \bibinfo{author}{Z.~Wang},
\newblock \bibinfo{title}{Relaxation-based importance sampling for structural reliability analysis},
\newblock \bibinfo{journal}{Structural Safety} \bibinfo{volume}{106} (\bibinfo{year}{2024}) \bibinfo{pages}{102393}.
\bibitem[{McLachlan and Peel(2000)}]{mclachlan2000finite}
\bibinfo{author}{G.~McLachlan}, \bibinfo{author}{D.~Peel}, \bibinfo{title}{Finite mixture models}, \bibinfo{publisher}{Wiley}, \bibinfo{year}{2000}.
\bibitem[{Dillon et~al.(2017)Dillon, Langmore, Tran, Brevdo, Vasudevan, Moore, Patton, Alemi, Hoffman, and Saurous}]{dillon2017tensorflow}
\bibinfo{author}{J.~V. Dillon}, \bibinfo{author}{I.~Langmore}, \bibinfo{author}{D.~Tran}, \bibinfo{author}{E.~Brevdo}, \bibinfo{author}{S.~Vasudevan}, \bibinfo{author}{D.~Moore}, \bibinfo{author}{B.~Patton}, \bibinfo{author}{A.~Alemi}, \bibinfo{author}{M.~Hoffman}, \bibinfo{author}{R.~A. Saurous},
\newblock \bibinfo{title}{Tensorflow distributions},
\newblock \bibinfo{journal}{arXiv preprint arXiv:1711.10604}  (\bibinfo{year}{2017}).
\bibitem[{{TensorFlow Probability}(2024)}]{tensorflow2024ess}
\bibinfo{author}{{TensorFlow Probability}}, \bibinfo{title}{Effective sample size}, \bibinfo{howpublished}{\url{https://www.tensorflow.org/probability/api_docs/python/tfp/mcmc/effective_sample_size}}, \bibinfo{year}{2024}. \bibinfo{note}{Accessed: 2024-10-03}.
\bibitem[{Alder and Wainwright(1959)}]{alder1959studies}
\bibinfo{author}{B.~J. Alder}, \bibinfo{author}{T.~E. Wainwright},
\newblock \bibinfo{title}{Studies in molecular dynamics. {I}. {General} method},
\newblock \bibinfo{journal}{The Journal of Chemical Physics} \bibinfo{volume}{31} (\bibinfo{year}{1959}) \bibinfo{pages}{459--466}.
\bibitem[{Betancourt(2017)}]{betancourt2017conceptual}
\bibinfo{author}{M.~Betancourt},
\newblock \bibinfo{title}{A conceptual introduction to {Hamiltonian Monte Carlo}},
\newblock \bibinfo{journal}{arXiv preprint arXiv:1701.02434}  (\bibinfo{year}{2017}).
\bibitem[{Hirt et~al.(2021)Hirt, Titsias, and Dellaportas}]{hirt2021entropy}
\bibinfo{author}{M.~Hirt}, \bibinfo{author}{M.~Titsias}, \bibinfo{author}{P.~Dellaportas},
\newblock \bibinfo{title}{Entropy-based adaptive {Hamiltonian Monte Carlo}},
\newblock \bibinfo{journal}{Advances in Neural Information Processing Systems} \bibinfo{volume}{34} (\bibinfo{year}{2021}) \bibinfo{pages}{28482--28495}.
\bibitem[{Qi and Minka(2002)}]{qi2002hessian}
\bibinfo{author}{Y.~Qi}, \bibinfo{author}{T.~P. Minka},
\newblock \bibinfo{title}{{Hessian-based Markov chain Monte-Carlo algorithms}},
\newblock in: \bibinfo{booktitle}{First Cape Cod Workshop on Monte Carlo Methods}, volume~\bibinfo{volume}{2}, \bibinfo{organization}{Massachusetts, USA}, \bibinfo{year}{2002}.
\bibitem[{Martin et~al.(2012)Martin, Wilcox, Burstedde, and Ghattas}]{martin2012stochastic}
\bibinfo{author}{J.~Martin}, \bibinfo{author}{L.~C. Wilcox}, \bibinfo{author}{C.~Burstedde}, \bibinfo{author}{O.~Ghattas},
\newblock \bibinfo{title}{A stochastic {Newton MCMC} method for large-scale statistical inverse problems with application to seismic inversion},
\newblock \bibinfo{journal}{SIAM Journal on Scientific Computing} \bibinfo{volume}{34} (\bibinfo{year}{2012}) \bibinfo{pages}{A1460--A1487}.
\bibitem[{Simsekli et~al.(2016)Simsekli, Badeau, Cemgil, and Richard}]{simsekli2016stochastic}
\bibinfo{author}{U.~Simsekli}, \bibinfo{author}{R.~Badeau}, \bibinfo{author}{T.~Cemgil}, \bibinfo{author}{G.~Richard},
\newblock \bibinfo{title}{{Stochastic quasi-Newton Langevin Monte Carlo}},
\newblock in: \bibinfo{booktitle}{International Conference on Machine Learning}, \bibinfo{organization}{PMLR}, \bibinfo{year}{2016}, pp. \bibinfo{pages}{642--651}.
\bibitem[{Leimkuhler et~al.(2018)Leimkuhler, Matthews, and Weare}]{leimkuhler2018ensemble}
\bibinfo{author}{B.~Leimkuhler}, \bibinfo{author}{C.~Matthews}, \bibinfo{author}{J.~Weare},
\newblock \bibinfo{title}{Ensemble preconditioning for {Markov chain Monte Carlo} simulation},
\newblock \bibinfo{journal}{Statistics and Computing} \bibinfo{volume}{28} (\bibinfo{year}{2018}) \bibinfo{pages}{277--290}.
\bibitem[{Brodlie et~al.(1973)Brodlie, Gourlay, and Greenstadt}]{brodlie1973rank}
\bibinfo{author}{K.~W. Brodlie}, \bibinfo{author}{A.~Gourlay}, \bibinfo{author}{J.~Greenstadt},
\newblock \bibinfo{title}{Rank-one and rank-two corrections to positive definite matrices expressed in product form},
\newblock \bibinfo{journal}{IMA Journal of Applied Mathematics} \bibinfo{volume}{11} (\bibinfo{year}{1973}) \bibinfo{pages}{73--82}.
\bibitem[{Gilks et~al.(1994)Gilks, Roberts, and George}]{gilks1994adaptive}
\bibinfo{author}{W.~R. Gilks}, \bibinfo{author}{G.~O. Roberts}, \bibinfo{author}{E.~I. George},
\newblock \bibinfo{title}{Adaptive direction sampling},
\newblock \bibinfo{journal}{Journal of the Royal Statistical Society: Series D (The Statistician)} \bibinfo{volume}{43} (\bibinfo{year}{1994}) \bibinfo{pages}{179--189}.
\bibitem[{Nocedal and Wright(2006)}]{nocedal2006numerical}
\bibinfo{author}{J.~Nocedal}, \bibinfo{author}{S.~J. Wright},
\newblock \bibinfo{title}{Numerical optimization},
\newblock \bibinfo{journal}{Springer}  (\bibinfo{year}{2006}).
\bibitem[{Liu and Nocedal(1989)}]{liu1989limited}
\bibinfo{author}{D.~C. Liu}, \bibinfo{author}{J.~Nocedal},
\newblock \bibinfo{title}{On the limited memory {BFGS} method for large scale optimization},
\newblock \bibinfo{journal}{Mathematical Programming} \bibinfo{volume}{45} (\bibinfo{year}{1989}) \bibinfo{pages}{503--528}.
\bibitem[{Beskos et~al.(2013)Beskos, Pillai, Roberts, Sanz-Serna, and Stuart}]{beskos2013optimal}
\bibinfo{author}{A.~Beskos}, \bibinfo{author}{N.~Pillai}, \bibinfo{author}{G.~Roberts}, \bibinfo{author}{J.-M. Sanz-Serna}, \bibinfo{author}{A.~Stuart},
\newblock \bibinfo{title}{Optimal tuning of the hybrid {Monte Carlo} algorithm},
\newblock \bibinfo{journal}{Bernoulli} \bibinfo{volume}{19} (\bibinfo{year}{2013}) \bibinfo{pages}{1501--1534}.
\bibitem[{Pasarica and Gelman(2010)}]{pasarica2010adaptively}
\bibinfo{author}{C.~Pasarica}, \bibinfo{author}{A.~Gelman},
\newblock \bibinfo{title}{Adaptively scaling the {Metropolis} algorithm using expected squared jumped distance},
\newblock \bibinfo{journal}{Statistica Sinica}  (\bibinfo{year}{2010}) \bibinfo{pages}{343--364}.
\bibitem[{Wang et~al.(2013)Wang, Mohamed, and Freitas}]{wang2013adaptive}
\bibinfo{author}{Z.~Wang}, \bibinfo{author}{S.~Mohamed}, \bibinfo{author}{N.~Freitas},
\newblock \bibinfo{title}{Adaptive {Hamiltonian} and {Riemann} manifold {Monte Carlo}},
\newblock in: \bibinfo{booktitle}{International Conference on Machine Learning}, \bibinfo{year}{2013}, pp. \bibinfo{pages}{1462--1470}.
\bibitem[{Wu et~al.(2018)Wu, Stoehr, and Robert}]{wu2018faster}
\bibinfo{author}{C.~Wu}, \bibinfo{author}{J.~Stoehr}, \bibinfo{author}{C.~P. Robert},
\newblock \bibinfo{title}{Faster {Hamiltonian Monte Carlo} by learning leapfrog scale},
\newblock \bibinfo{journal}{arXiv preprint arXiv:1810.04449}  (\bibinfo{year}{2018}).
\bibitem[{Chen et~al.(2020)Chen, Dwivedi, Wainwright, and Yu}]{chen2020fast}
\bibinfo{author}{Y.~Chen}, \bibinfo{author}{R.~Dwivedi}, \bibinfo{author}{M.~J. Wainwright}, \bibinfo{author}{B.~Yu},
\newblock \bibinfo{title}{Fast mixing of {Metropolized Hamiltonian Monte Carlo}: {Benefits} of multi-step gradients},
\newblock \bibinfo{journal}{The Journal of Machine Learning Research} \bibinfo{volume}{21} (\bibinfo{year}{2020}) \bibinfo{pages}{3647--3717}.
\bibitem[{Hoffman and Sountsov(2022)}]{hoffman2022tuning}
\bibinfo{author}{M.~D. Hoffman}, \bibinfo{author}{P.~Sountsov},
\newblock \bibinfo{title}{Tuning-free generalized {Hamiltonian Monte Carlo}},
\newblock in: \bibinfo{booktitle}{International Conference on Artificial Intelligence and Statistics}, \bibinfo{organization}{PMLR}, \bibinfo{year}{2022}, pp. \bibinfo{pages}{7799--7813}.
\bibitem[{Roberts and Stramer(2002)}]{roberts2002langevin}
\bibinfo{author}{G.~O. Roberts}, \bibinfo{author}{O.~Stramer},
\newblock \bibinfo{title}{Langevin diffusions and {Metropolis-Hastings} algorithms},
\newblock \bibinfo{journal}{Methodology and Computing in Applied Probability} \bibinfo{volume}{4} (\bibinfo{year}{2002}) \bibinfo{pages}{337--357}.
\bibitem[{Papakonstantinou et~al.(2022)Papakonstantinou, Nikbakht, and Eshra}]{PapakonICOSSAR2022}
\bibinfo{author}{K.~G. Papakonstantinou}, \bibinfo{author}{H.~Nikbakht}, \bibinfo{author}{E.~Eshra},
\newblock \bibinfo{title}{{Quasi-Newton Hamiltonian MCMC sampling for reliability estimation in high-dimensional non-Gaussian spaces}},
\newblock in: \bibinfo{booktitle}{13th {International Conference on Structural Safety \& Reliability (ICOSSAR), Shanghai, China}}, \bibinfo{year}{2022}.
\bibitem[{Nataf(1962)}]{nataf1962determination}
\bibinfo{author}{A.~Nataf},
\newblock \bibinfo{title}{Determination des distribution don't les marges sont donnees},
\newblock \bibinfo{journal}{Comptes rendus de l'Acad{\'e}mie des Sciences} \bibinfo{volume}{225} (\bibinfo{year}{1962}) \bibinfo{pages}{42--43}.
\bibitem[{Lebrun and Dutfoy(2009)}]{lebrun2009innovating}
\bibinfo{author}{R.~Lebrun}, \bibinfo{author}{A.~Dutfoy},
\newblock \bibinfo{title}{An innovating analysis of the {Nataf} transformation from the copula viewpoint},
\newblock \bibinfo{journal}{Probabilistic Engineering Mechanics} \bibinfo{volume}{24} (\bibinfo{year}{2009}) \bibinfo{pages}{312--320}.
\bibitem[{Pagani et~al.(2019)Pagani, Wiegand, and Nadarajah}]{pagani2019n}
\bibinfo{author}{F.~Pagani}, \bibinfo{author}{M.~Wiegand}, \bibinfo{author}{S.~Nadarajah},
\newblock \bibinfo{title}{An n-dimensional {Rosenbrock distribution} for {MCMC} testing},
\newblock \bibinfo{journal}{arXiv preprint arXiv:1903.09556}  (\bibinfo{year}{2019}).
\bibitem[{Neal(2003)}]{neal2003slice}
\bibinfo{author}{R.~M. Neal},
\newblock \bibinfo{title}{Slice sampling},
\newblock \bibinfo{journal}{The Annals of Statistics} \bibinfo{volume}{31} (\bibinfo{year}{2003}) \bibinfo{pages}{705--767}.
\bibitem[{Papakonstantinou et~al.(2023)Papakonstantinou, Eshra, and Nikbakht}]{PapakonICASP2023}
\bibinfo{author}{K.~G. Papakonstantinou}, \bibinfo{author}{E.~Eshra}, \bibinfo{author}{H.~Nikbakht},
\newblock \bibinfo{title}{Hamiltonian {MCMC} based framework for time-variant rare event uncertainty quantification},
\newblock in: \bibinfo{booktitle}{14th { International Conference on Applications of Statistics and Probability in Civil Engineering (ICASP), Dublin, Ireland}}, \bibinfo{year}{2023}.

\end{thebibliography}
\onecolumn
\section*{Appendix A. Connections between single-step QNp-HMCMC implementation and preconditioned Metropolis adjusted Langevin algorithm (MALA)}\label{Connection_MALA_Appendix}
\begin{proof}[Proof of \Cref{MALA_proposal_equivalence}]
First, HMCMC with single-step implementation can be written based on \cref{leapfrog} as:
\begin{equation}
\begin{aligned}\label{connection_1}
\tilde{\bm{x}} = \bm{x}\textsubscript{t} +\varepsilon\,\mathbf{M}^{-1}\big(\bm{z}\textsubscript{t} + \dfrac{\varepsilon}{2} 
\nabla_{\bm{X}}\mathcal{L}(\bm{x}\textsubscript{t})\big)= \bm{x}\textsubscript{t} + \dfrac{\varepsilon^2}{2}\,\mathbf{M}^{-1} \nabla_{\bm{X}}\mathcal{L}(\bm{x}\textsubscript{t})+\varepsilon\,\mathbf{M}^{-1}\bm{z}\textsubscript{t} 
\end{aligned}
\end{equation}
Since $\bm{Z} \sim \mathcal{N}(\textbf{\textit{0}}, \mathbf{M})$, it can be expressed as $\bm{Z}= \mathbf{M}^{1/2} \bm{Z}^\prime$ where $\bm{Z}^\prime \sim \mathcal{N}(\textbf{\textit{0}}, \mathbf{I})$, and $\mathbf{M}^{1/2}$ is computed using Cholesky decomposition, $\mathbf{M}=\mathbf{M}^{1/2}(\mathbf{M}^{1/2})^T$. \cref{connection_1} can then be written as:
\begin{equation}
\begin{aligned}\label{connection_2}
\tilde{\bm{x}} = \bm{x}\textsubscript{t} + \dfrac{\varepsilon^2}{2}\,\mathbf{M}^{-1} \nabla_{\bm{X}}\mathcal{L}(\bm{x}\textsubscript{t})+\varepsilon\,\mathbf{M}^{-1}\mathbf{M}^{1/2}\bm{z}^\prime_\text{t}
= \bm{x}\textsubscript{t} + \dfrac{\varepsilon^2}{2}\,\mathbf{M}^{-1} \nabla_{\bm{X}}\mathcal{L}(\bm{x}\textsubscript{t})+\varepsilon\,(\mathbf{M}^{-1/2})^T\bm{z}^\prime_\text{t}
\end{aligned}
\end{equation}
The right side is simplified following the Cholesky decomposition $\mathbf{M}^{-1}=\big(\mathbf{M}^{1/2}(\mathbf{M}^{1/2})^T\big)^{-1}=(\mathbf{M}^{-1/2})^T\mathbf{M}^{-1/2}$, and then
$\mathbf{M}^{-1}\mathbf{M}^{1/2}=(\mathbf{M}^{-1/2})^T\mathbf{M}^{-1/2}\mathbf{M}^{1/2}=(\mathbf{M}^{-1/2})^T$. Comparing \cref{connection_2} with \cref{LMC}, it is clear that HMCMC with single-step implementation is equivalent to a preconditioned Langevin Monte Carlo (LMC) where its preconditioning matrix equals the inverse of the mass matrix, $\mathbf{A}=\mathbf{M}^{-1}$.\par

The connection between the skew-symmetric preconditioned dynamics in the burn-in stage of single-step QNp-HMCMC and the preconditioned LMC can be studied similarly. The adaptive scheme in the single-step QNp-HMCMC can be written based on \cref{leapfrog_QNp} as:
\begin{equation}
\begin{aligned}\label{connection_QNp}
\tilde{\bm{x}} &= \bm{x}\textsubscript{t} +\varepsilon\,\mathbf{W}_\text{t}\mathbf{M}^{-1}\big(\bm{z}\textsubscript{t} + \dfrac{\varepsilon}{2} \mathbf{W}_\text{t}\nabla_{\bm{X}}\mathcal{L}(\bm{x}\textsubscript{t})\big)= \bm{x}\textsubscript{t} + \dfrac{\varepsilon^2}{2}\,\mathbf{W}_\text{t}\mathbf{M}^{-1}\mathbf{W}_\text{t} \nabla_{\bm{X}}\mathcal{L}(\bm{x}\textsubscript{t})+\varepsilon\,\mathbf{W}_\text{t}\mathbf{M}^{-1}\bm{z}\textsubscript{t} \\&=\bm{x}\textsubscript{t} + \dfrac{\varepsilon^2}{2}\,\mathbf{W}_\text{t}\mathbf{M}^{-1}\mathbf{W}_\text{t} \nabla_{\bm{X}}\mathcal{L}(\bm{x}\textsubscript{t})+\varepsilon\,\mathbf{W}_\text{t}(\mathbf{M}^{-1/2})^T\bm{z}^\prime_\text{t}
\end{aligned}
\end{equation}
Comparing \cref{connection_QNp} and \cref{LMC} shows that the burn-in stage in QNp-HMCM with a single-step implementation is equivalent to a preconditioned LMC with a preconditioning matrix $\mathbf{A}_\text{t}=\mathbf{W}_\text{t}\mathbf{M}^{-1}\mathbf{W}_\text{t}$, where $\mathbf{M}=\mathbf{I}$ in this work, at this burn-in stage.
\end{proof}

\begin{proof}[Proof of \Cref{MALA_acceptance_equivalence}]
The proposal of Langevin Monte Carlo in \cref{LMC} can be viewed as a Metropolis-Hastings update in which $\tilde{\bm{x}}$ is proposed from a Gaussian distribution, $\mathcal{N}\big( \bm{x}\textsubscript{t} +\dfrac{\varepsilon^2}{2}\,\mathbf{A}\nabla_{\bm{X}}\mathcal{L}(\bm{x}\textsubscript{t}),\,\varepsilon^2\mathbf{A}\big)$, and the acceptance probability of Metropolis adjusted Langevin algorithm (MALA) is then written as: 
\begin{equation}\label{acceptance_MALA}
\begin{aligned}
    \alpha&= \min\bigg\{1,\dfrac{\exp\big(\mathcal{L}(\tilde{\bm{x}})\big)\,\mathcal{N}\big(\bm{x}\textsubscript{t} \,\vert \tilde{\bm{x}} +\dfrac{\varepsilon^2}{2}\,\mathbf{A}\nabla_{\bm{X}}\mathcal{L}(\tilde{\bm{x}}),\,\varepsilon^2\mathbf{A}\big)}{\exp\big(\mathcal{L}(\bm{x}\textsubscript{t})\big)\,\mathcal{N}\big(\tilde{\bm{x}}\,\vert \bm{x}\textsubscript{t} +\dfrac{\varepsilon^2}{2}\,\mathbf{A}\nabla_{\bm{X}}\mathcal{L}(\bm{x}\textsubscript{t}),\,\varepsilon^2\mathbf{A}\big)}\bigg\}
    \\&= \min\bigg\{1,\dfrac{\exp\big(\mathcal{L}(\tilde{\bm{x}})\big)\,-\dfrac{1}{2\varepsilon^2}\big(\bm{x}\textsubscript{t}- \tilde{\bm{x}} -\dfrac{\varepsilon^2}{2}\,\mathbf{A}\nabla_{\bm{X}}\mathcal{L}(\tilde{\bm{x}})\big)^T\,\mathbf{A}^{-1}\,\big(\bm{x}\textsubscript{t}- \tilde{\bm{x}} -\dfrac{\varepsilon^2}{2}\,\mathbf{A}\nabla_{\bm{X}}\mathcal{L}(\tilde{\bm{x}})\big)}
    {\exp\big(\mathcal{L}(\bm{x}\textsubscript{t})\big)\,- \dfrac{1}{2\varepsilon^2}\big(\tilde{\bm{x}}-\bm{x}\textsubscript{t} -\dfrac{\varepsilon^2}{2}\,\mathbf{A}\nabla_{\bm{X}}\mathcal{L}(\bm{x}\textsubscript{t})\big)^T\,\mathbf{A}^{-1}\,\big(\tilde{\bm{x}}- \bm{x}\textsubscript{t}-\dfrac{\varepsilon^2}{2}\,\mathbf{A}\nabla_{\bm{X}}\mathcal{L}(\bm{x}\textsubscript{t})\big)}\bigg\}
\end{aligned}
\end{equation}

To check the equivalence of the acceptance probability of preconditioned MALA and single-step HMCMC, we write the resampled momentum $\bm{z}\textsubscript{t}$ and the simulated one $\tilde{\bm{z}}$  based on \cref{leapfrog} as:
\begin{equation} 
\bm{z}\textsubscript{t}=\bm{z}\textsubscript{t+$\varepsilon$/2} - \dfrac{\varepsilon}{2} \nabla_{\bm{X}}\mathcal{L}(\bm{x}\textsubscript{t}) = \dfrac{1}{\varepsilon}\mathbf{M}(\tilde{\bm{x}}-\bm{x}\textsubscript{t}) - \dfrac{\varepsilon}{2} \nabla_{\bm{X}}\mathcal{L}(\bm{x}\textsubscript{t})= \dfrac{1}{\varepsilon}\mathbf{M}\bigg(\tilde{\bm{x}}-\bm{x}\textsubscript{t}-\dfrac{\varepsilon^2}{2}\textbf{M}^{-1} \nabla_{\bm{X}}\mathcal{L}(\bm{x}\textsubscript{t})\bigg)
\end{equation}

\begin{equation}
\tilde{\bm{z}} = \bm{z}\textsubscript{t+$\varepsilon$/2} + \dfrac{\varepsilon}{2}\nabla_{\bm{X}}\mathcal{L}(\tilde{\bm{x}}) =\dfrac{1}{\varepsilon}\mathbf{M}(\tilde{\bm{x}}-\bm{x}\textsubscript{t})+ \dfrac{\varepsilon}{2}\nabla_{\bm{X}}\mathcal{L}(\tilde{\bm{x}})= -\dfrac{1}{\varepsilon}\mathbf{M}\bigg(\bm{x}\textsubscript{t}-\tilde{\bm{x}}-\dfrac{\varepsilon^2}{2}\textbf{M}^{-1} \nabla_{\bm{X}}\mathcal{L}(\tilde{\bm{x}})\bigg)
\end{equation} 
The acceptance probability of HMCMC can then be expressed as:
\begin{equation}\label{acceptance_HMC}
\begin{aligned}
    \alpha &=\min \bigg\{1,\dfrac{\exp\big(\mathcal{L}(\tilde{\bm{x}})-\dfrac{1}{2} \tilde{\bm{z}}^{T}\ \mathbf{M}^{-1}\tilde{\bm{z}}\big)}{\exp\big(\mathcal{L}(\bm{x}_{\text{t}})-\dfrac{1}{2}{\bm{z}_{\text{t}}}^{T}\ \mathbf{M}^{-1}\bm{z}_{\text{t}}\big)}\bigg\}\\&=\min \bigg\{1,\dfrac{\exp\bigg(\mathcal{L}(\tilde{\bm{x}})-\dfrac{1}{2\varepsilon^2}\big(\bm{x}\textsubscript{t}-\tilde{\bm{x}}-\dfrac{\varepsilon^2}{2}\textbf{M}^{-1} \nabla_{\bm{X}}\mathcal{L}(\tilde{\bm{x}})\big)^{T}\mathbf{M}^T\ \mathbf{M}^{-1}\mathbf{M}\big(\bm{x}\textsubscript{t}-\tilde{\bm{x}}-\dfrac{\varepsilon^2}{2}\textbf{M}^{-1} \nabla_{\bm{X}}\mathcal{L}(\tilde{\bm{x}})\big)\bigg)}{\exp\bigg(\mathcal{L}(\bm{x}_{\text{t}})-\dfrac{1}{2\varepsilon^2} \big(\tilde{\bm{x}}-\bm{x}\textsubscript{t}-\dfrac{\varepsilon^2}{2}\textbf{M}^{-1} \nabla_{\bm{X}}\mathcal{L}(\bm{x}\textsubscript{t})\big)^{T}\mathbf{M}^T\ \mathbf{M}^{-1}\mathbf{M}\big(\tilde{\bm{x}}-\bm{x}\textsubscript{t}-\dfrac{\varepsilon^2}{2}\textbf{M}^{-1} \nabla_{\bm{X}}\mathcal{L}(\bm{x}\textsubscript{t})\big)\bigg)}\bigg\}\\&=\min \bigg\{1,\dfrac{\exp\bigg(\mathcal{L}(\tilde{\bm{x}})-\dfrac{1}{2\varepsilon^2}\big(\bm{x}\textsubscript{t}-\tilde{\bm{x}}-\dfrac{\varepsilon^2}{2}\textbf{M}^{-1} \nabla_{\bm{X}}\mathcal{L}(\tilde{\bm{x}})\big)^{T}\mathbf{M}\big(\bm{x}\textsubscript{t}-\tilde{\bm{x}}-\dfrac{\varepsilon^2}{2}\textbf{M}^{-1} \nabla_{\bm{X}}\mathcal{L}(\tilde{\bm{x}})\big)\bigg)}{\exp\bigg(\mathcal{L}(\bm{x}_{\text{t}})-\dfrac{1}{2\varepsilon^2} \big(\tilde{\bm{x}}-\bm{x}\textsubscript{t}-\dfrac{\varepsilon^2}{2}\textbf{M}^{-1} \nabla_{\bm{X}}\mathcal{L}(\bm{x}\textsubscript{t})\big)^{T}\mathbf{M}\big(\tilde{\bm{x}}-\bm{x}\textsubscript{t}-\dfrac{\varepsilon^2}{2}\textbf{M}^{-1} \nabla_{\bm{X}}\mathcal{L}(\bm{x}\textsubscript{t})\big)\bigg)}\bigg\}
\end{aligned}
\end{equation}
Comparing \cref{acceptance_HMC} and \cref{acceptance_MALA}, it is now obvious that the acceptance ratio of the single-step HMCMC is equivalent to that of the preconditioned MALA when using a preconditioning matrix $\mathbf{A}=\mathbf{M}^{-1}$, with $\mathbf{M}$ being a symmetric positive definite mass matrix. 

The equivalence of the acceptance probabilities of the preconditioned MALA and the burn-in stage in QNp-HMCMC can also be shown by rewriting \cref{leapfrog_QNp} as:
\begin{equation} 
\begin{aligned}
    \bm{z}\textsubscript{t}&=\bm{z}\textsubscript{t+$\varepsilon$/2} -\dfrac{\varepsilon}{2} \mathbf{W}_\text{t}\nabla_{\bm{X}}\mathcal{L}(\bm{x}\textsubscript{t}) = \dfrac{1}{\varepsilon}\mathbf{M}\mathbf{W}_{\text{t}}^{-1}(\tilde{\bm{x}}-\bm{x}\textsubscript{t}) -\dfrac{\varepsilon}{2} \mathbf{W}_\text{t}\nabla_{\bm{X}}\mathcal{L}(\bm{x}\textsubscript{t}) \\&= \dfrac{1}{\varepsilon}\mathbf{M}\mathbf{W}_\text{t}^{-1}\bigg(\tilde{\bm{x}}-\bm{x}\textsubscript{t}-\dfrac{\varepsilon^2}{2}\mathbf{W}_\text{t}\textbf{M}^{-1}\mathbf{W}_\text{t}\nabla_{\bm{X}}\mathcal{L}(\bm{x}\textsubscript{t})\bigg)
\end{aligned}
\end{equation}

\begin{equation}
\begin{aligned}
    \tilde{\bm{z}} &= \bm{z}\textsubscript{t+$\varepsilon$/2} + \dfrac{\varepsilon}{2}\mathbf{W}_\text{t}\nabla_{\bm{X}}\mathcal{L}(\tilde{\bm{x}}) =\dfrac{1}{\varepsilon}\mathbf{M}\mathbf{W}_{\text{t}}^{-1}(\tilde{\bm{x}}-\bm{x}\textsubscript{t})+ \dfrac{\varepsilon}{2}\mathbf{W}_\text{t}\nabla_{\bm{X}}\mathcal{L}(\tilde{\bm{x}})\\&= -\dfrac{1}{\varepsilon}\mathbf{M}\mathbf{W}_\text{t}^{-1}\bigg(\bm{x}\textsubscript{t}-\tilde{\bm{x}}-\dfrac{\varepsilon^2}{2}\mathbf{W}_\text{t}\textbf{M}^{-1}\mathbf{W}_\text{t} \nabla_{\bm{X}}\mathcal{L}(\tilde{\bm{x}})\bigg)
\end{aligned}
\end{equation} 
The corresponding acceptance probability can then be expressed as:
\begin{equation}\label{acceptance_HMC_single}
\begin{aligned}
    &\alpha=\min \bigg\{1,\dfrac{\exp(\mathcal{L}(\tilde{\bm{x}}))}{\exp(\mathcal{L}(\bm{x}_{\text{t}}))}\cdot R^{\prime}\bigg\},\\&
    R^{\prime}=\dfrac{\exp\bigg(-\dfrac{1}{2\varepsilon^2}\big(\bm{x}\textsubscript{t}-\tilde{\bm{x}}-\dfrac{\varepsilon^2}{2}\mathbf{W}_\text{t}\textbf{M}^{-1}\mathbf{W}_\text{t} \nabla_{\bm{X}}\mathcal{L}(\tilde{\bm{x}})\big)^{T}(\mathbf{W}_\text{t}^{-1}\mathbf{M}\mathbf{W}_\text{t}^{-1})
    \big(\bm{x}\textsubscript{t}-\tilde{\bm{x}}-\dfrac{\varepsilon^2}{2}\mathbf{W}_\text{t}\textbf{M}^{-1}\mathbf{W}_\text{t} \nabla_{\bm{X}}\mathcal{L}(\tilde{\bm{x}})\big)\bigg)}
    {\exp\bigg(-\dfrac{1}{2\varepsilon^2} \big(\tilde{\bm{x}}-\bm{x}\textsubscript{t}-\dfrac{\varepsilon^2}{2}\mathbf{W}_\text{t}\textbf{M}^{-1}\mathbf{W}_\text{t}\nabla_{\bm{X}}\mathcal{L}(\bm{x}\textsubscript{t})\big)^{T}(\mathbf{W}_\text{t}^{-1}\mathbf{M}\mathbf{W}_\text{t}^{-1})
    \big(\tilde{\bm{x}}-\bm{x}\textsubscript{t}-\dfrac{\varepsilon^2}{2}\mathbf{W}_\text{t}\textbf{M}^{-1}\mathbf{W}_\text{t}\nabla_{\bm{X}}\mathcal{L}(\bm{x}\textsubscript{t})\big)\bigg)}
    \end{aligned}
\end{equation}
Consequently, the acceptance probability in the burn-in stage of QNp-HMCMC is equivalent to that of the preconditioned MALA with a preconditioning matrix $\mathbf{A}_\text{t}=\mathbf{W}_\text{t}\mathbf{M}^{-1}\mathbf{W}_\text{t}$, where $\mathbf{M}=\mathbf{I}$ in the burn-in stage in this work.
\end{proof}

\end{document}